\documentclass{jfm}

\usepackage{caption}
\usepackage{subcaption}

\usepackage{graphicx}
\usepackage{newtxtext}
\usepackage{newtxmath}
\usepackage{natbib}
\usepackage{hyperref}
\hypersetup{
    colorlinks = true,
    urlcolor   = blue,
    citecolor  = black,
}

\newcommand{\RomanNumeralCaps}[1]
\linenumbers

\title{Connecting transonic buffet with incompressible low-frequency oscillations on aerofoils}

\author{Pradeep Moise\aff{1,2}\corresp{\email{pradeep890@gmail.com}},
  Markus Zauner\aff{1}
 \and Neil Sandham\aff{1}}

\affiliation{\aff{1}University of Southampton, Southampton, Hampshire, SO17 1BJ, UK\aff{2}Indian Institute of Technology Kanpur, Kalyanpur, Kanpur, Uttar Pradesh, 208016, India}

\begin{document}
\maketitle

\begin{abstract}
Self-sustained low-frequency flow unsteadiness over rigid aerofoils in the transonic regime near stall is referred to as transonic buffet. Although the exact physical mechanisms underlying this phenomenon are unclear, it is generally assumed to be unique to the transonic regime. This assumption is shown to be incorrect here by performing large-eddy simulations of flow over a NACA0012 profile for a wide range of flow conditions. At zero incidence and sufficiently high freestream Mach numbers, $M$, transonic buffet occurs with shock waves present in the flow. However, self-sustained, periodic oscillations that occur at similar frequencies are observed at lower $M$ for which shock waves are absent and the entire flow field remains subsonic at all times. At higher incidence angles, the oscillations are sustained at progressively lower $M$. Oscillations were observed for $M$ as low as 0.3, where compressibility effects are small. A spectral proper orthogonal decomposition shows that the spatial structure of these oscillations (\textit{i.e.}, mode shapes) are essentially the same for all cases. These results indicate that buffet on aerofoils does not necessarily require the presence of shock waves. Furthermore, the trend seen with increasing incidence angles suggests that transonic buffet on aerofoils and low-frequency oscillations reported in the incompressible regime (Zaman \textit{et al.}, 1989, J. Fluid Mech., vol. 202, pp. 403--442) have similar origins. Thus, physical models which rely specifically on shock waves to explain the sustenance of transonic buffet are incorrect. The insights gained from this study could be useful in understanding the origins of ``transonic" buffet and reformulating mitigation strategies by shifting the focus away from shock waves. 
\end{abstract}


\section{Introduction}
\label{secIntro}


Transonic flows over wings can exhibit self-sustained, coherent, low-frequency oscillations referred to as transonic buffet \citep{Helmut1974}. These oscillations can lead to strong variations in lift and can cause structural fatigue, failure, or loss of control of aircraft. Thus, transonic buffet is detrimental to aircraft performance and can limit their flight envelope \citep{Lee2001}. For these reasons, transonic buffet has been extensively studied, with many studies focusing on the simple configuration of flow over aerofoils to understand its essential characteristics \citep[\textit{e.g.}][]{Crouch2007}. Although the exact physical mechanisms underlying this flow phenomenon on aerofoils remain unclear, one commonly-accepted feature which has been assumed to play an essential role in its origin is the shock wave in the transonic flow field. This assumption has prevailed from the first few reported observations of transonic buffet \citep{Hilton1947} till date \citep{Giannelis2017}. Indeed, \cite{McDevitt1985}, based on an extensive experimental study of this phenomenon noted that transonic buffet is ``shock-induced" (see their p. 4), while most physical models of transonic buffet require shock waves to be present in the flow field \citep[\textit{e.g.}][]{TijdemanReport, Gibb1988, Lee1990}. More importantly, this phenomenon has been identified only in the transonic regime (hence the epithet) with the fore-aft motion of shock waves considered as its defining feature. In this study, we show that coherent flow oscillations with the same features as transonic buffet can occur on aerofoils even when shock waves are absent and the entire spatial flow field remains subsonic at all times. Furthermore, by appropriately changing flow conditions, a link is established between transonic buffet and low-frequency oscillations (LFO) which occur in the incompressible regime at high incidence angles \citep{Zaman1989}, suggesting that the name `transonic buffet' is misleading.   

Several classifications of transonic buffet exist. One such is based on whether the flow is over a three-dimensional wing (or a swept infinite-wing section), as opposed to an unswept infinite-wing section. These are respectively referred to as wing and aerofoil buffet. Although there are some interesting differences between these two types (\textit{e.g.}, dominance of modes with strong spanwise variations, \citep{Crouch2019, Timme2020}), insights gained from the latter have been useful in understanding the former \citep[\textit{e.g.},][]{Crouch2007}. In this study, we focus only on aerofoil buffet. Another classification of interest here is that of Type I and Type II transonic buffet \citep{Giannelis2017}. Type I is associated with fore-aft motion of shock waves occurring on both sides of an aerofoil and is generally observed for symmetric aerofoils at zero incidence. By contrast, Type II is characterised by shock waves present only on the suction side with shock motion and flow oscillations of significant amplitude restricted to the same. It is generally observed at high incidence angles. This classification goes back to \citet{Lee2001}, where it was noted that ``There is some difference in the mechanisms of periodic shock motion between a lifting airfoil at incidence and a symmetrical one at zero incidence" and has been highlighted in several other studies \citep{Iovnovich2012, Giannelis2017}. Different mechanisms have been proposed to govern the two types \citep{Gibb1988, Lee1990}, but it was shown recently \citet{Moise2022, Moise2022Trip} that the dynamical features of the two types are similar. Note that most recent studies focus on Type II transonic buffet \citep{Fukushima2018, Garbaruk2021}, although there are several earlier studies that scrutinise Type I transonic buffet \citep{Gibb1988, McDevitt1976}. 

Transonic buffet is also categorised as laminar or turbulent transonic buffet based on the boundary layer characteristics at the shock wave's foot \citep{Brion2020, Dandois2018}. The former type is characterised by the boundary layer remaining laminar from the leading edge till approximately the foot of the shock wave, while the latter has the transition to turbulence occurring well upstream of the shock foot. Turbulent transonic buffet can be observed at high Reynolds numbers \citep{Lee1989}, when transition is triggered artificially \citep[\textit{e.g.}, boundary layer tripping,][]{Roos1980, Brion2020}, or when a turbulent viscosity model is assumed to be active in the case of Reynolds Averaged Navier Stokes (RANS) simulations \citep[\textit{e.g.}][]{Xiao2006a, Crouch2007, Sartor2015}. Most studies focus on this type due to its relevance to flows at high Reynolds numbers and/or to avoid difficulties in modelling natural transition in RANS simulations. By contrast, laminar transonic buffet remains relatively unexplored. Interest in studying laminar aerofoils with applications towards improving aircraft performance and reducing skin friction drag has led to several recent studies \citep{Dandois2018, Brion2020, Zauner2020PRF}. Although initially considered to have a distinct mechanism as opposed to turbulent transonic buffet, it has been shown recently in several studies \citep{Moise2022, Moise2022Trip, Zauner2022} that the essential spatio-temporal characteristics are the same for both types, suggesting similar underlying mechanisms. It was also shown in these studies that in contrast to a single shock wave that is commonly observed in transonic flows around aerofoils, multiple shock waves can develop if the freestream Reynolds number is sufficiently low and transition occurs naturally. Interestingly, it was shown that laminar transonic buffet can also occur in the latter situation with multiple shock waves.

An important insight into the mechanism underlying transonic buffet on aerofoils comes from \citet{Crouch2007}, where it was shown using a Reynolds-Averaged Navier-Stokes (RANS) framework that the flow becomes globally unstable under certain conditions leading to transonic buffet. However, the physical reasons behind the origins of this instability remain unresolved. In this regard, there are several models that attempt to further explain the flow physics which are compatible with a global linear instability framework \citep{Gibb1988, Lee1990, Hartmann2013}. The most popular among these appears to be the feedback loop model proposed in \citet{Lee1990} for Type II transonic buffet. Lee suggested that shock wave motion can lead to waves that travel downstream from the shock foot to the trailing edge along the boundary layer. These waves are scattered at the trailing edge and lead to upstream-travelling `Kutta' waves \citep{TijdemanReport} which in turn interact with the shock wave and induce its motion, leading to a self-sustained oscillation loop. The essential requirement for all these models is the presence of a shock wave in the flow field.

While shock-wave-based models of transonic buffet remain popular, there is evidence that suggests that shock waves might not be crucial for transonic buffet. Based on a sensitivity analysis of a transonic-buffet flow field, \citet{Paladini2019a} suggested that the shock wave plays only a secondary role and noted that any feedback loop sustaining the oscillations must exist within the separated boundary layer. While this study reports transonic buffet only in the presence of a shock wave, it can be inferred from other studies that there might be situations wherein oscillations resembling transonic buffet could occur even without a shock wave. For example, in the ``Type B" category of transonic buffet \citep{Tijdeman1980}, the shock wave vanishes during parts of the oscillation cycle. More importantly, in two-dimensional simulations of flow (laminar) around a NACA0012 aerofoil at zero incidence angle, \citet{Bouhadji2003} and \citet{Jones2006} have reported oscillations at subsonic conditions, although they have not explored the relationship between these oscillations and transonic buffet. Also, in transonic flows over infinite or three-dimensional swept wings, spanwise-varying unsteady features referred to as buffet cells have been reported \citep{Iovnovich2015, Timme2020}, which are closely associated with transonic buffet on aerofoils. Recently, these have been linked with stall cells that occur for similar flow conditions, but in the incompressible regime \citep{Plante2020}. 

Coherent low-frequency oscillations have also been observed in incompressible flows on aerofoils -- the eponymous LFO \citep{Zaman1989, Bragg1993, Sandham2008, Almutairi2010, Busquet2021}. Interestingly, Zaman \textit{et al.}, who were among the first to experimentally and numerically study this phenomenon, noted that these oscillations are hydrodynamic in nature, distinct from vortex shedding, and involve a quasi-periodic switching between stalled and unstalled conditions. Using panel methods coupled with integral boundary layer equations and a transition model, \citet{Sandham2008} showed that LFO arise due to interactions between the potential flow and boundary layer. LFO were also examined using direct numerical simulations in \citet{Almutairi2010} and several follow-up studies \citep[\textit{e.g.}][]{Almutairi2013, AlMutairi2017} by performing simulations at low freestream Mach numbers using the same compressible flow solver as the present study. By performing global linear stability analysis of RANS results, \citet{Iorio2016} have shown that LFO arise as unstable modes while \citet{Busquet2021} have studied hysteresis features. 
It is interesting to note that LFO has been observed for a wide range of freestream Reynolds numbers, $Re$, based on aerofoil chord. For example, \citet{Zaman1989} examined LFO at $Re \sim O(10^5)$, while \citet{Bragg1993}, \citet{Busquet2021} and \citet{Iorio2016} studied LFO at $Re \sim O(10^6)$, with the latter exploring a Reynolds number as high as $Re = 6\times 10^6$.
Although the characteristics of LFO, (i) low-frequency, (ii) occurrence close to stall conditions and a behaviour of periodic switching between stalled and unstalled states and (iii) origins as a global instability, are also characteristic of transonic buffet, the connection between the two phenomena has not been clearly explored. In \citet{Iorio2016}, which is the only study to examine both transonic buffet and LFO together, it was suggested that the two are distinct phenomena. However, the simulations of the two phenomena were carried out at flow conditions which were highly dissimilar, implying that direct comparisons of flow features might not be appropriate.

Motivated by these considerations, we perform here three-dimensional large-eddy simulations of the flow around the symmetric NACA0012 aerofoil and examine the coherent features of the flow using a spectral proper orthogonal decomposition. The simulations are performed for free-transition conditions at different flow conditions so as to examine laminar transonic buffet and oscillations resembling the same occurring under subsonic conditions. Furthermore, by varying the incidence angle, freestream Mach number and Reynolds number in small steps, we find the link between transonic buffet and LFO. The methodology adopted for this is provided in \S\ref{secMethod}. Results at a low freestream Reynolds number and zero incidence angle are presented in \S\ref{secNaCA}, following which connections are made to LFO in \S\ref{secLFO}. The implications of these results are discussed in \S\ref{secDisc}, and \S\ref{secConc} concludes the study. 

\section{Methodology}
\label{secMethod}
The methodology adopted here is similar to that used and extensively discussed in previous studies \citep{Zauner2020PRF, Moise2022, Zauner2022}, highlights of which are provided below. 

\subsection{Numerical simulations}
\label{subsecMethodLES}
The simulations are performed using SBLI, an in-house, scalable, high-order, multi-block, compressible flow solver with shock-capturing capabilities \citep{Yao2009}. This solver has been extensively used to study transonic buffet \citep{Zauner2019, Zauner2020PRF, Moise2022, Moise2022Trip, Zauner2022} and LFO \citep{Almutairi2010, Almutairi2013} for various flow conditions. Fourth-order and third-order finite-difference schemes are used for spatial and temporal discretisation, respectively. The spectral-error-based implicit approach is used for LES \citep{Jacobs2018, Zauner2020PRF}, which involves the application of a weak, low-pass, sixth-order spatial filter when required. 

The configuration considered is that of an unconfined flow undergoing free transition over a NACA0012 infinite-wing section with a blunt trailing edge of thickness 0.5$\%$ chord at a specified incidence angle, $\alpha$ (results for a few cases of Dassault Aviation's V2C profile are also reported in appendix~\ref{secV2C}). The profile is extruded in the spanwise direction for a width of $L_z$, where $z$ denotes the spanwise direction. The streamwise and the third orthonormal Cartesian directions are denoted as $x$ and $y$, respectively. For non-zero incidence angles, the chord-based coordinate directions are denoted by $x'$ and $y'$, respectively. The aerofoil is treated as an isothermal wall, whereas periodic boundary conditions are applied in the spanwise direction. Non-reflecting integral characteristic boundary conditions are used on the inflow boundaries, while zonal characteristic boundary conditions \citep{Sandberg2006} are applied on the outflow boundaries. These boundaries are located sufficiently far from the aerofoil to not affect the near field characteristics (\textit{i.e.}, the same distances as used in \citet{Moise2022}, where the further extension of the domain was found to have no significant effects). Shock waves in the flow field are captured using a total-variation diminishing scheme. 

The length, velocity, density and temperature scales used to solve the Navier-Stokes equations in dimensionless form are the aerofoil chord and corresponding freestream values. The freestream Reynolds number based on these scales is $Re = 5\times 10^4$ for all cases except those reported in \S\ref{subSecLFOVsHighReTB}, where two additional cases of $Re = 5\times10^5$ and $1.5\times10^6$ are reported. A wide range of freestream Mach numbers ($0.3 \leq M \leq 0.8$) and incidence angles ($0^\circ \leq \alpha < 10^\circ$) were examined here. The parameters studied in each of the following sections and buffet features are provided at their start (tables~\ref{tableA0MEffects}, \ref{tableDomainEffects}, \ref{tableHighAoA} and \ref{tableHighRe}).

The fluid is assumed to be a perfect gas with a specific heat ratio of 1.4 and satisfying Fourier’s law of heat conduction. The Prandtl number is 0.72. It is also assumed to be Newtonian and satisfying Sutherland’s law for viscosity, with the Sutherland constant as 110.4 K for a reference temperature of 268.67 K. The time step used is $3.2\times 10^{-5}$ (implying approximately $5\times10^5$ iterations in a buffet cycle). For the lowest freestream Mach number considered, $M = 0.3$, the time step is lowered to $1.6\times 10^{-5}$ which was found to be required to capture the  relative increase in the speed of acoustic waves.

\begin{figure} 
\centerline{
\includegraphics[width=0.495\textwidth]{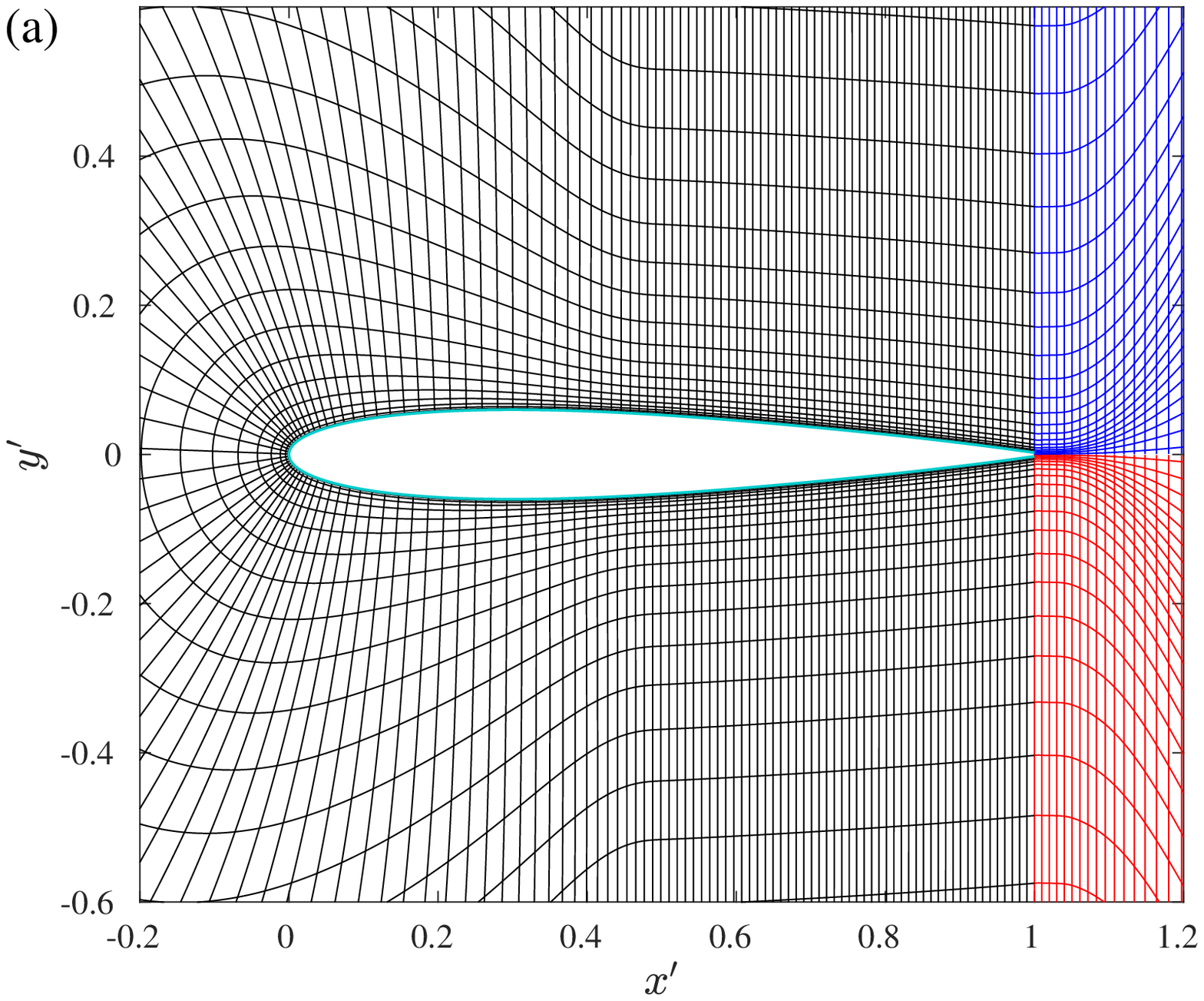}
\includegraphics[width=0.495\textwidth]{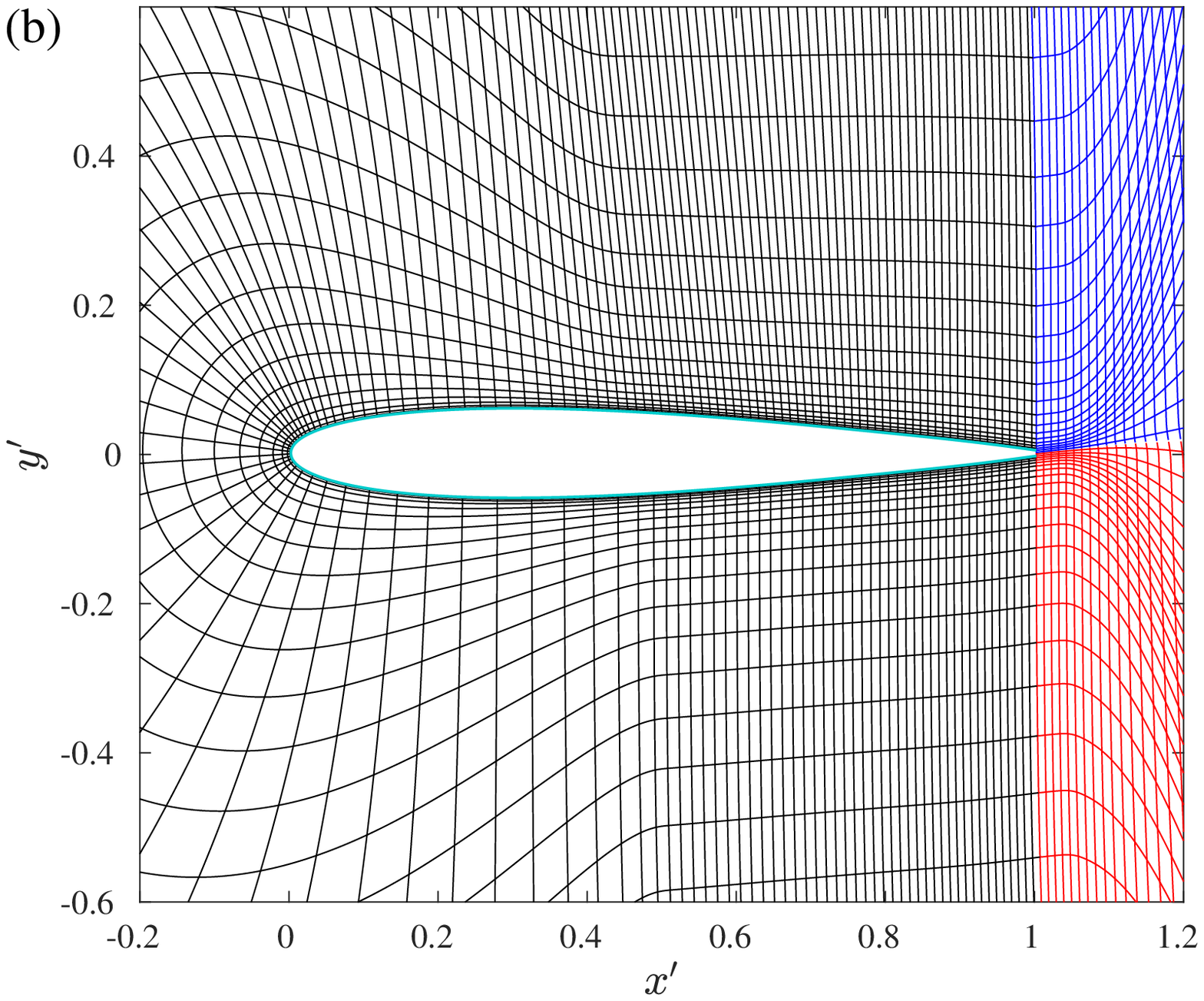}
}
\caption{Features of (a) Grid 0 and (b) Grid 1 in the vicinity of the aerofoil in the $x-y$ plane (only every 10$^\mathrm{th}$ point shown for clarity).}
\label{fGrids}
\end{figure}

Multi-block C-H grids are used for the simulations. The grid points are first generated in the $x-y$ plane, following which the domain is extruded in the spanwise direction by generating points in that direction with a constant grid spacing. Given the large variations in $\alpha$ considered, two different grids are used. Grid 0 (figure~\ref{fGrids}\textit{a}) is used for zero and low incidence angles ($\alpha \leq 4^\circ$) and contains a symmetric distribution of grid points on the suction and pressure side. By contrast, Grid 1 (figure~\ref{fGrids}\textit{b}), which is used for higher $\alpha$ (at which flow dynamics vary strongly on the suction side), has a lower resolution on the pressure side. These grids were generated using an open-source code \citep{Zauner2018}. Since this code generates the grid-point positions in the $x-y$ plane based on spacing requirements, curvature, \textit{etc.}, it can be easily adapted to generate grids with similar characteristics for different aerofoils. Here, for Grid 1, we use a similar distribution in the $x-y$ plane as that used for LES of flow over Dassault Aviation's V2C aerofoil and ONERA's OALT25 in previous studies \citep{Zauner2020PRF, Moise2022, Zauner2022}. Note that the LES result at $Re = 5\times 10^5$ using V2C's equivalent grid to Grid 1 has been shown to match well with a direct numerical simulation employing a refined grid \citep{Zauner2020PRF}. Additionally, the grid for OALT25 which is equivalent to Grid 1 has also been shown to capture buffet frequency accurately for $Re = 3\times 10^6$ (see the comparison with LES of \citet{Dandois2018} and experiments of \citet{Brion2020} reported in \citet{Zauner2022}). Thus, the present choices of grid distributions in the $x-y$ plane associated with Grid 1 and Grid 0 (which is even further resolved) are considered more than adequate to capture features at a relatively lower Reynolds number of $Re = 5\times 10^4$. The spanwise resolution is chosen as $\Delta z = 0.002$ for cases where $Re = 5\times 10^4$ and $\Delta z = 0.001$ for $Re \geq 5\times10^5$, the latter being the same as that used in the LES of \citet{Zauner2020PRF}.

\subsection{Spectral proper orthogonal decomposition}
\label{subSecSPODMethod}
Spectral orthogonal decomposition (SPOD) is employed in the present study to examine spatio-temporally coherent features in the LES flow-field \citep{Lumley1970, Glauser1987, Towne2018}. The methodology adopted is the same as that extensively discussed in \citet{Moise2022} and only a brief summary is given here. For a zero-mean, stationary, stochastic process, the ideal basis that represents a given ensemble of its realisations consists of the eigenfunctions, $\boldsymbol{\psi}$, of the cross-spectral density tensor, $\boldsymbol{S}$, satisfying
\begin{equation}
    \int_\Omega \boldsymbol{S}(\boldsymbol{x},\boldsymbol{x'},St)\boldsymbol{W}\boldsymbol{\psi}(\boldsymbol{x},St)d\Omega =  \lambda(St) \boldsymbol{\psi}(\boldsymbol{x'},St). \label{eqnSPODEVP}
\end{equation}
Here, $\boldsymbol{W}$ is a weight associated with the appropriate inner product on the spatial domain, $\Omega$, $\boldsymbol{x}$ and $\boldsymbol{x'}$ are any two points in the domain, $St$ is the Strouhal number based on aerofoil chord and freestream velocity, and $\lambda$ represents the eigenvalue. The eigenvalues are indexed in decreasing order (\textit{i.e.}, $\lambda_1 > \lambda_2 > ... \:\lambda_i > ...$), where $\lambda_i$ is referred to as the $i$-th eigenvalue. The corresponding eigenfunctions are also indexed accordingly, implying that the expected value of the projection of $\boldsymbol{\psi}_1$ with the realisations is the maximum. Note that $\boldsymbol{\psi}_i(\boldsymbol{x},St_0)$ gives the spatial structure of the $i$-th SPOD mode at a specific Strouhal number, $St_0$. The temporal behaviour of this SPOD mode is given by 
\begin{equation}
    \boldsymbol{\phi}_i(\boldsymbol{x},t) = \mathrm{Re}\left\{\boldsymbol{\psi}_i(\boldsymbol{x},St_0) \exp(2i\pi St_0 t) \right\}, 
\end{equation}
where $t$ represents time and $\mathrm{Re}\{\}$ denotes the real part. This can be rewritten as 
\begin{equation}
    \boldsymbol{\phi}_i(\boldsymbol{x},\phi) = \mathrm{Re}\left\{\boldsymbol{\psi}_i(\boldsymbol{x},St_0) \exp(i\phi) \right\}, 
    \label{eqnSPODSpatioTemp}
\end{equation}
where $\phi = 2\pi St_0 t$, is the phase within a oscillation cycle of frequency, $St_0$.

In this study, we use the streaming algorithm and the numerical code provided in \citet{SCHMIDT201998} to perform SPOD. Flow-field data based on the $z=0$ plane are stored at intervals of 0.16 (\textit{i.e.}, sampling frequency of 6.25). Each snapshot consists of the density, pressure and velocity fields. These snapshots are grouped into blocks such that at least three buffet cycles are captured in the time interval, $T_B$, associated with each block (\textit{i.e.}, for buffet frequency, $St_b \approx 0.025$, $T_B = 120$, implying 750 snapshots per block). To compute $\boldsymbol{S}$, Welch's approach was employed with a Hamming window function and 50\% overlap of snapshots between blocks. The approximated cell area associated with each grid point was used to compute the weight matrix, $\boldsymbol{W}$. The SPOD algorithm used reduces the computational expense by computing only a subset of the eigenvalues and corresponding SPOD modes. Here, only the first two dominant eigenvalues (indices 1 and 2) were computed. No significant changes in the dominant eigenvalue and eigenfunction occurred when the parameters governing SPOD (block size, sampling frequency, \textit{etc.}) were changed, indicating the robustness of the approach. Additionally, in our previous study \citep{Moise2022Trip}, we have matched the SPOD modes from LES with modes from global linear stability analysis based on RANS results for transonic buffet on the V2C aerofoil at similar flow conditions as the present study. 

\section{Zero-incidence results}
\label{secNaCA}
We report unsteady flow features at $\alpha = 0^\circ$ and $Re = 5\times10^4$ in this section. At this incidence, a Type I laminar transonic buffet is expected beyond a threshold freestream Mach number. The effect of varying this parameter is explored first for a narrow spanwise width of $L_z = 0.05$ in \S\ref{subSecNarrowFreeMach}, following which the effect of increasing the span is examined in \S\ref{subSecDomain}.

\subsection{Effect of freestream Mach number}
\label{subSecNarrowFreeMach}

\begin{figure} 
\centerline{
\includegraphics[width=0.495\textwidth]{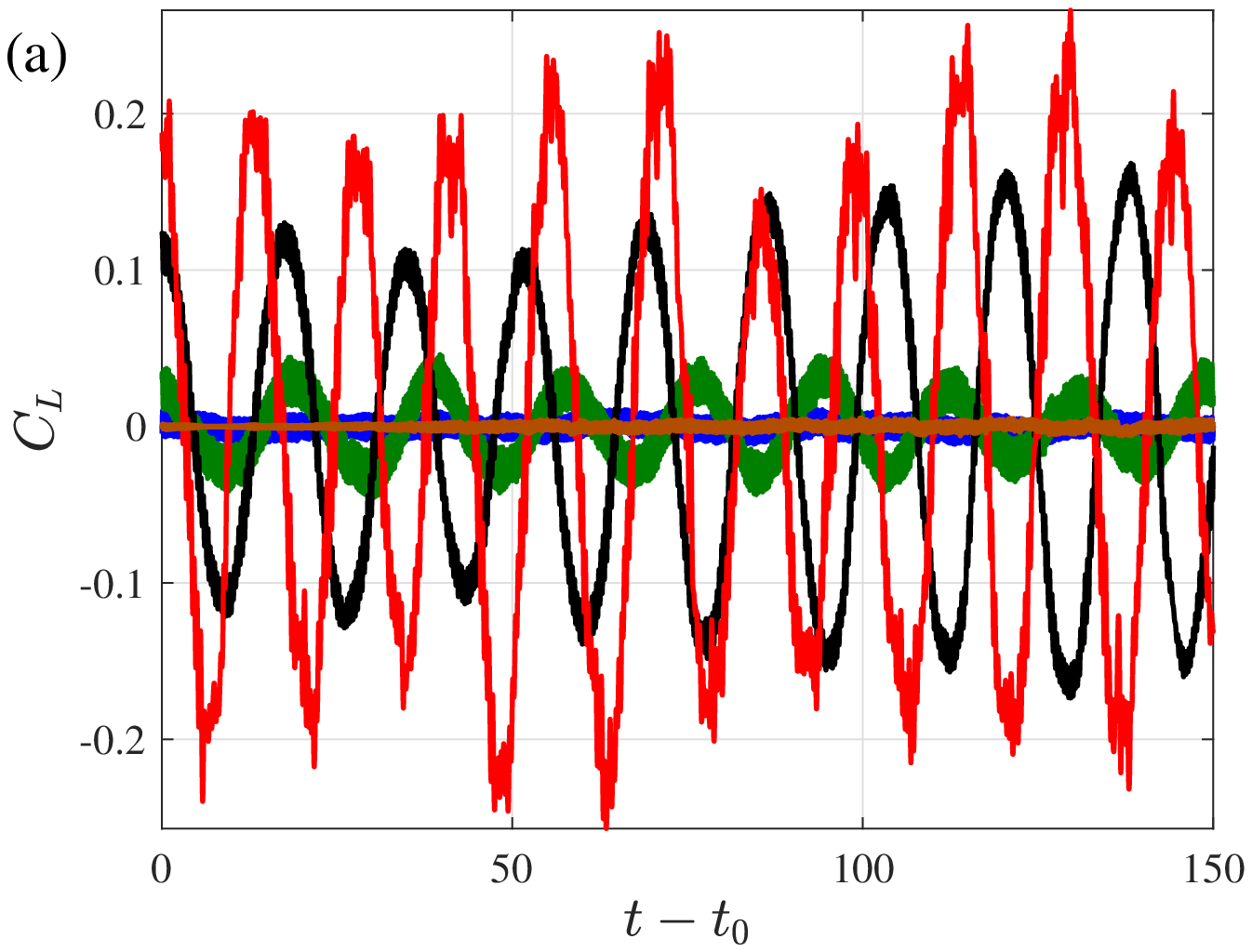}
\includegraphics[width=0.495\textwidth]{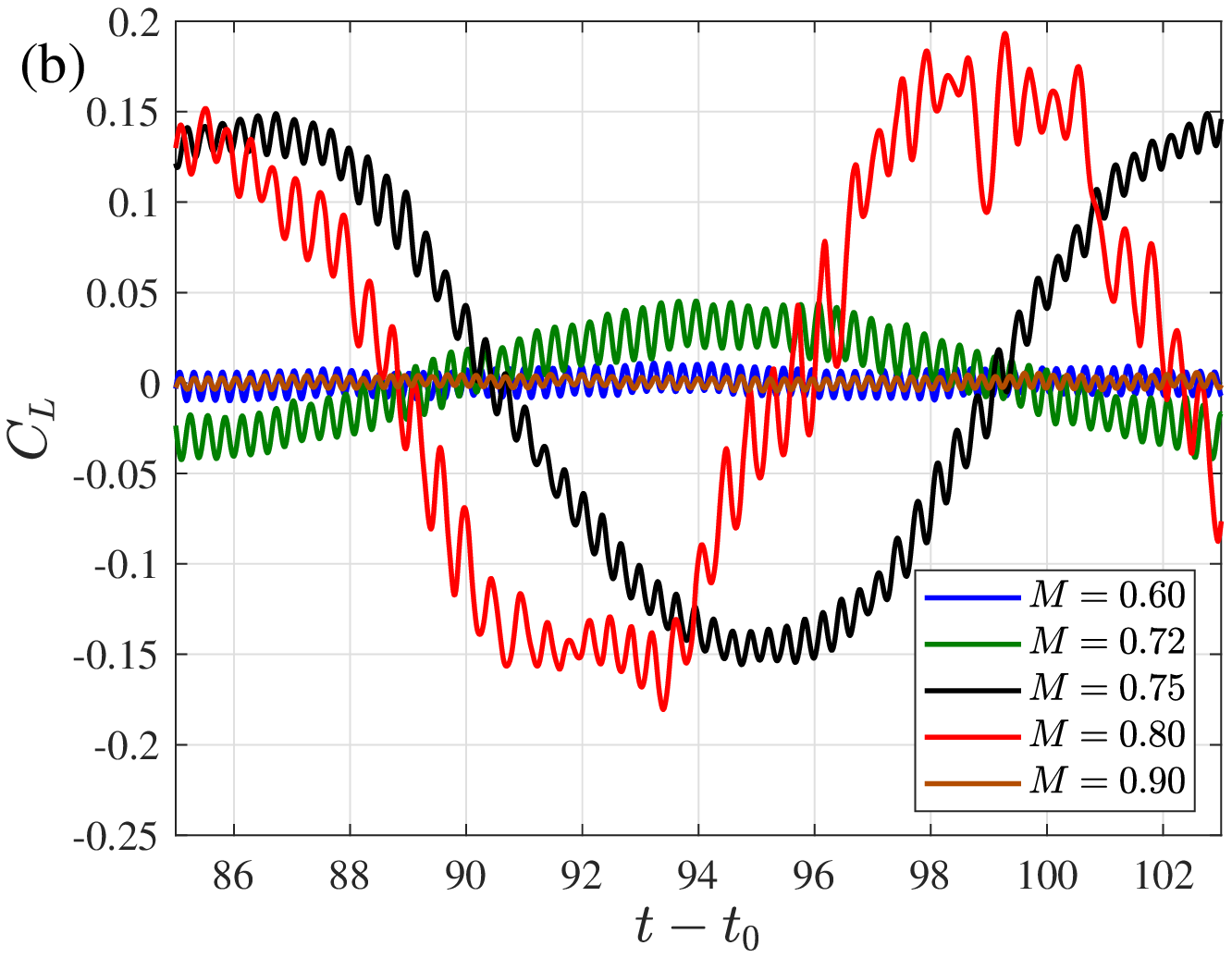}
}
\caption{Temporal variation of lift coefficient past transients for NACA0012 aerofoil at zero incidence and different freestream Mach numbers (a) till the end of simulation indicating oscillations at a low frequency associated with buffet and (b) for a shorter time interval, highlighting oscillations at a higher frequency associated with a von~K\'arm\'an vortex street.}
\label{fNACAClMach}
\end{figure}

\begin{figure} 
\centerline{
\includegraphics[width=0.45\textwidth]{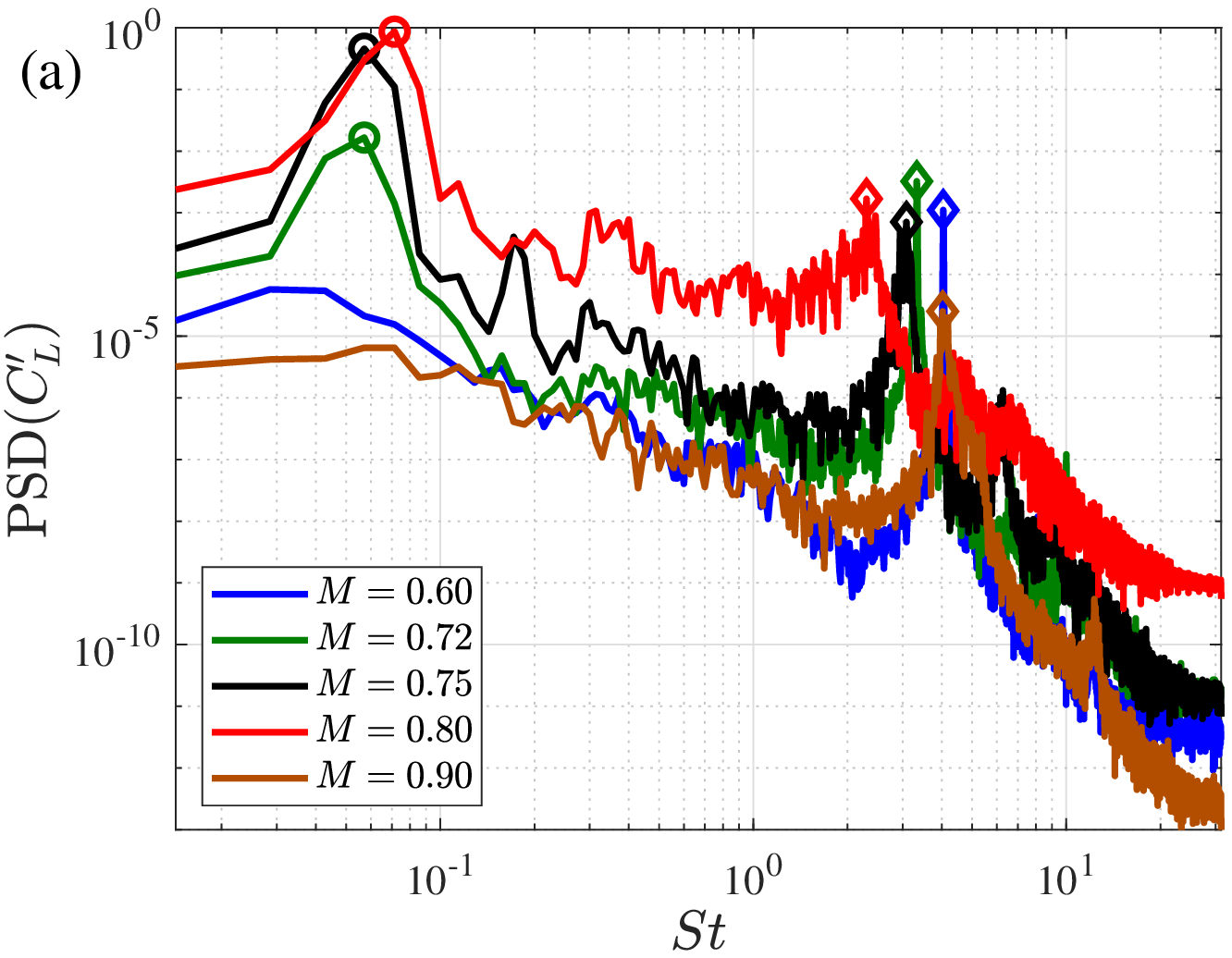}
\includegraphics[width=0.495\textwidth]{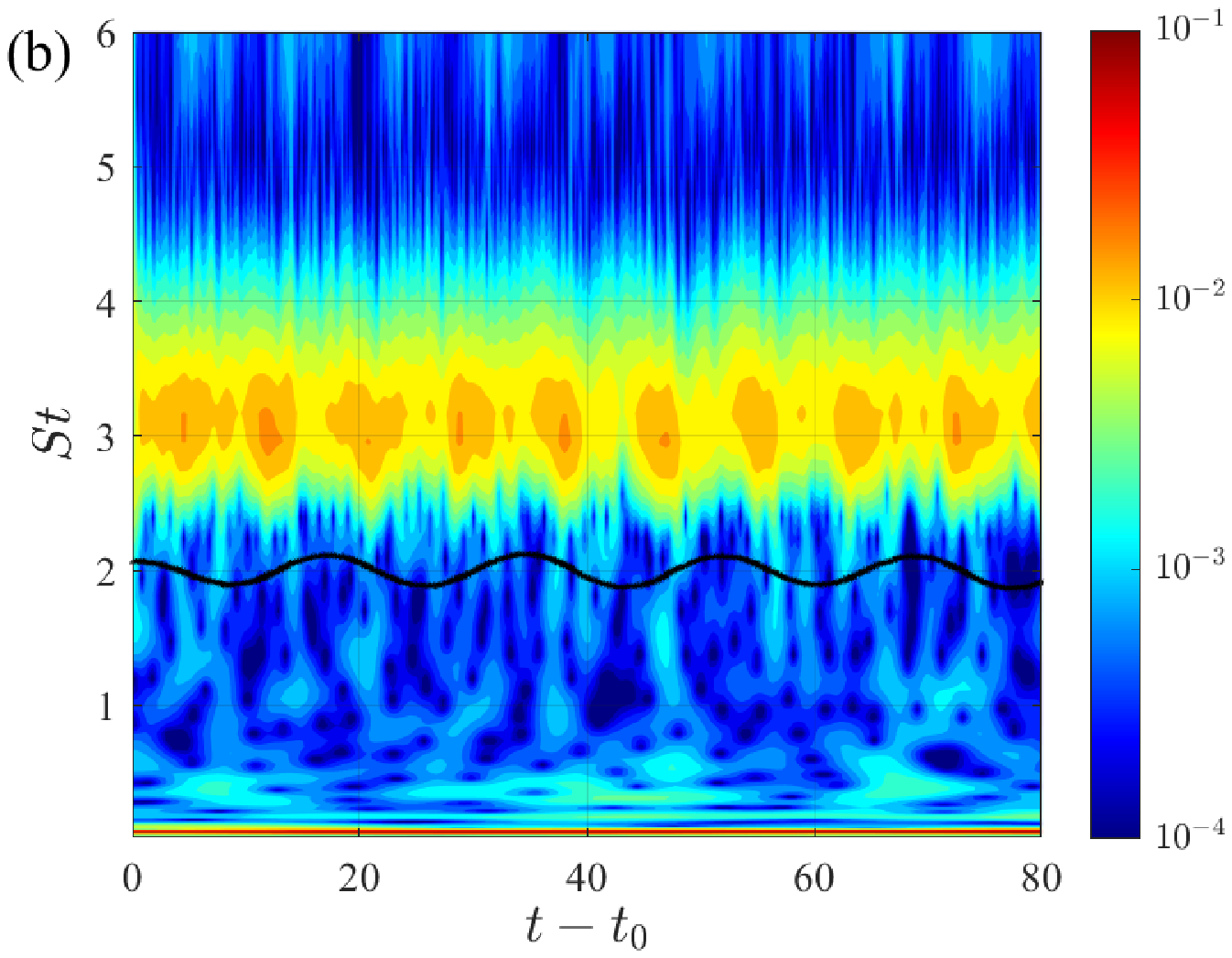}
}
\caption{(a) Power spectral density (PSD) of the fluctuating component of lift coefficient as a function of the Strouhal number, $St$ for $\alpha = 0^\circ$ and different $M$. Circles and diamonds highlight the peaks at Strouhal numbers associated with buffet and vortex shedding, respectively. (b) Scalogram for $M = 0.75$ based on the fluctuating lift coefficient. For reference, $C_L'(t)+2$ is overlaid on the scalogram as a black curve.}
\label{fNACAClPSD}
\end{figure}

The temporal variation of the lift coefficient past transients is shown for different freestream Mach numbers, $M$, in figure~\ref{fNACAClMach}\textit{a}. Periodic oscillations at a low frequency (time period, $T \sim O(10)$) are evident for all cases except $M = 0.6$ and 0.9. These oscillations will be shown to be related to transonic buffet. The variation of the lift coefficient within a shorter time interval is provided in figure~\ref{fNACAClMach}\textit{b}, indicating oscillations at a higher frequency ($T \sim O(1)$) for all $M$. These will be shown to be related to a von~K\'arm\'an vortex street. The power spectral densities of the fluctuating component of the lift coefficient for different $M$ are provided in figure~\ref{fNACAClPSD}\textit{a}. The peaks associated with these low- and high-frequency oscillations are highlighted using circles and diamonds, respectively. The amplitudes and frequency associated with these peaks are also documented in table~\ref{tableA0MEffects}. The energy associated with the latter is found to be approximately the same except for the case of $M = 0.9$. By contrast, the energy is negligible for the former at $M = 0.6$ and $M=0.9$, indicating buffet onset and offset, respectively. Buffet frequency is seen to increase monotonically with $M$, a trend which matches that reported in other studies on Type II transonic buffet \citep{Dor1989, Jacquin2009, Brion2020, Moise2022}. A scalogram based on the fluctuating component of the lift coefficient is plotted in figure~\ref{fNACAClPSD}\textit{b} for a representative case of $M = 0.75$. The temporal variation of the lift coefficient is overlaid on the plot for reference (black curve). It can be inferred from the figure that there are temporal variations in the intensity of the high-frequency oscillations within a period of the low-frequency cycle, indicating that the wake mode behaviour is modulated by buffet oscillations, similar to the results reported in \citet{Moise2022} for a different aerofoil and flow conditions.

\begin{table}
\begin{center}
\def~{\hphantom{0}}
\begin{tabular}{lcccc}
$M$  & $St_b$ & $St_w$   &  PSD$_b$ & PSD$_w$ \\[6pt]
0.6  & --     & 4.042    &   0     & 0.0011 \\
0.72 & 0.057  & 3.329    &   0.0166 & 0.0032 \\
0.75 & 0.057  & 3.085    &   0.4566 & 0.0007 \\
0.8  & 0.071  & 2.300    &   0.8636 & 0.0016 \\
0.9  & --     & 4.028    &   0     & 0.00003 \\
\end{tabular}
\caption{Comparison of buffet and wake mode features for $L_z = 0.05$, $\alpha = 0^\circ$, $Re = 5\times10^4$ and different $M$ for the NACA0012 aerofoil (all cases reported in \S\ref{subSecNarrowFreeMach}).}
\label{tableA0MEffects}
\end{center}
\end{table}

\begin{figure} 
\centering
\includegraphics[trim={0.9cm 0cm 0cm 0cm},clip,width=.32\textwidth]{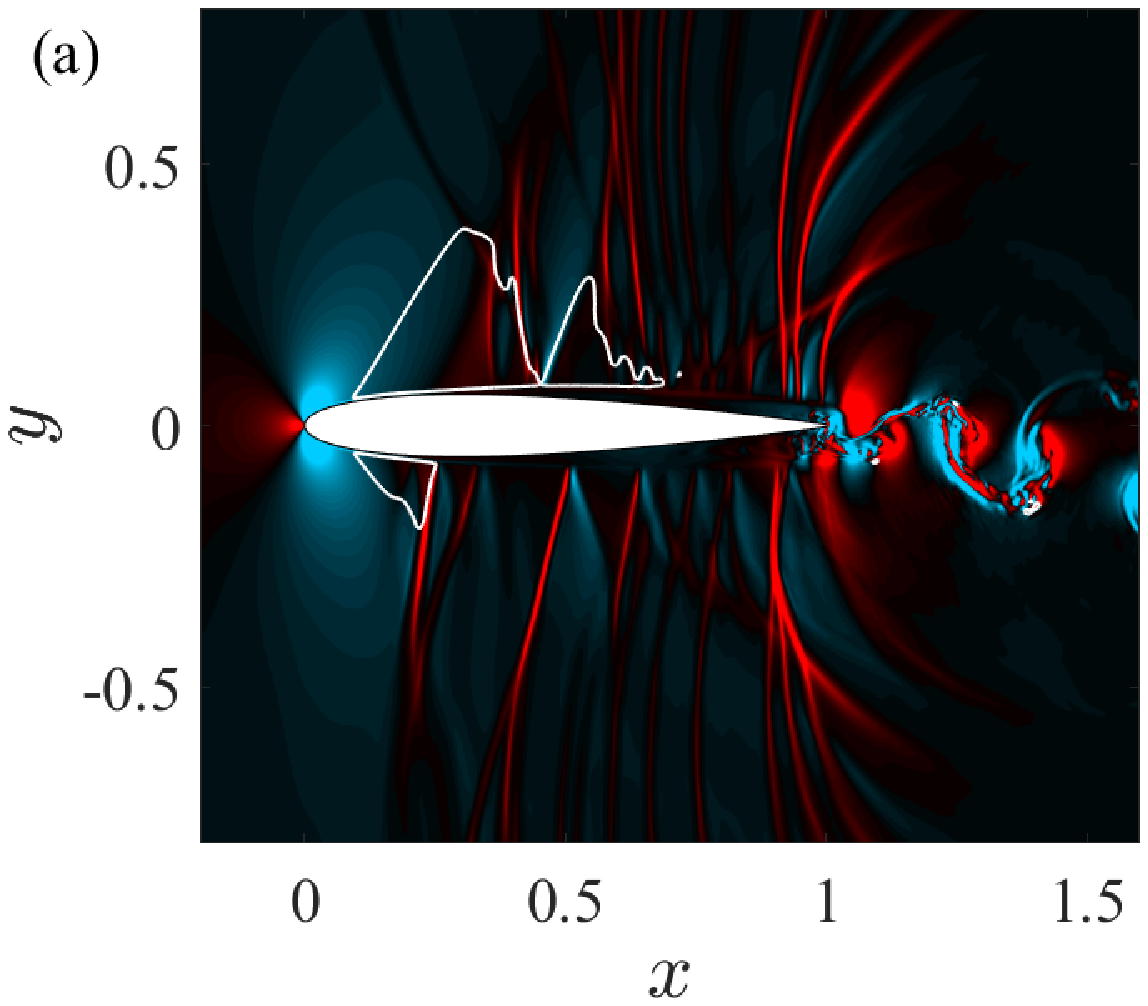}
\includegraphics[trim={0.9cm 0cm 0cm 0cm},clip,width=.32\textwidth]{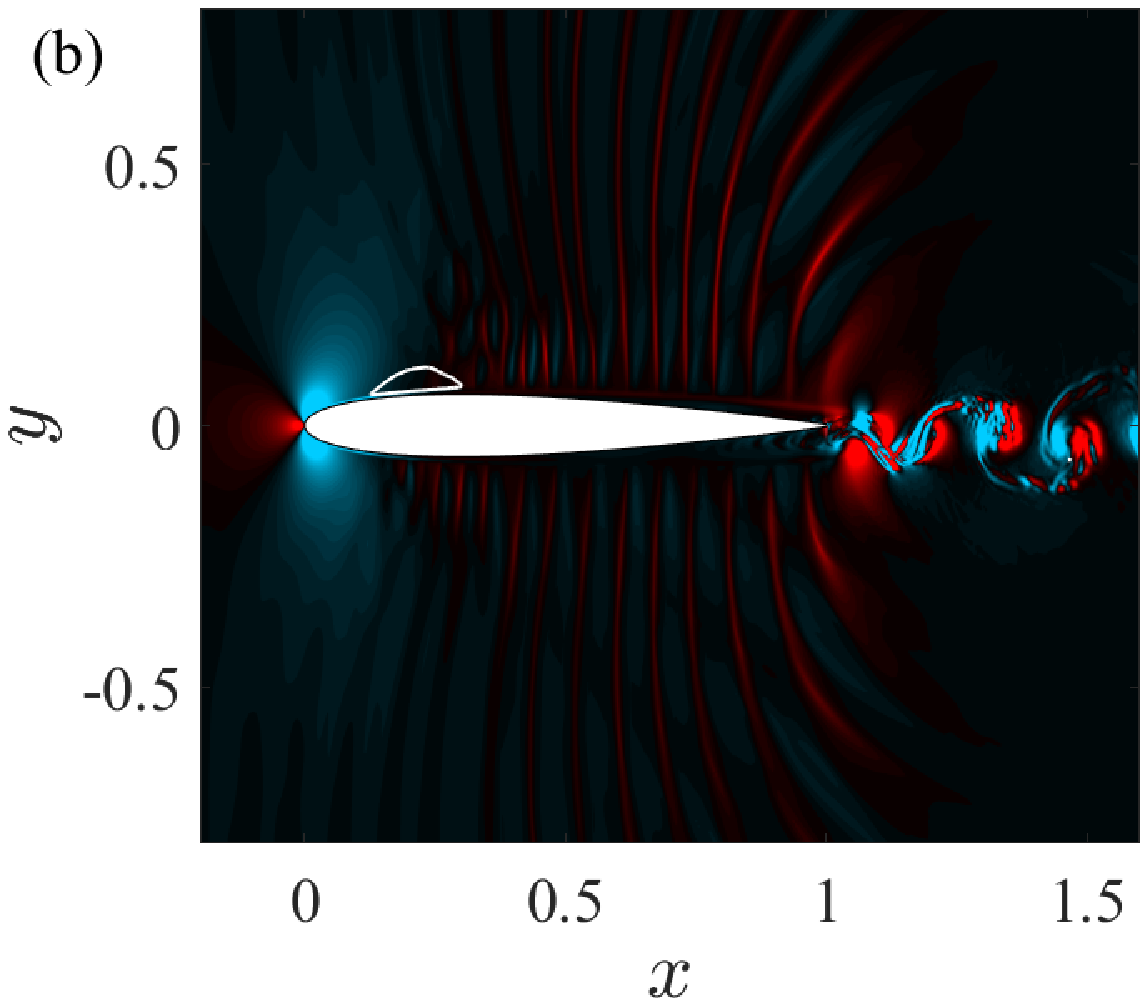}
\includegraphics[trim={0.9cm 0cm 0cm 0cm},clip,width=.32\textwidth]{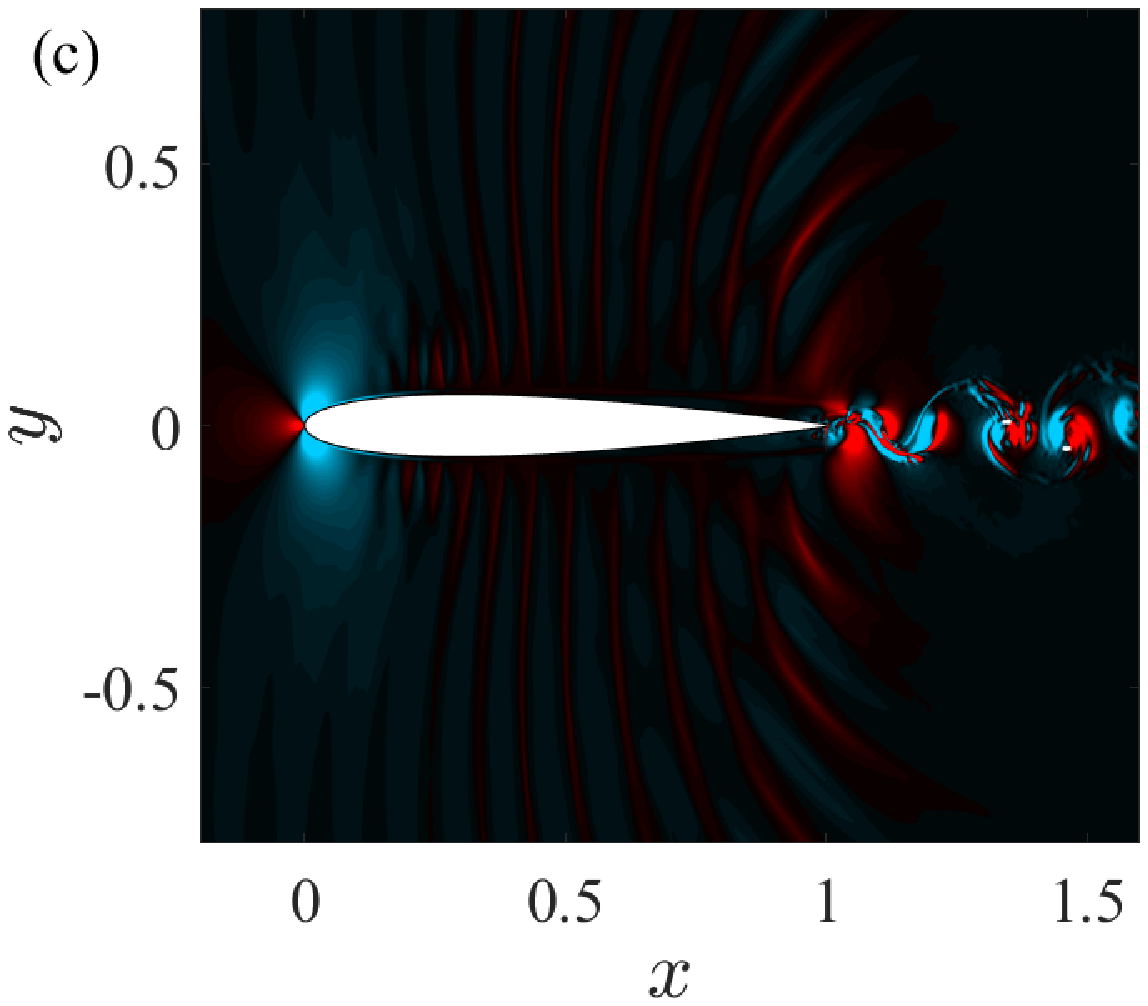}
\includegraphics[trim={0.9cm 0cm 0cm 0cm},clip,width=.32\textwidth]{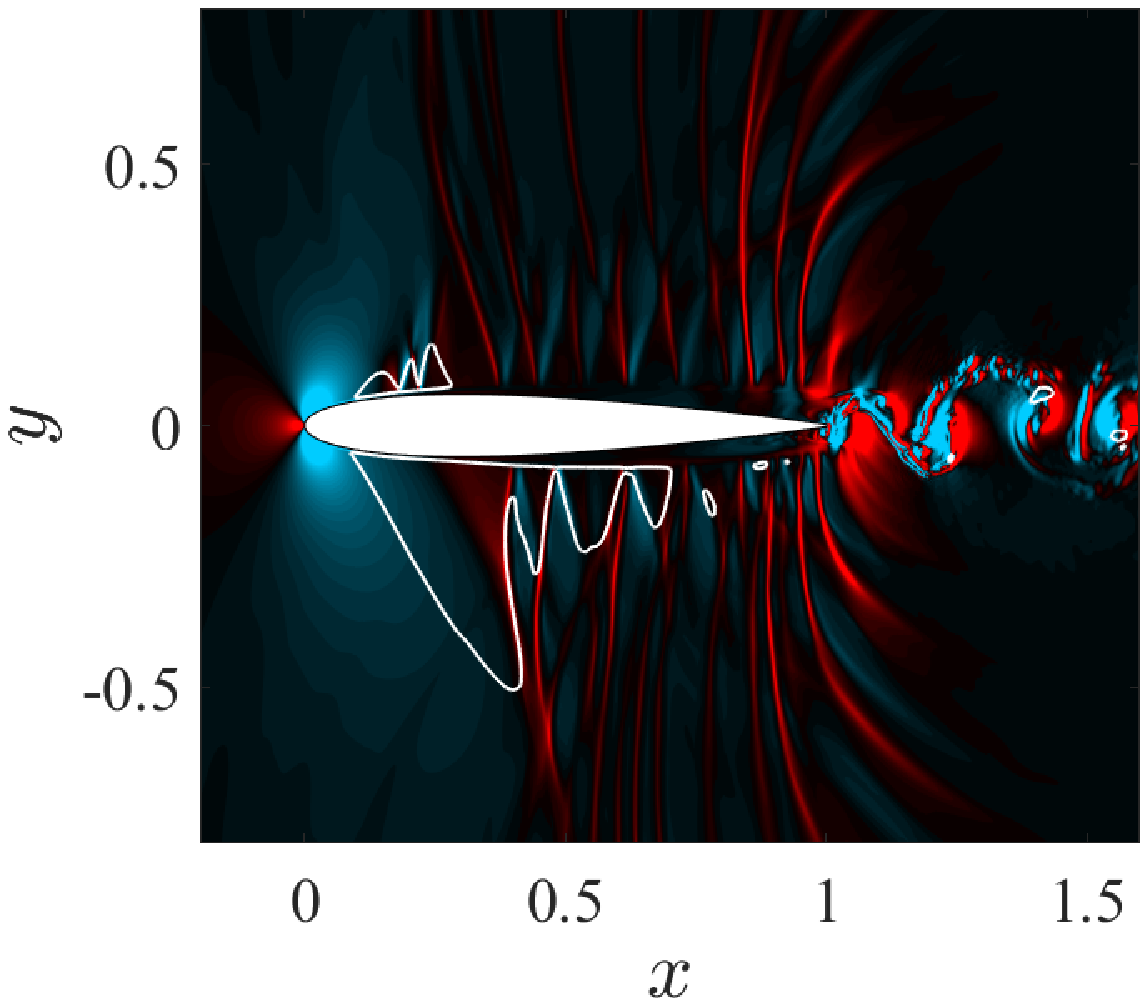}
\includegraphics[trim={0.9cm 0cm 0cm 0cm},clip,width=.32\textwidth]{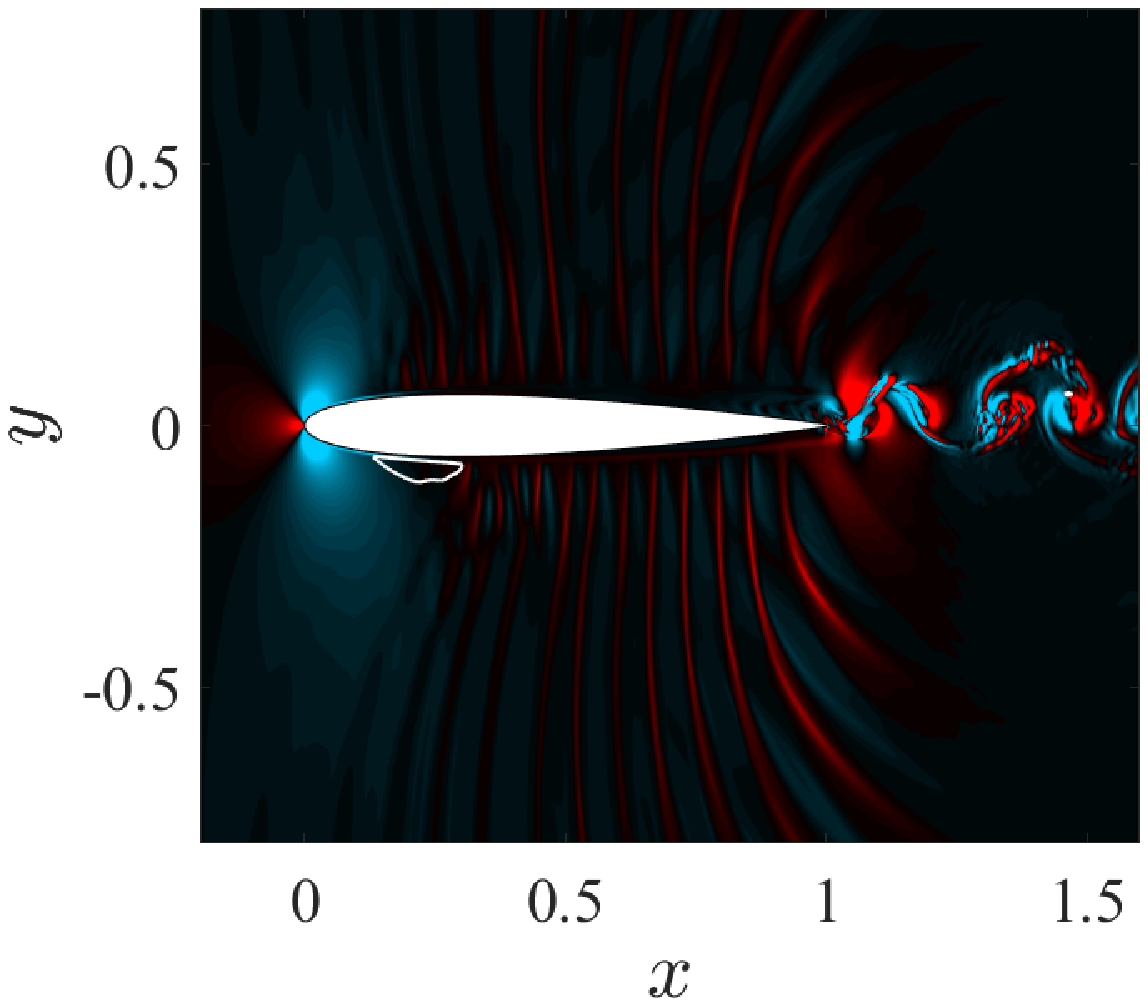}
\includegraphics[trim={0.9cm 0cm 0cm 0cm},clip,width=.32\textwidth]{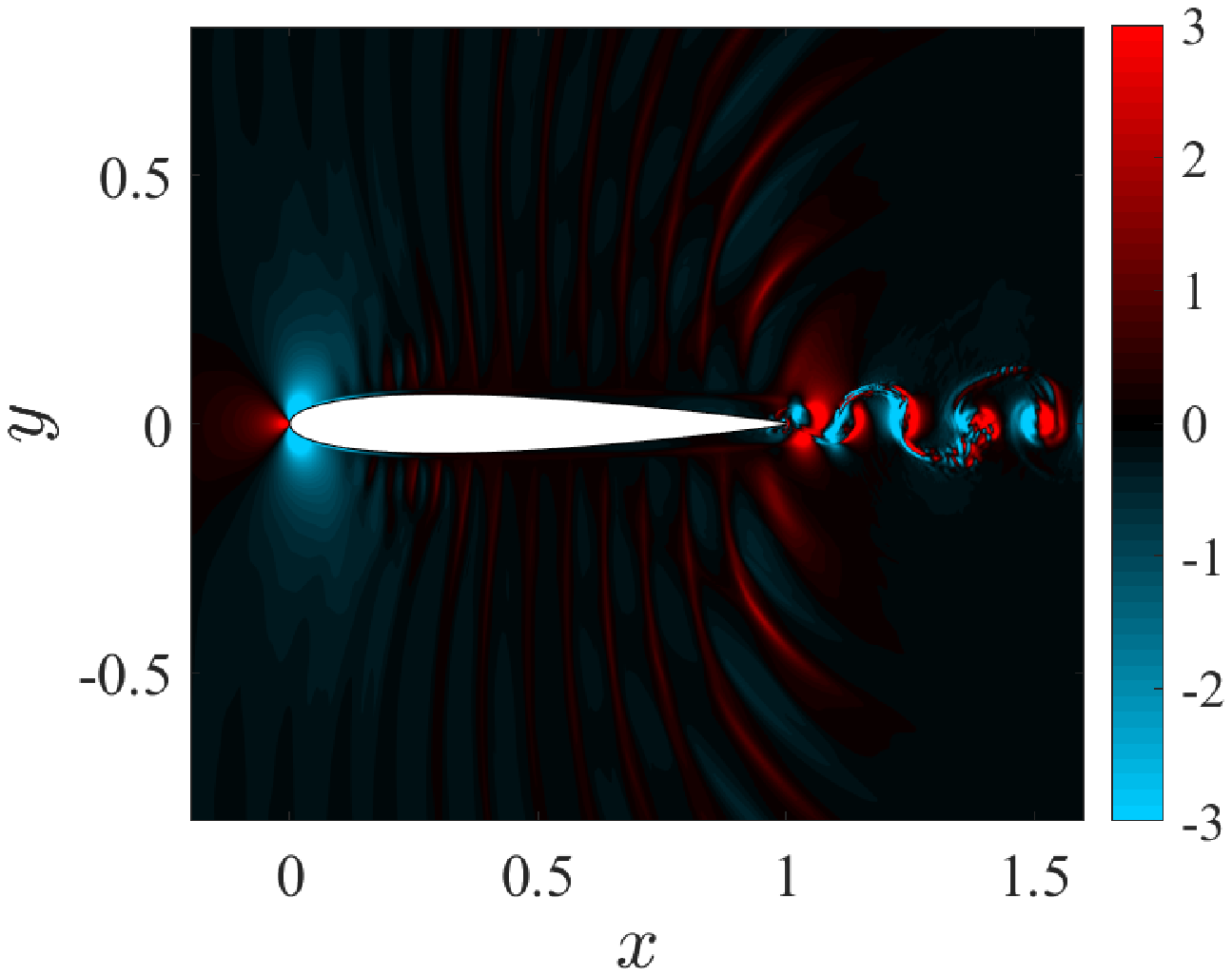}
\caption{Streamwise density gradient contours on the $x-y$ plane shown at the approximate high-(top) and low-(bottom) lift phases of the low-frequency cycle for $\alpha = 0^\circ$ and (a) $M = 0.8$, (b) 0.75 and (c) 0.72. The sonic line is highlighted using a white curve.}
    \label{fNACADensGrad}
\end{figure}

Henceforth, we focus on the three cases of $M = 0.8$, $0.75$ and 0.72. The instantaneous spatial flow field at approximately the high- and low-lift phases of the low-frequency cycle for these cases are shown in figure~\ref{fNACADensGrad} using contours of the streamwise density gradient. Note that this field is similar to Schlieren visualisations in experiments. The sonic line (white curve), \textit{i.e.}, the isoline based on the instantaneous local Mach number, $M_\mathrm{loc} = 1$, is overlaid for reference and delineates the supersonic region in the flow. Due to the use of a symmetric aerofoil, the flow field in the low-lift phase approximately mirrors (about the $x'$-axis) that in the high-lift phase (\textit{i.e.}, pressure-side features observed on the suction side and vice versa). The case, $M = 0.8$, is shown in figure~\ref{fNACADensGrad}\textit{a}, where there are supersonic regions on both sides of the aerofoil which are terminated by shock waves. In the high-lift phase, the supersonic region on the suction side is significantly larger with multiple shock waves, whereas these features are inverted in the low-lift phase. This implies that the shock waves traverse the range $0.3 \leq x \leq 0.7$ on both sides. This is also seen in the temporal variation of these contours visualised in movie 1 provided as Supplementary material. Thus, this case of $M = 0.8$ can be categorised as a Type I laminar transonic buffet. We emphasise that, whereas the presence of a single shock wave is typical for transonic buffet under forced-transition conditions or sufficiently high $Re$, laminar transonic buffet at lower $Re$ can have multiple shock waves present \citep{Zauner2019}. Irrespective of the number of shock waves or transition type, transonic buffet features such as frequency and the SPOD mode's spatial structure have been shown to be essentially the same \citep{Moise2022Trip, Zauner2022}.

For a lower freestream Mach number of $M = 0.75$ (figure~\ref{fNACADensGrad}\textit{b}), the supersonic region is observed in the high-lift phase only on the suction side and is drastically reduced in area. Indeed, there are large time intervals within the low-frequency cycle in which the flow remains subsonic, as seen from movie 2 provided as Supplementary material. Furthermore, in the high- and low-lift phases, the transition from supersonic to subsonic flow is gradual, suggesting the absence of shock waves in the flow field. Nevertheless, the power spectral density of the low-frequency peak in figure~\ref{fNACAClPSD}\textit{a} does not significantly reduce, implying strong flow oscillations. From these results, it can be inferred that oscillations resembling transonic buffet occur even in the absence of shock waves occur for these flow conditions. At an even lower $M = 0.72$, (figure~\ref{fNACADensGrad}\textit{c}), the flow was observed to be subsonic at all times (see movie 3) suggesting that both shock waves and supersonic regions are not essential for the sustenance of these oscillations. 

\begin{figure} 
\centering
\includegraphics[width=.45\textwidth]{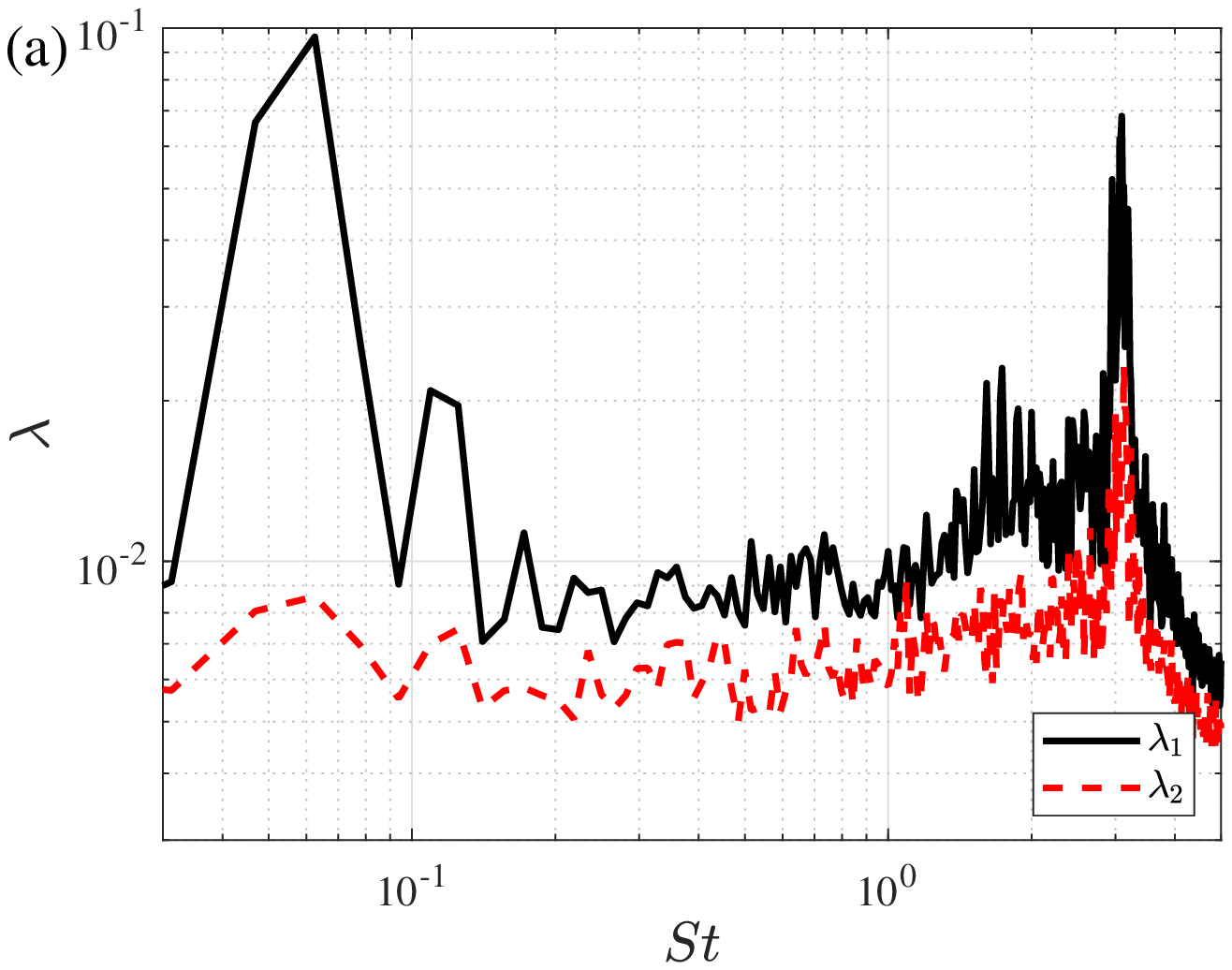}
\includegraphics[width=.45\textwidth]{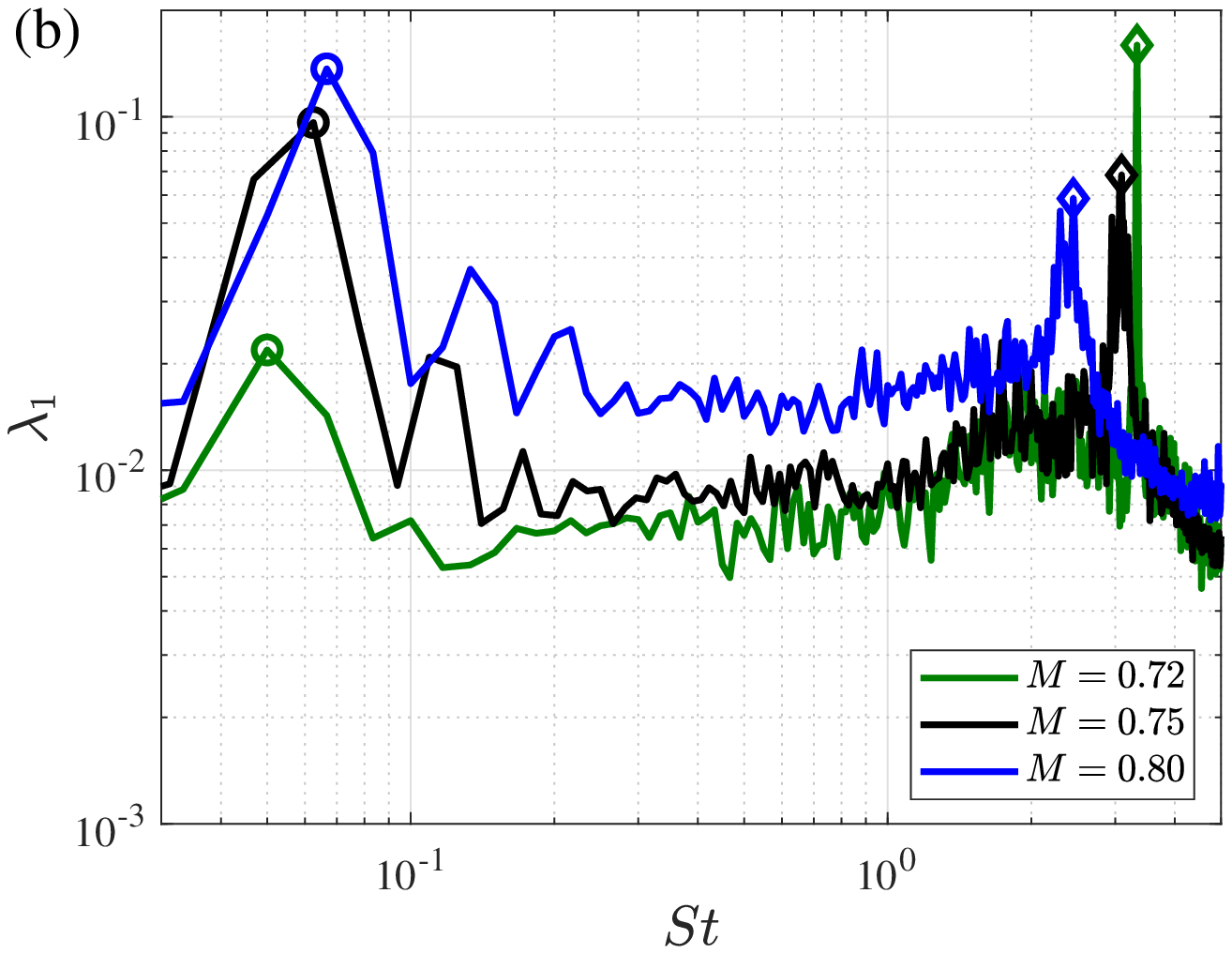}
\caption{(a) Eigenvalue spectra (logarithmic scale) from SPOD shown for (a) the first (dominant) and second eigenvalue for the case $M = 0.75$ and (b) only the dominant eigenvalue for different $M$ at $\alpha = 0^\circ$. Circles and diamonds highlight the peaks associated with the buffet and wake modes, respectively.}
\label{fNACASPODSpectrum}
\end{figure}

\begin{figure} 
\centering
\includegraphics[trim={3cm 0cm 3.5cm 0cm},clip,width=.32\textwidth]{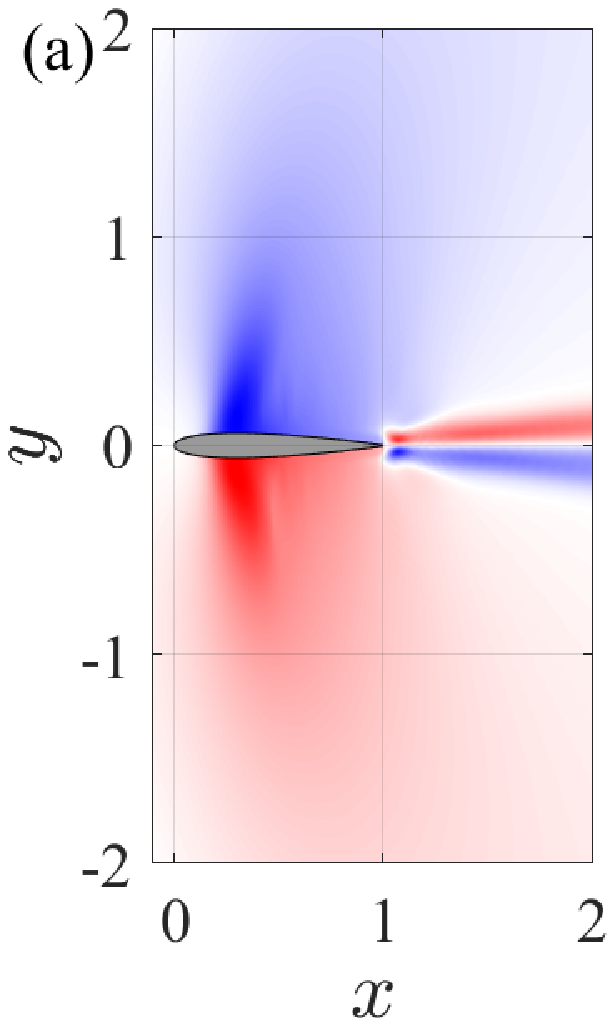}
\includegraphics[trim={3cm 0cm 3.5cm 0cm},clip,width=.32\textwidth]{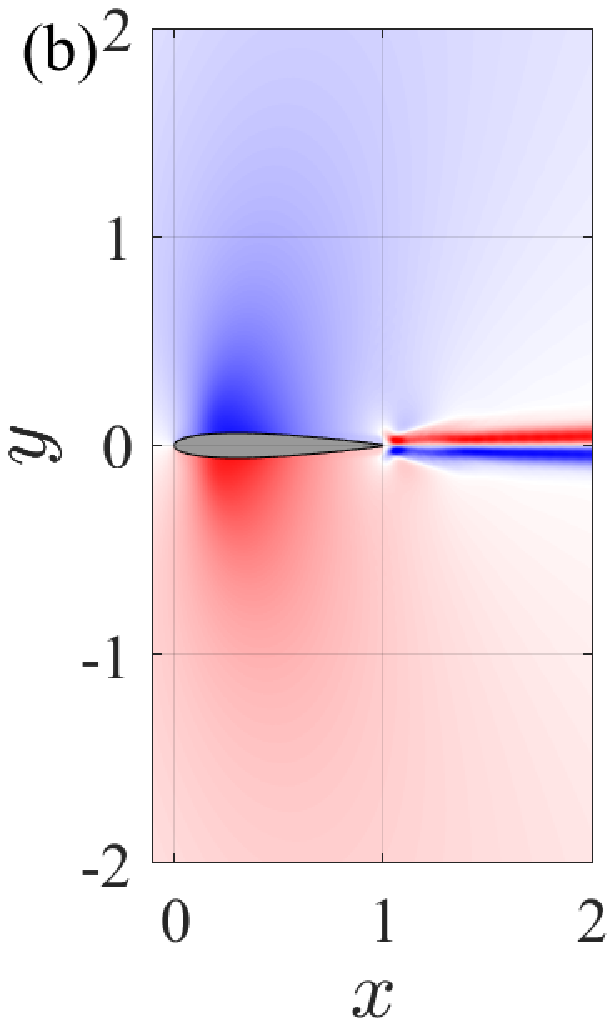}
\includegraphics[trim={3cm 0cm 3.5cm 0cm},clip,width=.32\textwidth]{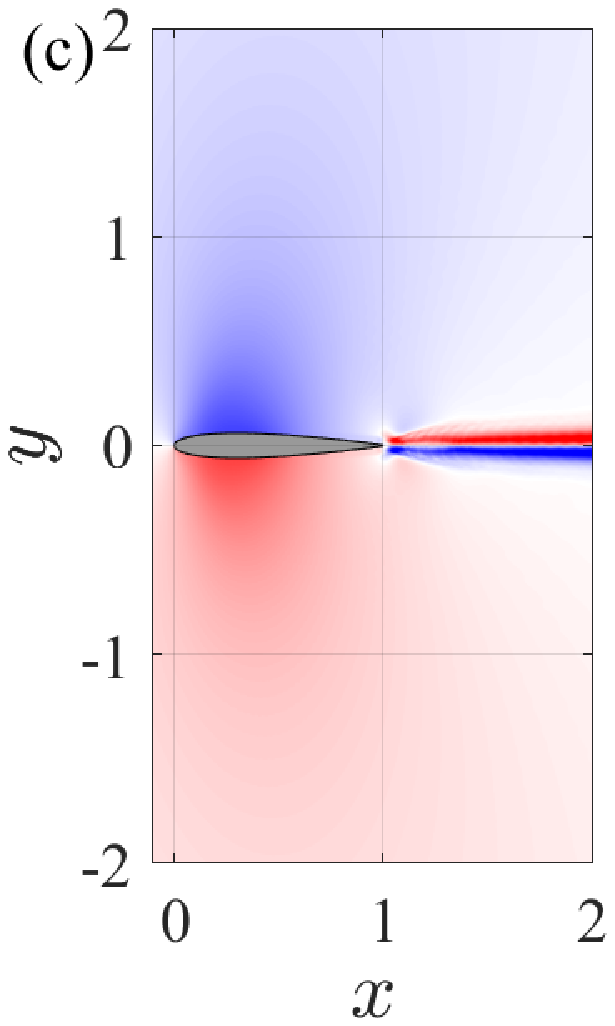}
\includegraphics[trim={3cm 0cm 3.5cm 0cm},clip,width=.32\textwidth]{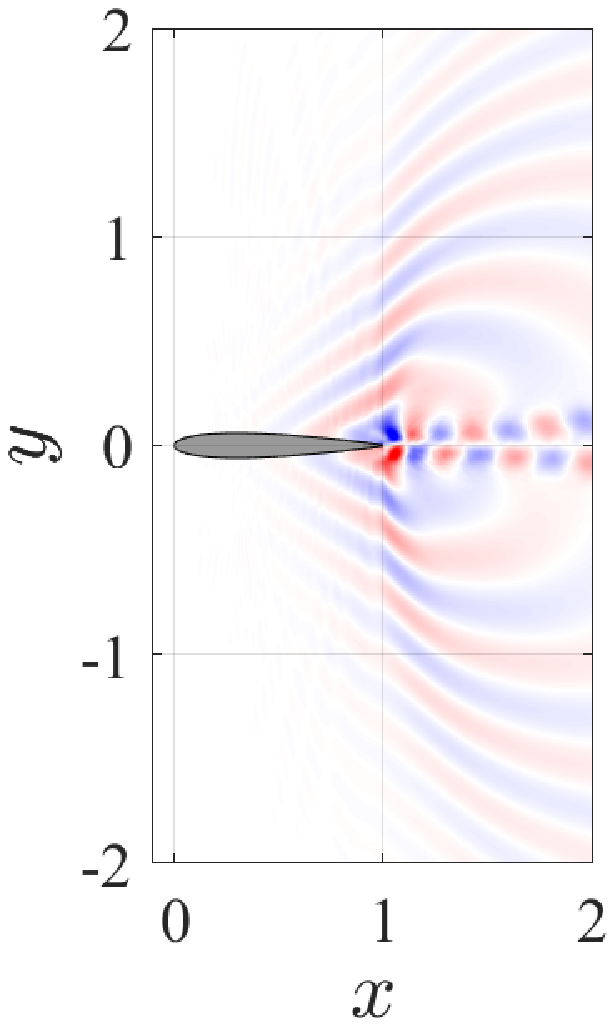}
\includegraphics[trim={3cm 0cm 3.5cm 0cm},clip,width=.32\textwidth]{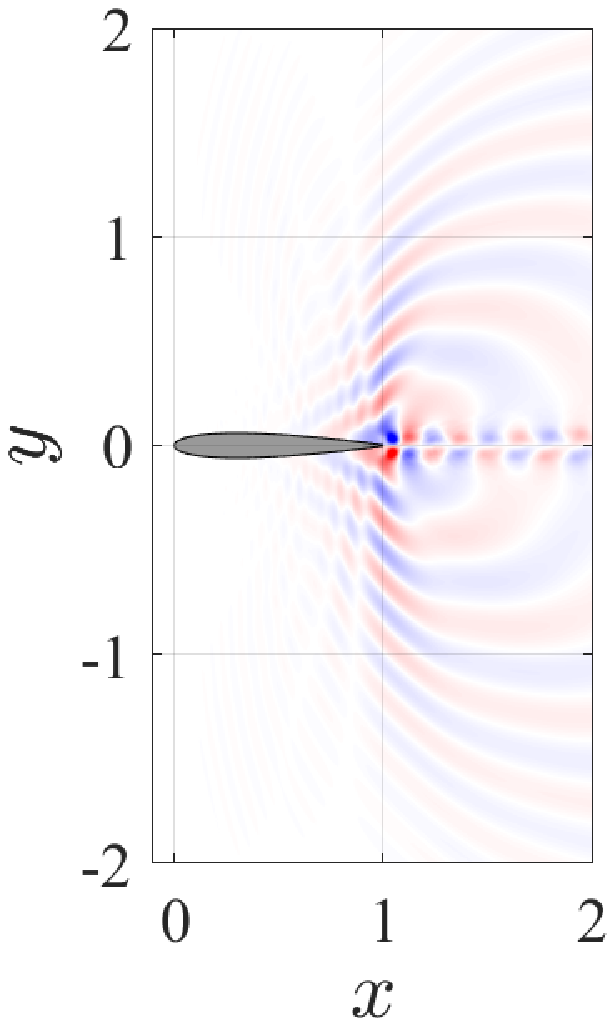}
\includegraphics[trim={3cm 0cm 3.5cm 0cm},clip,width=.32\textwidth]{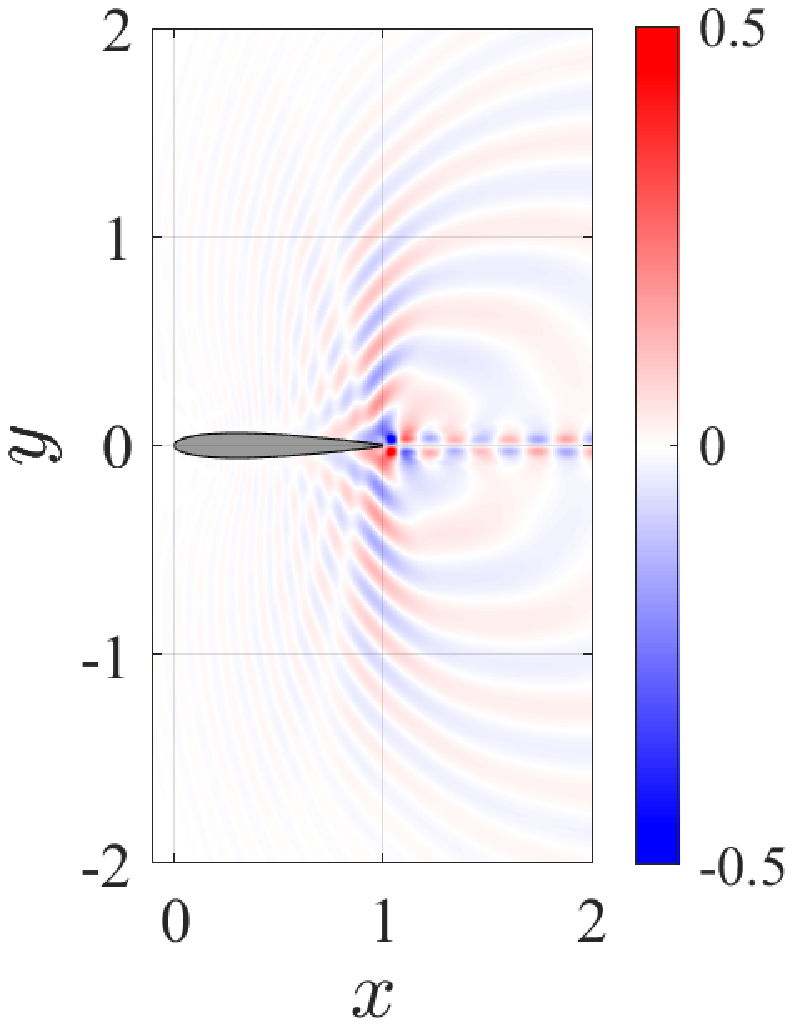}
\caption{Buffet (top) and wake (bottom) modes from SPOD are shown using contour plots of the pressure field at $\alpha = 0^\circ$ and (a) $M = 0.8$, (b) $M = 0.75$ and (c) $M = 0.72$.}
    \label{fNACA0SPODModes}
\end{figure}

The frequency variation of the first two eigenvalues obtained using a SPOD is shown in figure~\ref{fNACASPODSpectrum}\textit{a} for the representative case of $M = 0.75$. At the buffet frequency, it is seen that the energy content associated with the second SPOD mode is an order of magnitude smaller compared to the first, while there is significant energy content in the second SPOD mode at higher frequencies associated with vortex shedding ($St \approx 3$). Since the focus of this study is on buffet, only the dominant eigenvalue and its eigenfunction from SPOD (\textit{i.e.}, $\lambda_1$ and $\boldsymbol{\psi}_1$) are further considered. The spectra associated with the dominant eigenvalue for various $M$ are compared in figure~\ref{fNACASPODSpectrum}\textit{b}. Similarities with the spectra based on the lift coefficient (figure~\ref{fNACAClPSD}\textit{a}) are evident, and we refer to the SPOD modes associated with the low- and high-frequency peaks as buffet and wake modes, respectively. The spatial structure of these modes is compared for different $M$ in figure~\ref{fNACA0SPODModes} using contours of pressure. Note that to facilitate comparisons, a reference phase in the oscillation cycle of a SPOD mode from equation~\ref{eqnSPODSpatioTemp} needs to be chosen. In this study, we have chosen the phase relative to the instant when the lift fluctuation (computed using the SPOD mode's pressure field) reaches a maximum in an oscillation cycle. Additionally, the entire temporal variation within a buffet cycle is provided for the cases of $M = 0.75$ and 0.8 (movies 4 and 5, Supplementary material).

The pressure field of the wake modes for all $M$ consist of a von~K\'arm\'an vortex street pattern and pressure waves that originate from near the trailing edge. The buffet modes for all $M$ also resemble each other. Note that for a Type I buffet, the same features seen on the suction side are mirrored on the pressure side, but of opposite sign. Focusing on the suction side, it is seen that the pressure reduction above the aerofoil (large blue region) is coupled with a pressure increase in a small region in the wake (narrow red stripe). This spatial structure is an essential characteristic of transonic buffet as discussed later in \S\ref{subSecOtherTB}. Thus, in addition to the similarity in frequency, it is seen that the spatial structure of the coherent oscillations is similar irrespective of whether the flow is transonic or not, corroborating previous conclusions. We also note that while the symmetric NACA0012 aerofoil is the focus of this study (motivated by results reported in \citet{Bouhadji2003} and \citet{Jones2006}), oscillations resembling transonic buffet can also occur for supercritical aerofoils, as shown in appendix~\ref{secV2C} for Dassault Aviation's V2C aerofoil at $\alpha=0^\circ$. This is seen for the V2C aerofoil for parametric values almost the same as NACA0012 aerofoil and the SPOD modes also have a similar spatial structure. 

In summary, self-sustained oscillations at a low frequency of $St \approx 0.1$ can be observed for the NACA0012 aerofoil at zero incidence for a freestream Reynolds number of $Re = 50,000$. These resemble transonic buffet, but appear irrespective of whether the flow is transonic or not. Thus, it is established in this section that shock waves are not essential for Type I transonic buffet to occur. In conjunction with the result that transonic buffet offset occurs at $M = 0.9$, this implies that shock waves are neither necessary nor sufficient for these oscillations to sustain. Henceforth, we will use `transonic buffet' and `transonic-buffet-like oscillations' (TBLO) to distinguish between cases for which the flow remains always transonic and for which it does not, respectively, and `buffet' to collectively refer to either. 

\subsection{Effect of spanwise extent}
\label{subSecDomain}
\begin{figure} 
\centerline{
\includegraphics[width=0.45\textwidth]{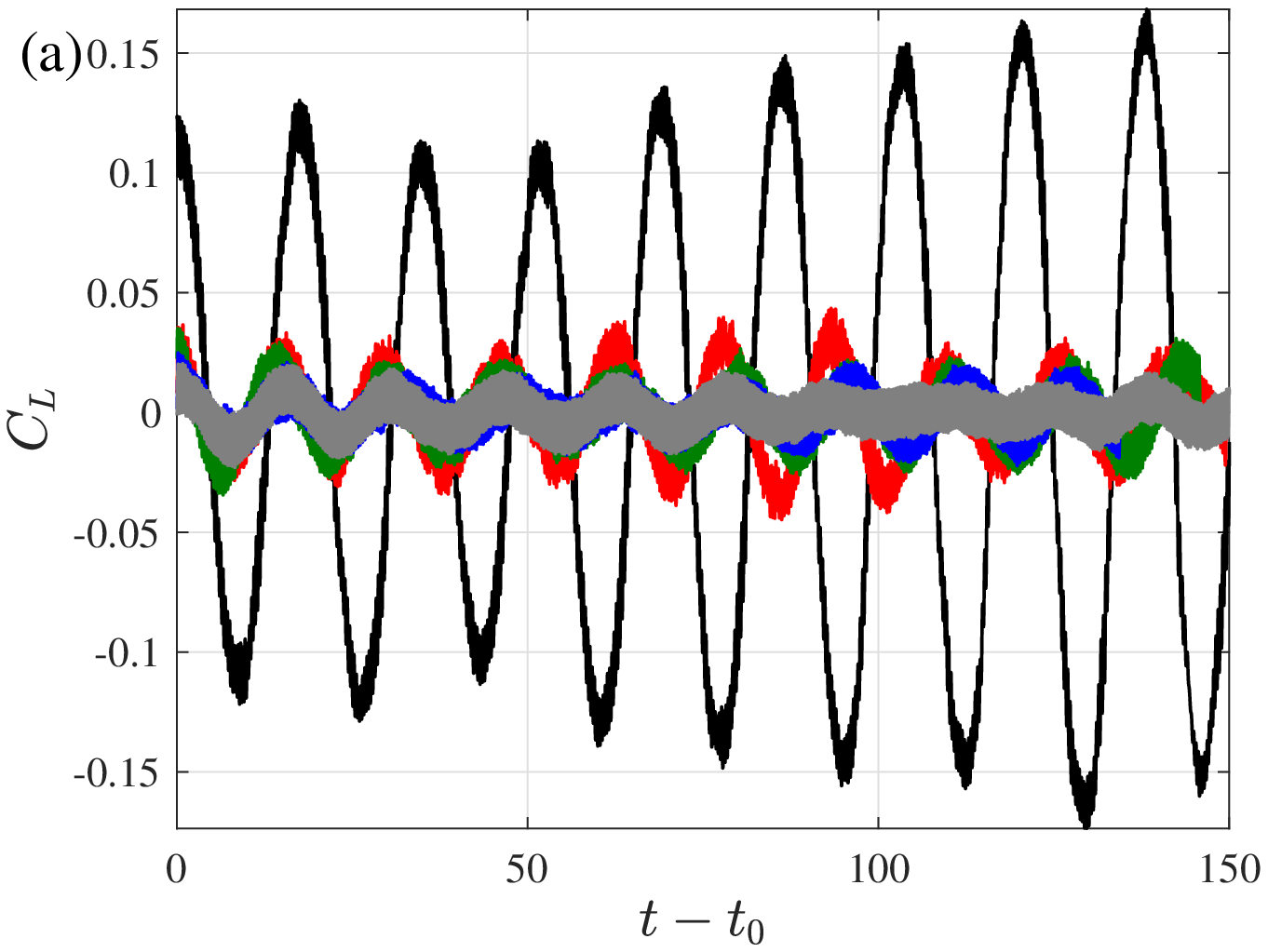}
\includegraphics[width=0.45\textwidth]{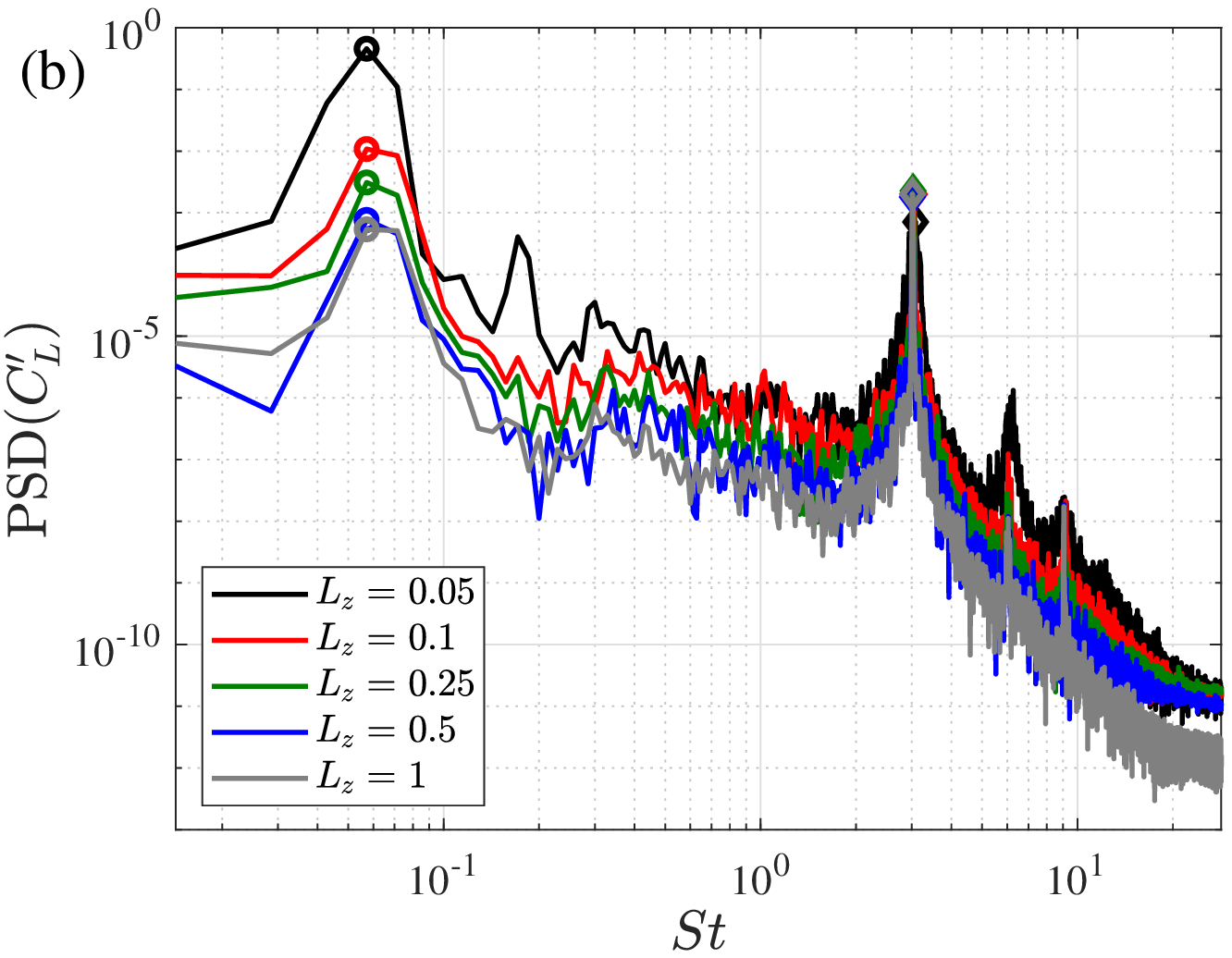}
}
\caption{(a) Temporal variation of lift coefficient past transients and (b) PSD of its fluctuating component as a function of the Strouhal number for $M = 0.75$, $\alpha = 0^\circ$ and different spanwise domain widths. Circles and diamonds highlight the buffet and wake mode Strouhal numbers, respectively.}
\label{fNACAClPSDLz}
\end{figure}

\begin{table}
\begin{center}
\def~{\hphantom{0}}
\begin{tabular}{lcccc}
$L_z$ & $St_b$ & $St_w$ & PSD$_b$  & PSD$_w$ \\[6pt]
0.05  & 0.057  & 3.1    &   0.4566 & 0.0007 \\
0.1   & 0.057  & 3.1    &   0.0108 & 0.0020 \\
0.25  & 0.057  & 3.0    &   0.0031 & 0.0023 \\
0.5   & 0.057  & 3.0    &   0.0008 & 0.0018 \\
1     & 0.057  & 3.0    &   0.0005 & 0.0020 \\
\end{tabular}
\caption{Comparison of buffet and wake mode features for $M = 0.75$, $\alpha = 0^\circ$, $Re = 5\times10^4$ and different spanwise widths for the NACA0012 aerofoil (all cases reported in \S\ref{subSecDomain}).}
\label{tableDomainEffects}
\end{center}
\end{table}

Although transonic buffet on unswept infinite-wing sections is essentially two-dimensional, it has been shown in \citet{Zauner2020PRF} that the spanwise width can have a significant effect on its amplitude (but not frequency). Motivated by this, we examined the amplitudes of the TBLO that occur at $M = 0.75$ for wider spanwise widths of $L_z = 0.1$, 0.25, 0.5 and 1. A similar study for transonic buffet at $M = 0.8$ is presented in appendix~\ref{appLzEffectOnM8}. The temporal variation of the lift coefficient and its power spectral density are shown in figure~\ref{fNACAClPSDLz}. The Strouhal number and the power spectral densities of the peaks in the spectra (buffet and wake modes denoted by subscripts `$b$' and `$w$') are also documented in table~\ref{tableDomainEffects}. The effect of $L_z$ on both $St_b$ and $St_w$ is negligible. However, the spanwise width has a strong effect on the amplitude of the oscillations, with the sharpest drop observed when $L_z$ is doubled from 0.05 to 0.1. Further increase in $L_z$ leads to smaller reductions in the amplitude, although sinusoidal oscillations are evident for all cases examined, including $L_z = 1$ (grey curve). Note that for all cases, there are variations in the amplitudes between buffet cycles (\textit{i.e.}, irregular cycles), although the oscillations are persistent. 

\begin{figure} 
\centering
\includegraphics[trim={2.8cm 0cm 3.5cm 0cm},clip,width=.32\textwidth]{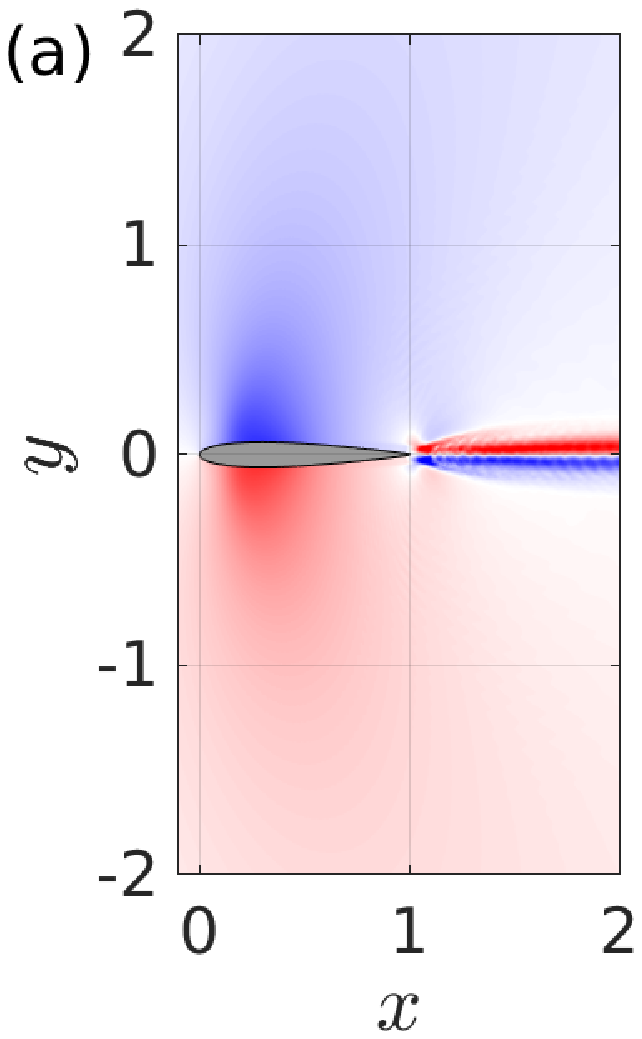}
\includegraphics[trim={2.8cm 0cm 3.5cm 0cm},clip,width=.32\textwidth]{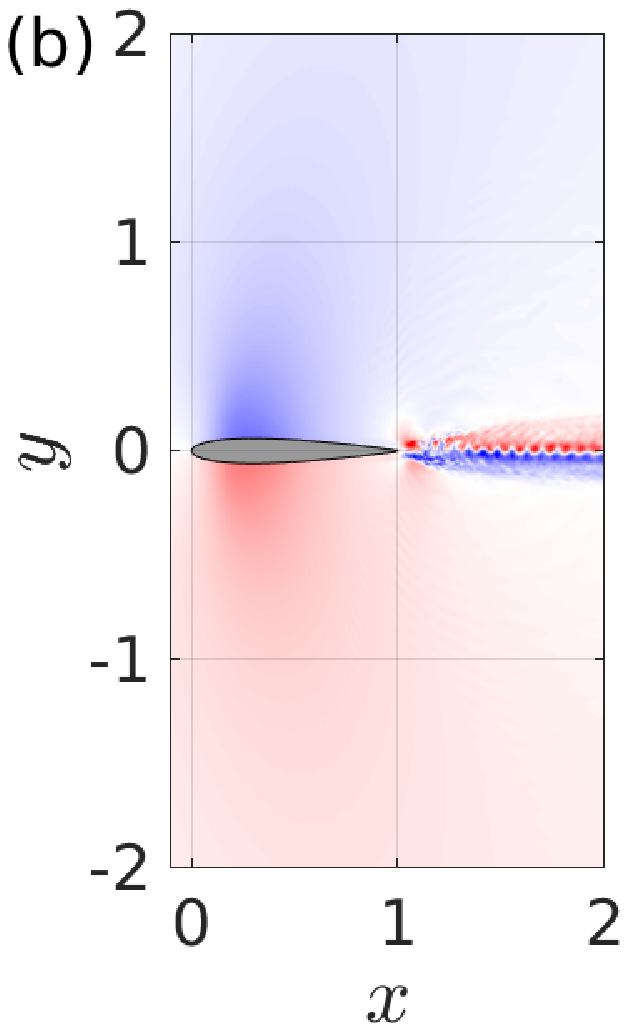}
\includegraphics[trim={2.8cm 0cm 3.5cm 0cm},clip,width=.32\textwidth]{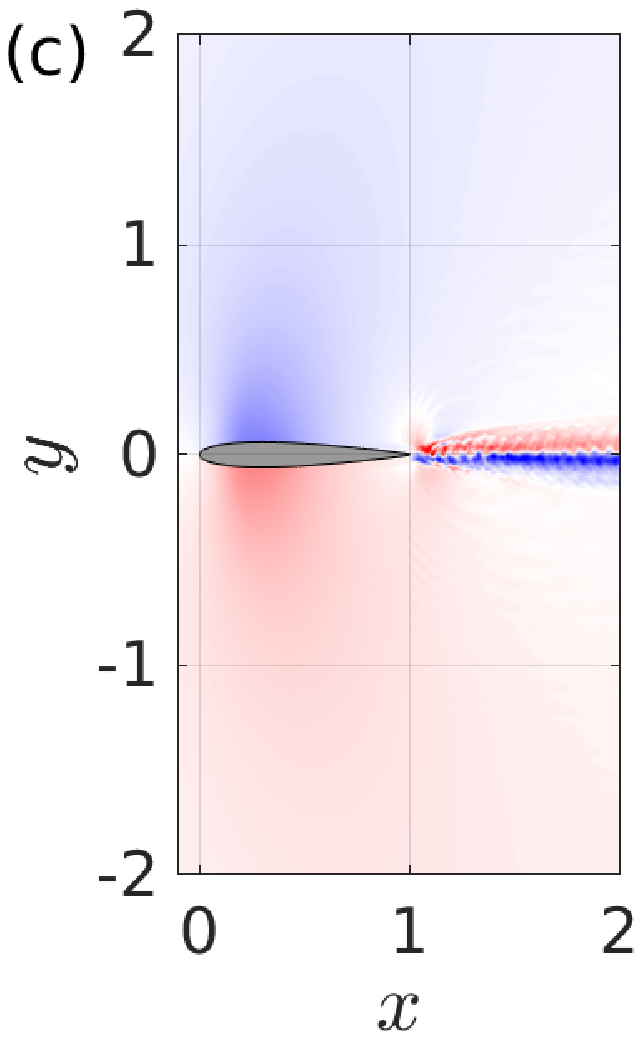}
\includegraphics[trim={2.8cm 0cm 3.5cm 0cm},clip,width=.32\textwidth]{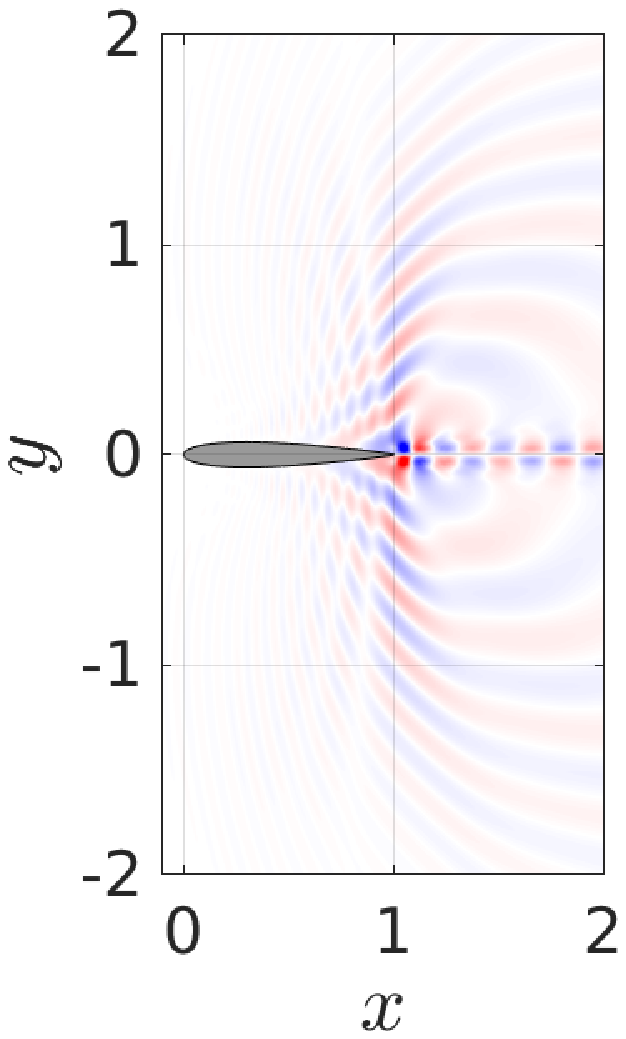}
\includegraphics[trim={2.8cm 0cm 3.5cm 0cm},clip,width=.32\textwidth]{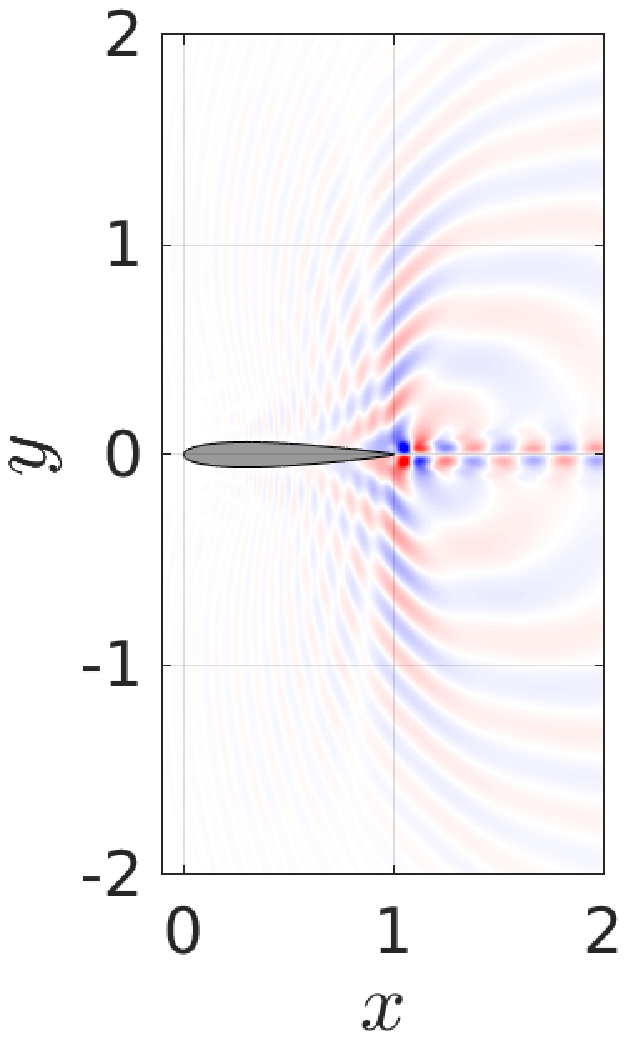}
\includegraphics[trim={2.8cm 0cm 3.5cm 0cm},clip,width=.32\textwidth]{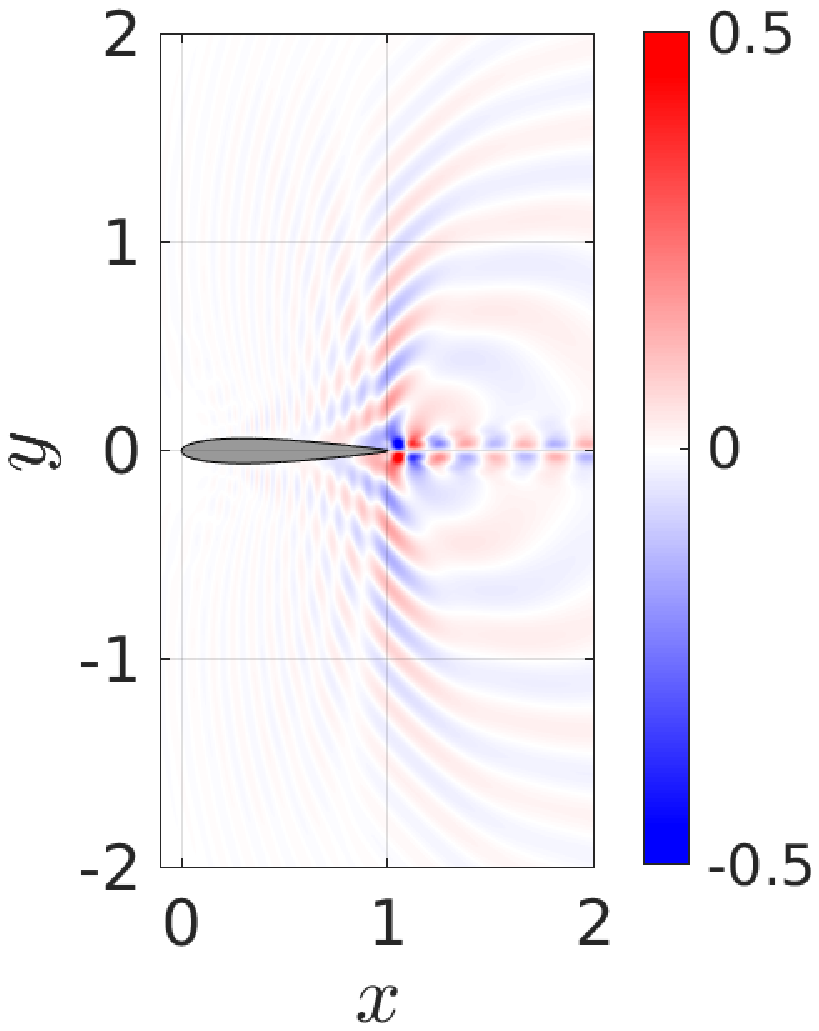}
\caption{Buffet (top) and wake (bottom) modes from SPOD are shown using contour plots of the pressure field at $M = 0.75$ and $\alpha = 0^\circ$ for (a) $L_z = 0.1$, (b) $L_z = 0.5$ and (c) $L_z = 1$.}
    \label{fNACADomainSPODModes}
\end{figure}

\begin{figure} 
\centerline{
\includegraphics[width=0.45\textwidth]{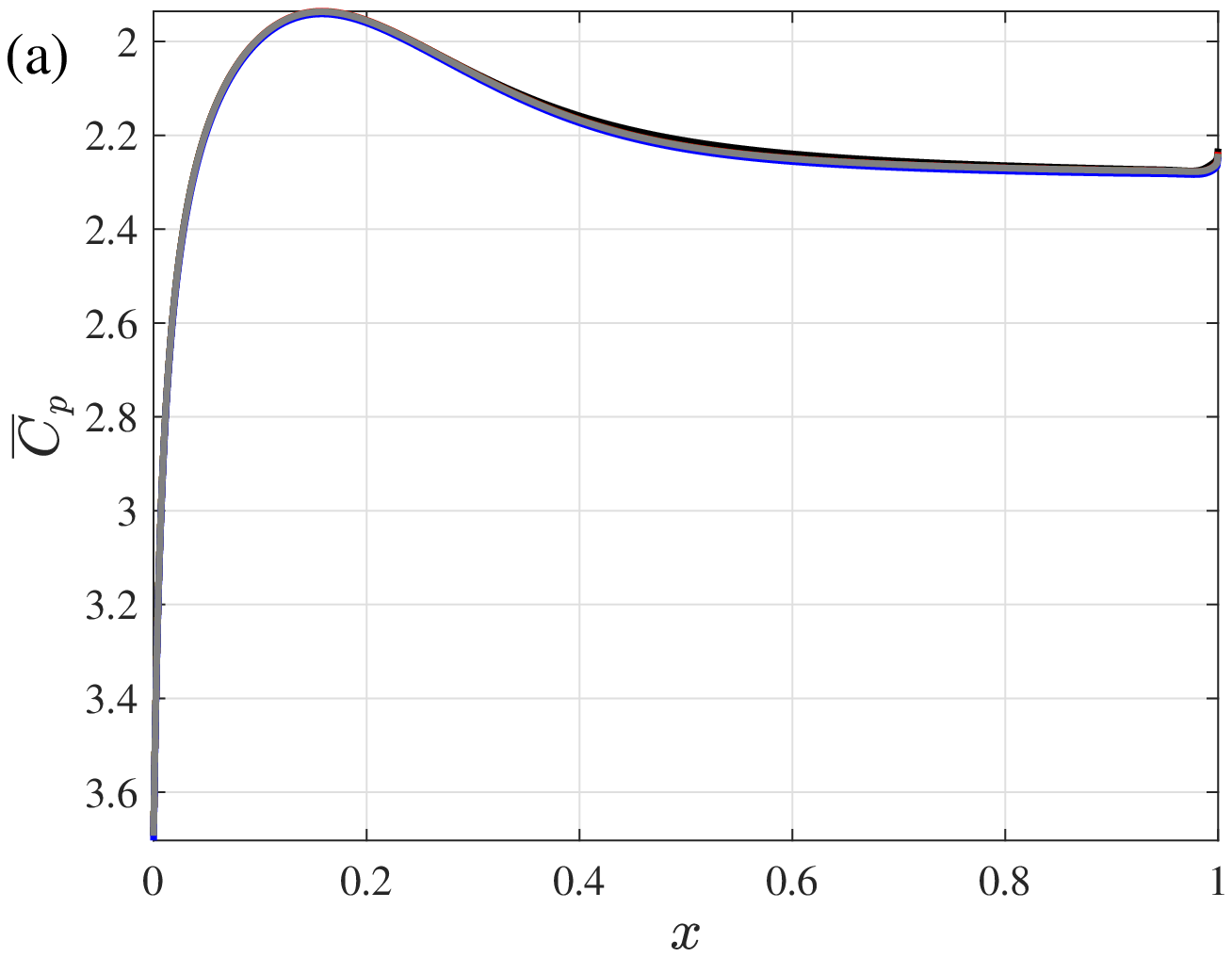}
\includegraphics[width=0.45\textwidth]{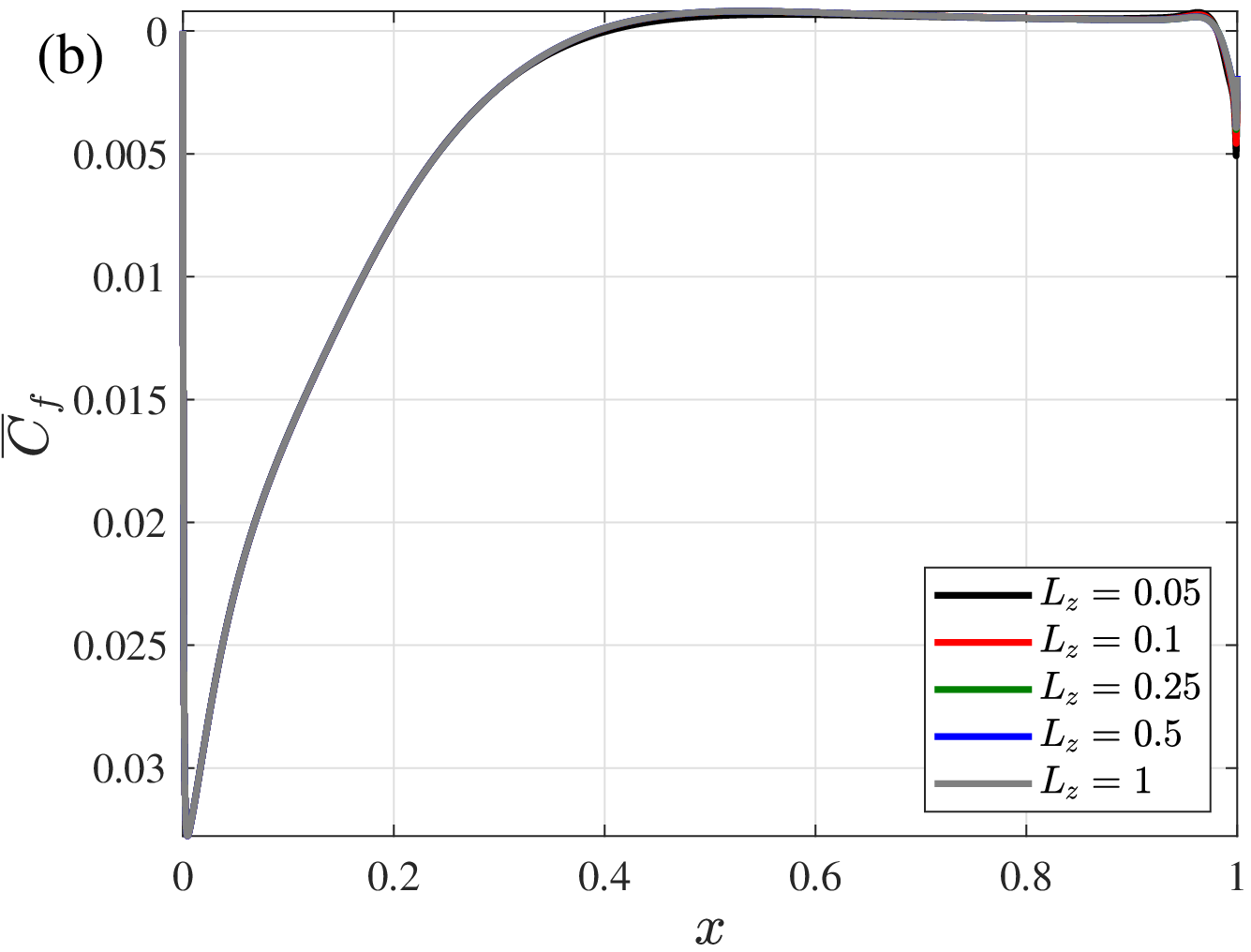}
}
\caption{Time- and span-averaged (a) pressure coefficient and (b) skin-friction coefficient variation along aerofoil surface for $M = 0.75$, $\alpha = 0^\circ$ and different domain widths.}
\label{fNACACpCfLz}
\end{figure}

Examining the spatio-temporal flow field in the $x-y$ plane, it was found that TBLO features are qualitatively similar for all $L_z$ (see movie 6 for the case $L_z = 1$ provided as Supplementary material). The SPOD spectra were also found to be similar, with peaks present at buffet and wake mode frequencies (not shown for brevity). The spatial features of these modes are shown for a few select cases in figure~\ref{fNACADomainSPODModes} and can be seen to be qualitatively similar. Overall, the only major difference observed was that, unlike the case of $L_z = 0.05$, where the flow has a small supersonic region in the high- and low-lift phases (see white curve highlighting sonic line in figure~\ref{fNACADensGrad}\textit{b}), the flow remains subsonic at all times for wider domains (see movie 6). Thus, the conclusion that TBLO can be sustained in the subsonic regime holds true even for the largest span considered. Nevertheless, these results suggest that the spanwise extent somehow plays a crucial role in determining buffet amplitude. This is further confirmed by examining the mean flow features. The time- and span-averaged pressure and skin-friction coefficients are compared for different $L_z$ in figure~\ref{fNACACpCfLz}. It can be inferred from these plots that the mean flow is not significantly altered by the variation in spanwise width, even for the cases of $L_z = 0.05$ and $L_z = 0.1$ for which the power spectral density changes by a factor of approximately four. Assuming that buffet arises as a global instability \citep{Crouch2007}, this implies that the growth and/or saturation features of this instability are strongly affected by the spanwise width. This trend is also observed in the presence of shock waves (\textit{i.e.} Type I transonic buffet), as shown in appendix~\ref{appLzEffectOnM8}. However, these results cannot be generalised -- for Type II transonic buffet on the supercritical V2C aerofoil, \citet{Zauner2020PRF} showed that amplitudes of oscillations increase with an increase in span, which is opposite to the trend observed here. 

Although the imposition of periodic boundary conditions in the spanwise direction can artificially confine the flow, nevertheless, it provides a controlled numerical experiment that highlights the crucial role that three-dimensional features can play in determining buffet amplitude. This effect could arise from span-varying disturbances affecting boundary layer or wake transition characteristics, but the connection is not obvious. 
Further analysis towards understanding this effect is required, but we have restricted the scope of this study so as to focus on exploring the links between TBLO and transonic buffet.

\section{The link between transonic buffet and low-frequency oscillations}
\label{secLFO}

\subsection{Link between Type I transonic buffet and LFO}
\label{subSecLFOLowRe}
The results at $\alpha = 0^\circ$ and $Re = 5\times 10^4$ discussed in the preceding section establish the link between Type I transonic buffet and TBLO. To extend this further and relate with LFO which occur in the incompressible regime (\textit{i.e.}, low $M$) and close to stall (\textit{i.e.}, high $\alpha$), the $\alpha-M$ parameter space is further explored here by both increasing $\alpha$ and decreasing $M$ simultaneously at this $Re$. It is emphasised that each simulation is performed with the same type of initial conditions of uniform flow and constant boundary conditions and incidence angle (\textit{i.e.}, this is not a hysteresis study). The incidence angles considered are $0^\circ$ to $8^\circ$ in steps of $2^\circ$, and additionally, $\alpha = 9.4^\circ$. For each $\alpha$, the freestream Mach number required for the simulation was estimated based on the oscillation features observed at lower $\alpha$. Due to the numerical expense involved, no attempts were made to further vary $M$ at a given $\alpha$ as long as sustained oscillations were present at the estimated $M$. Also, based on the results reported in the previous section on domain extent, $L_z = 0.1$ was chosen for all the simulations in this section as a compromise between accuracy and numerical expense (see figure~\ref{fNACAClPSDLz} and table~\ref{tableDomainEffects}). All cases reported in this section and their buffet characteristics are summarised in table~\ref{tableHighAoA}.
\begin{table}
\begin{center}
\def~{\hphantom{0}}
\begin{tabular}{cccc}
$\alpha$    & $M$   &  $St_b$ &  PSD$_b$ \\[6pt]
$0^\circ$   & 0.75  &  0.057  &  0.0108  \\
$2^\circ$   & 0.70  &  0.018  &  0.0163  \\
$4^\circ$   & 0.70  &  0.033  &  0.0214  \\
$6^\circ$   & 0.50  &  0.033  &  0.0816  \\
$8^\circ$   & 0.40  &  0.033  &  0.1203  \\
$9.4^\circ$  & 0.30  &  0.033  &  0.2024  \\
\end{tabular}
\caption{Comparison of buffet and wake mode features for $L_z = 0.1$, $Re = 5\times10^4$ and different ($M$,$\alpha$) for the NACA0012 aerofoil (all cases considered in \S\ref{subSecLFOLowRe}; case, $\alpha = 0^\circ$, $M = 0.75$, included for reference).}
\label{tableHighAoA}
\end{center}
\end{table}

\begin{figure} 
\centerline{
\includegraphics[width=0.45\textwidth]{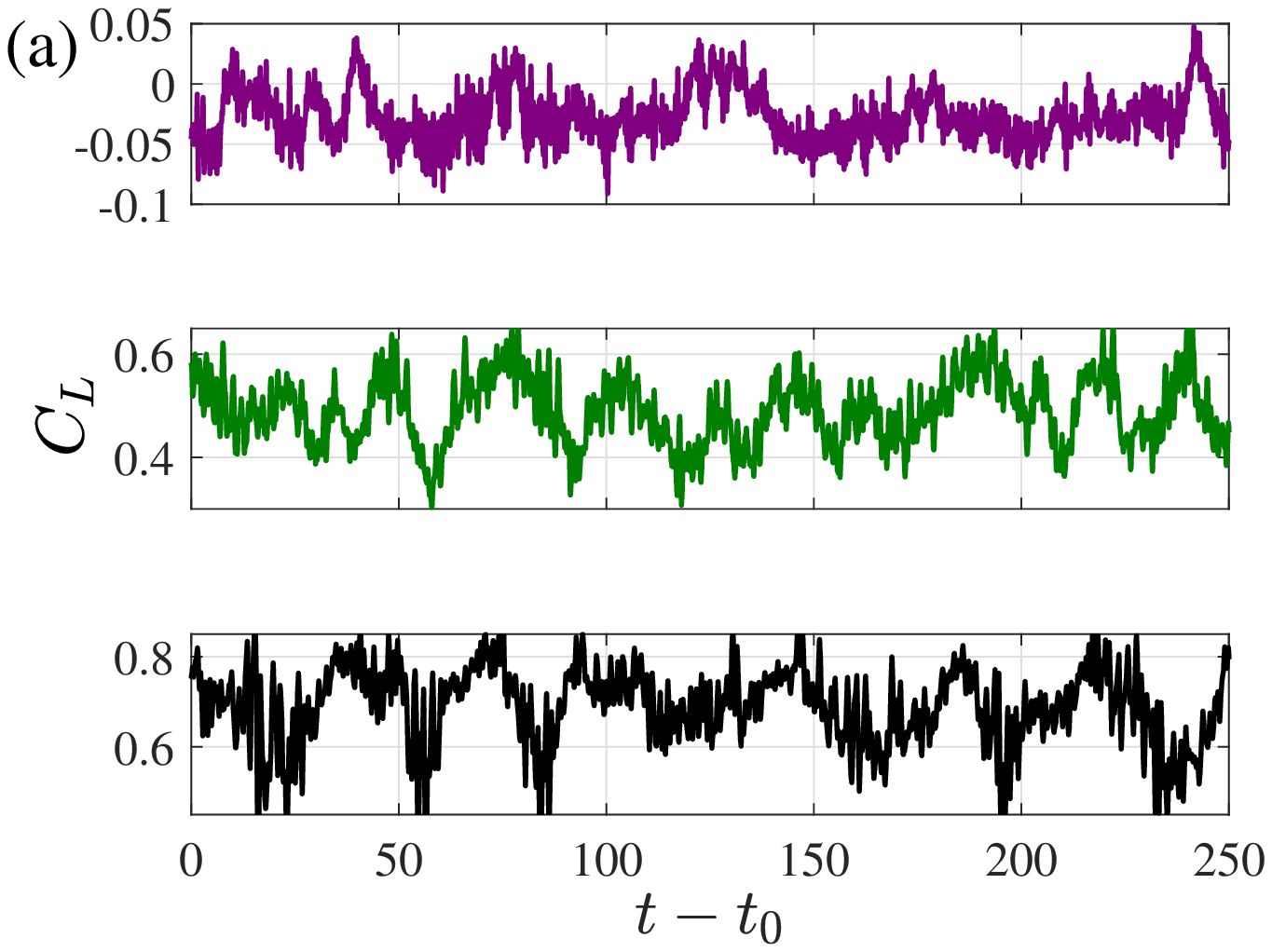}
\includegraphics[width=0.45\textwidth]{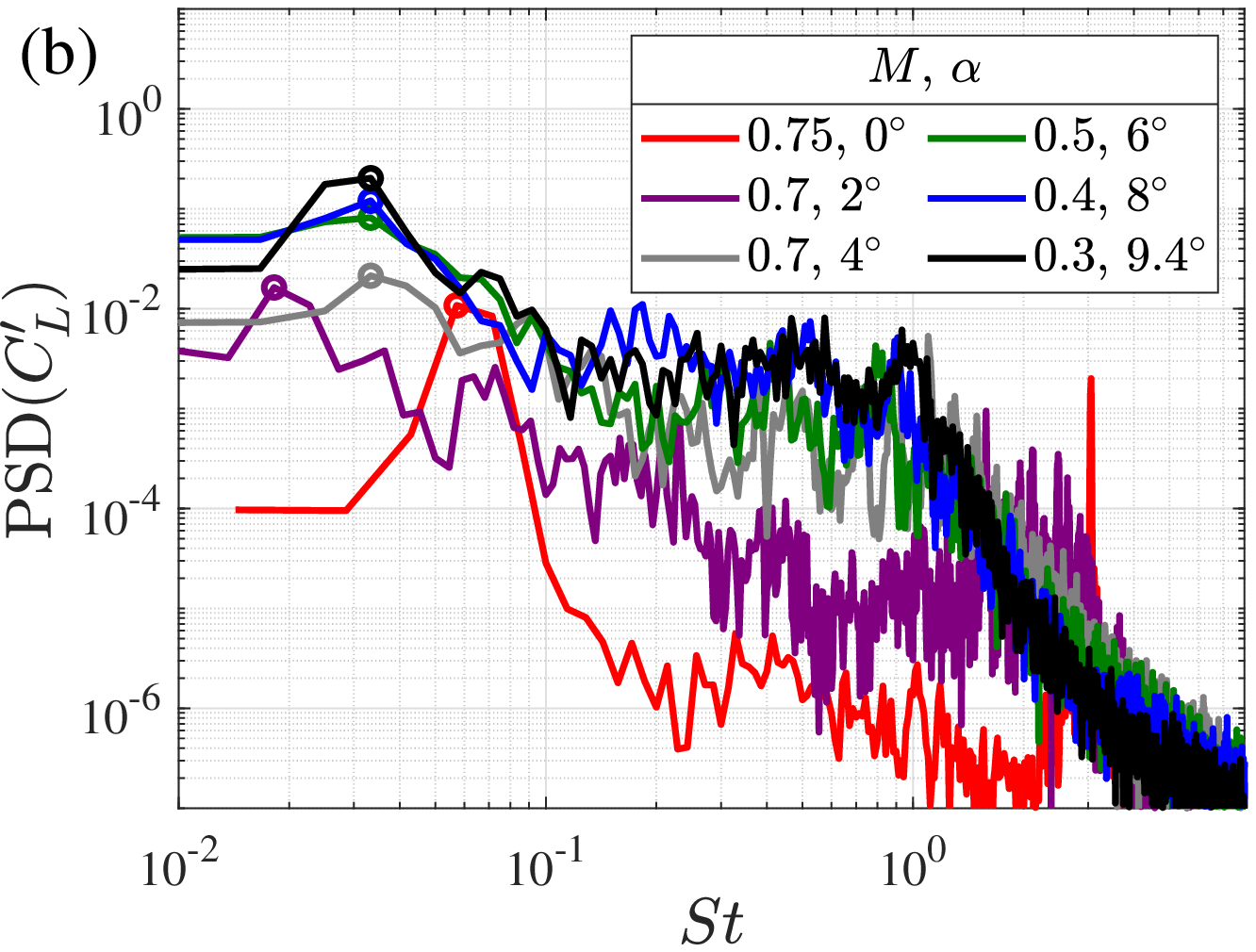}
}
\caption{(a) Temporal variation of lift coefficient past transients for $M = 0.7$, $\alpha = 2^\circ$ (top), $M = 0.5$, $\alpha = 6^\circ$ (middle) and $M = 0.3$, $\alpha = 9.4^\circ$ (bottom), and (b) PSD of its fluctuating component as a function of the Strouhal number, $St$ when $\alpha$ and $M$ varied simultaneously at $Re = 5\times10^4$.}
\label{fNACAClPSDLFO}
\end{figure}

The lift variations for a few of these cases are shown in figure~\ref{fNACAClPSDLFO}\textit{a}. The variations are strongly irregular and this makes it difficult to discern if buffet occurs. Even for the case of $\alpha = 9.4$ and $M = 0.3$ (black curve) for which buffet can be visually inferred, there are periods when these oscillations at a low frequency are almost completely damped out, as seen for example, in $100 \leq t-t_0 \leq 150$. However, these oscillations recover and remain persistent at later times of $t > 150$. The power spectral densities of lift fluctuations for all cases studied are shown in figure~\ref{fNACAClPSDLFO}\textit{b}. A case at zero-incidence ($M=0.75$, $L_z = 0.1$, discussed in \S\ref{subSecDomain}) is additionally provided for reference. It is evident that there are peaks in the spectra at a low frequency (highlighted by circles) indicating buffet for all cases, although there is wide variability in both $St_b$ and PSD$_b$. Nevertheless, as shown later, SPOD confirms that the spatial structure of the modes associated with these peaks is the same (figure~\ref{fNACALFOSPODModes}). Unlike the monotonic increase in $St_b$ seen in \S\ref{subSecNarrowFreeMach} when only $M$ is varied, the variation here does not have any specific trend. This is because although $M$ is increased, $\alpha$ is reduced simultaneously and transonic buffet frequency can vary in a non-monotonic fashion when $\alpha$ is changed \citep[cf.][figure~17\textit{b}]{Moise2022}. Similarly, note that the buffet amplitude at a given $\alpha$ has a nonlinear variation with $M$ (given that it is negligible at onset and offset values of $M$). Thus, monotonic trends in the frequency or amplitude of buffet are not expected when both $M$ and $\alpha$ are varied simultaneously. Note that, unlike the zero-incidence case, there is no discrete peak associated with a wake mode for the others. Instead, there is a broadband bump in the spectra centred about $St \approx 1$. A similar bump associated with wake modes is commonly reported for transonic buffet at non-zero incidences \citep[\textit{e.g.}][]{Moise2022}). Shock waves were absent in the flow field for all non-zero incidence angles studied, implying that the sustained oscillations can be categorised as TBLO for these cases. For $\alpha \geq 4^\circ$, the energy content on the suction side was dominant and thus, these cases will also be referred to as Type II TBLO.   

\begin{figure} 
\centerline{
\includegraphics[width=0.495\textwidth]{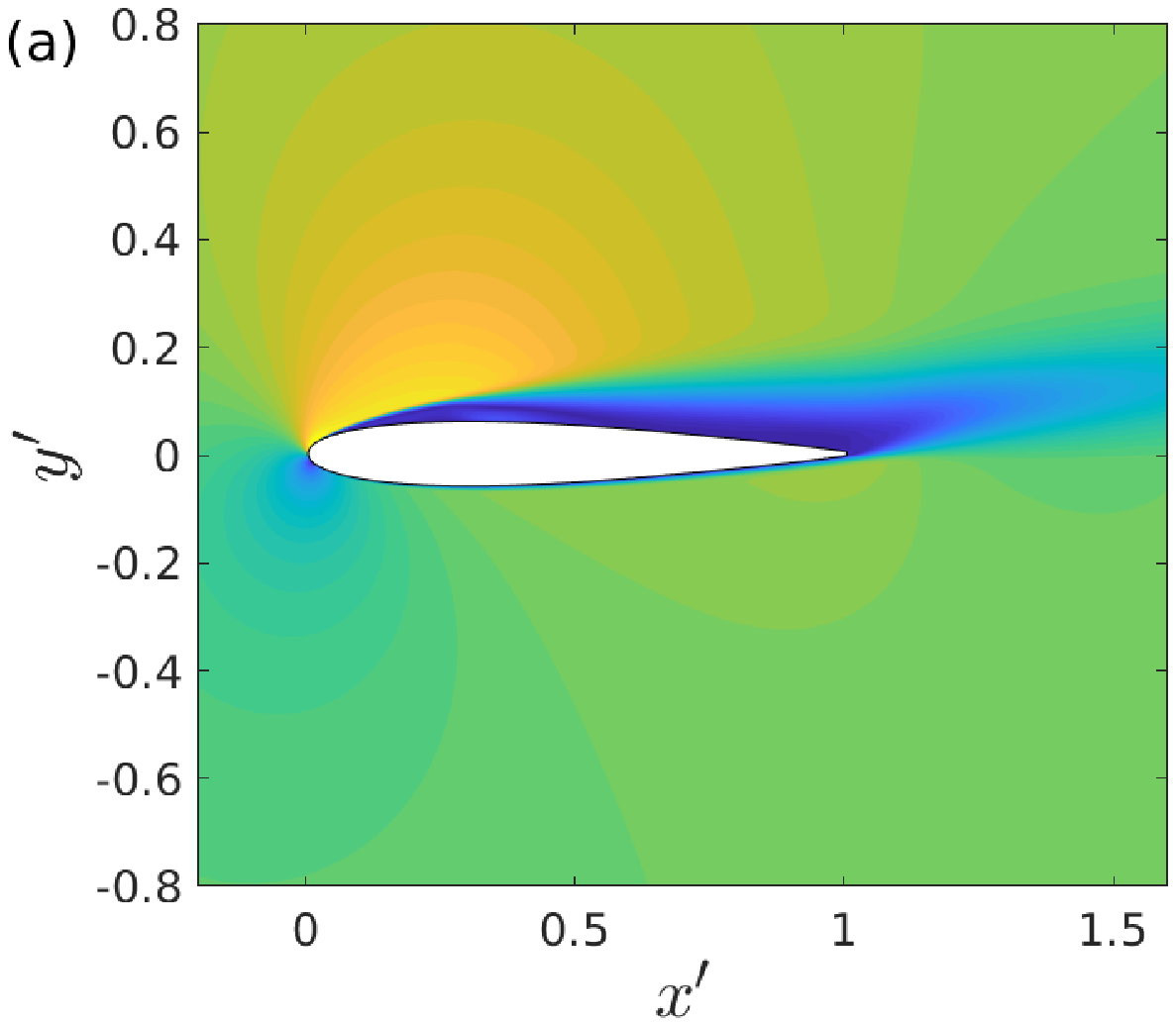}
\includegraphics[width=0.495\textwidth]{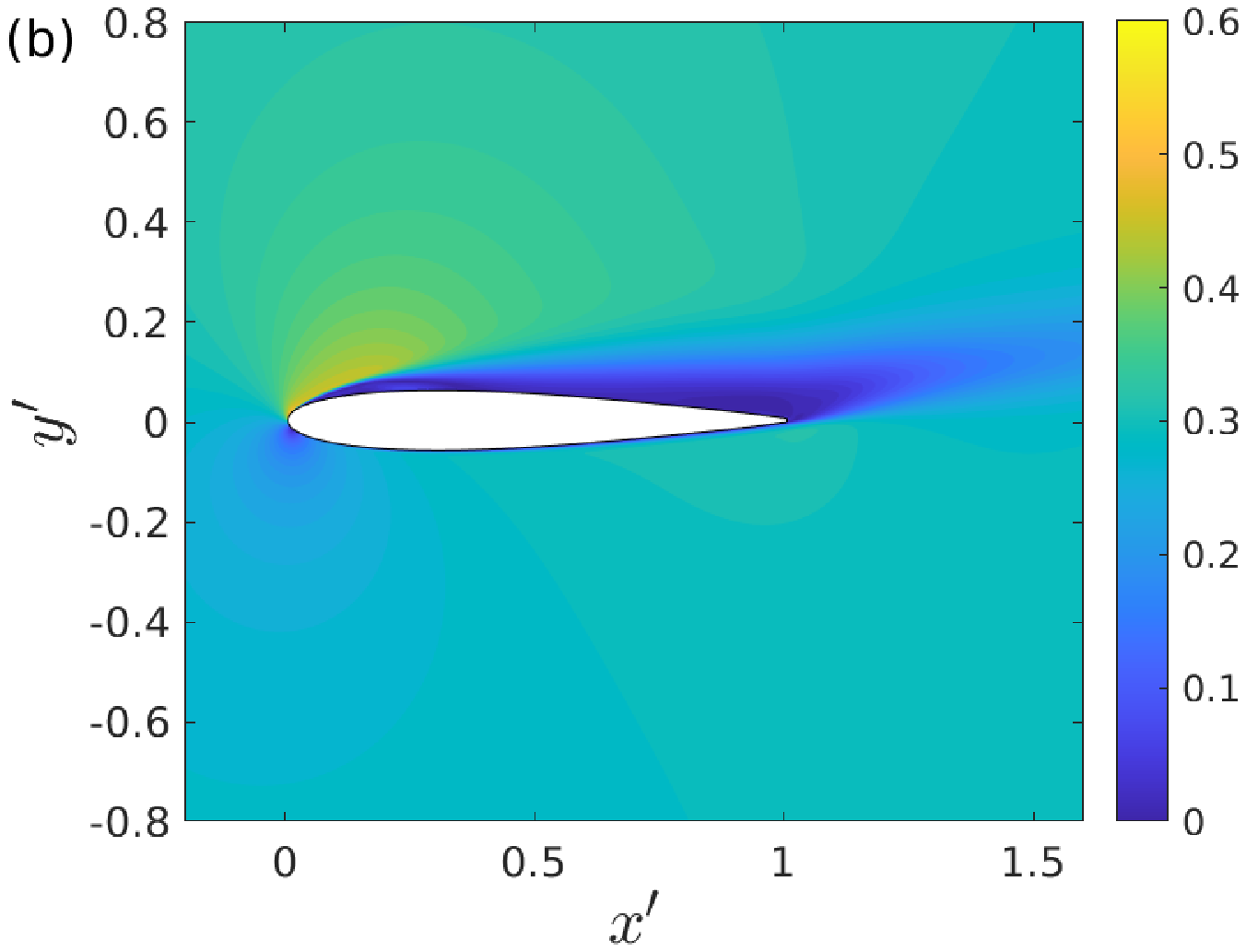}
}
\caption{Contours of $\overline{M}_\mathrm{loc}$ for $Re = 5\times10^4$ and (a) $\alpha = 8^\circ$, $M = 0.4$ and (b) $\alpha = 9.4^\circ$, $M = 0.3$.}
\label{fNACAContMachLFO}
\end{figure}

\begin{figure} 
\centering
\includegraphics[trim={2.5cm 0cm 3.5cm 0cm},clip,width=.32\textwidth]{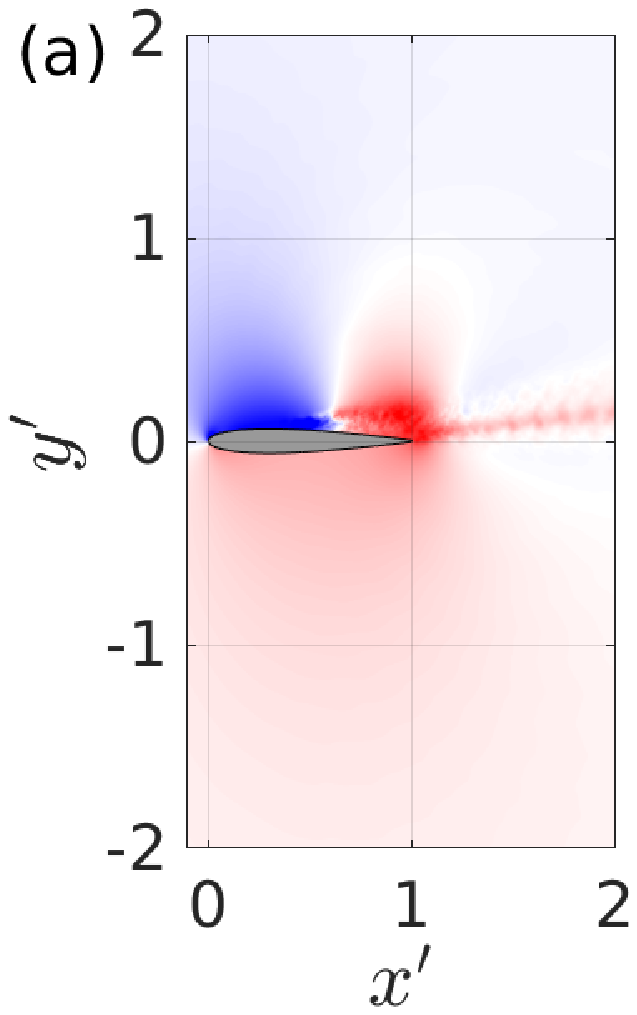}
\includegraphics[trim={2cm 0cm 3.5cm 0cm},clip,width=.32\textwidth]{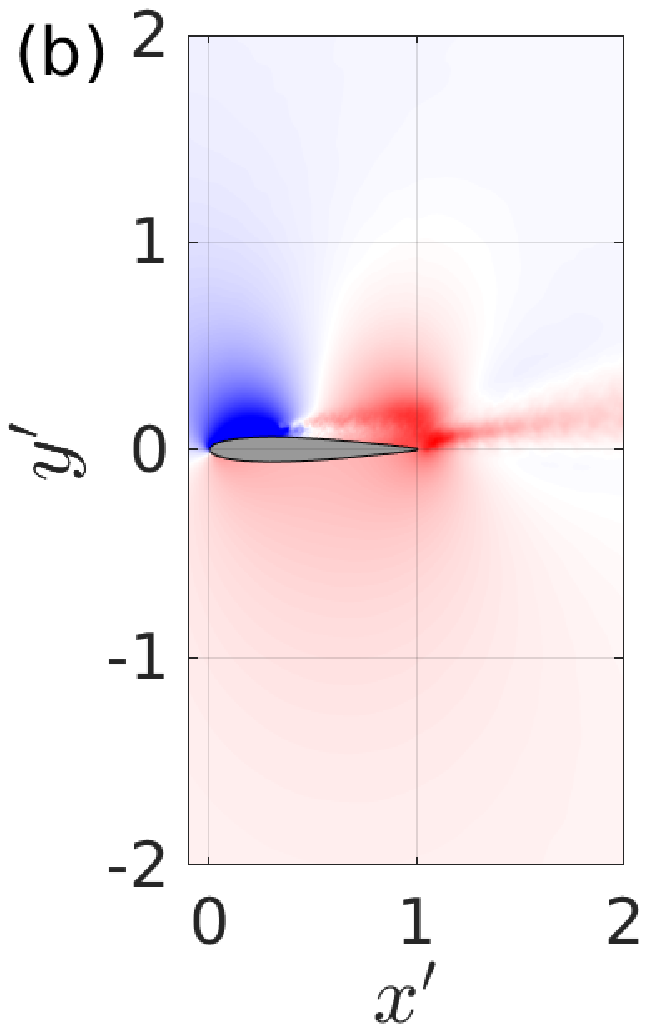}
\includegraphics[trim={2.5cm 0cm 3.5cm 0cm},clip,width=.32\textwidth]{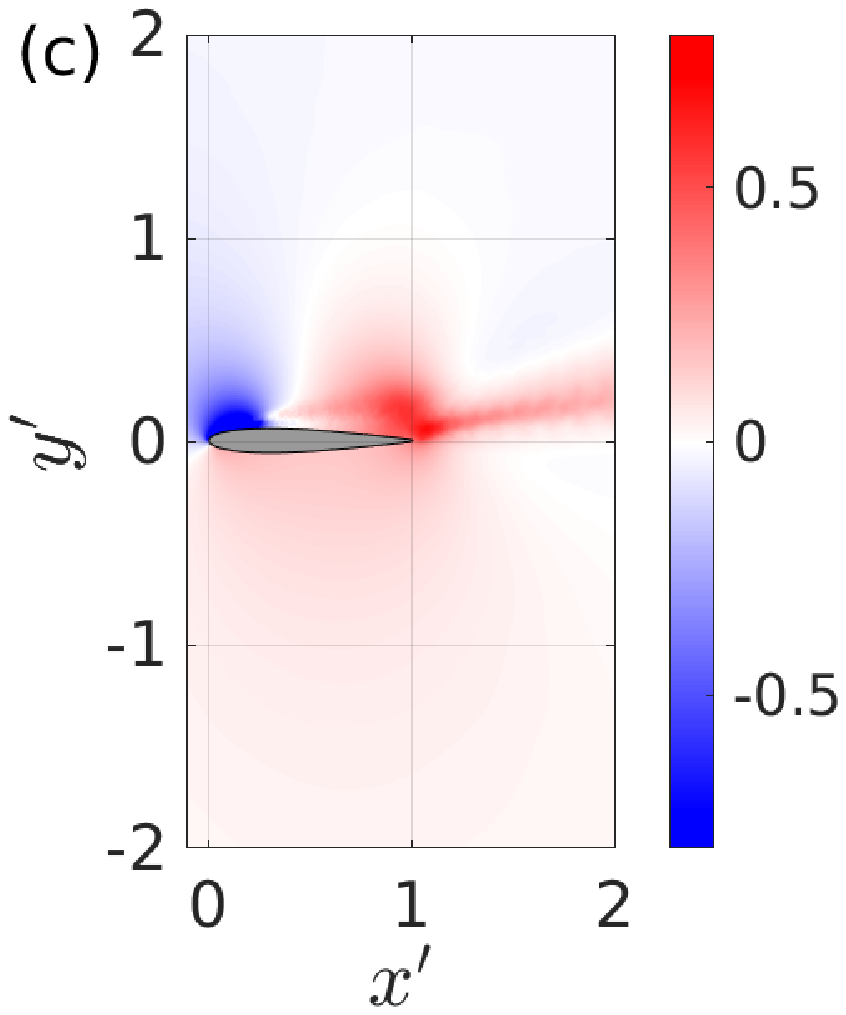}
\caption{Buffet modes from SPOD are shown using contour plots of the pressure field for  at $Re = 5\times10^4$ and (a) $\alpha = 6^\circ$, $M = 0.5$, (b) $\alpha = 8^\circ$, $M = 0.4$ and (c) $\alpha = 9.4^\circ$, $M = 0.3$.} 
    \label{fNACALFOSPODModes}
\end{figure}

Contours of time- and span-averaged local Mach number, $\overline{M}_\mathrm{loc}$, are shown for the cases of $\alpha = 8^\circ$, $M = 0.4$ and $\alpha = 9.4^\circ$, $M = 0.3$ in figure~\ref{fNACAContMachLFO}. The maximum $\overline{M}_\mathrm{loc}$ attained is 0.65 for the former and 0.51 for the latter and most regions of the flow are found to have $\overline{M}_\mathrm{loc} \leq 0.4$ for the latter case. Thus, it can be inferred that compressibility effects are insignificant in most of the flow field, especially when $\alpha = 9.4^\circ$ and $M = 0.3$. Although the requirement for a reduced time-step prevented exploration at even lower $M$, the present trend of sustained oscillations occurring for decreasing maximum values, $\max(\overline{M}_\mathrm{loc})$, with increasing $\alpha$ and decreasing $M$ suggests that these oscillations can also sustain in the incompressible regime. The peak in the spectrum for the lowest $M$ simulated occurs at a frequency similar to those reported in experiments studying LFO. For example, \citet{rinoie_takemura_2004} reported $St_h = 0.008$, where $St_h$ is the Strouhal number based on the projected height ($h = c\sin\alpha$) and freestream velocity. This result is for the NACA0012 profile at $Re = 1.3\times10^5$ for $M \ll 0.3$ and is of the same order of magnitude as seen in the present study, with $St_h = St\times \sin\alpha \approx 0.005$ for $Re = 5\times10^4$, $M = 0.3$ and $\alpha = 9.4^\circ$. Thus, we will also refer to the oscillations occurring at the lowest $M$ simulated in the present study as LFO.

The SPOD modes associated with the low-frequency peaks are shown in figure~\ref{fNACALFOSPODModes} for the higher incidences simulated. 
Additionally, an animation showing the temporal variation of the SPOD mode is provided for the highest incidence simulated (movie 7, Supplementary material). The spatial structure for all cases resembles that of Type II transonic buffet. 
A closer look shows that they are also qualitatively similar to the Type I buffet modes shown previously (cf. figure~\ref{fNACA0SPODModes}) when the suction side's features alone are considered. That is, for the present modes, a reduction of pressure on the fore part of the aerofoil (blue region) is coupled with an increase in pressure in the aft part, which extends into the wake (red region). The parametric variation indicates that the extent of the blue region spreads closer to the trailing edge as $\alpha$ is reduced. For the Type I modes at zero incidence, the blue region extends to the trailing edge, but the wake region (red) is similar to that seen here. 

In summary, these results demonstrate that both buffet frequency and spatio-temporal modal features (based on SPOD) are essentially the same for Type I transonic buffet ($\alpha = 0^\circ$ and $M = 0.8$), TBLO (all cases considered in the current section) and LFO ($\alpha = 9.4^\circ$ and $M = 0.3$). Furthermore, the results directly link these phenomena by showing that TBLO can be sustained for intermediate values of $\alpha$ and $M$ by reducing $M$ appropriately when increasing $\alpha$. We again emphasise that these low-frequency TBLO are distinct from vortex shedding (\textit{i.e.} wake modes). The latter accompanies TBLO for all the cases studied, but at a higher frequency (bump in spectra at $St \approx 1$) and is not the focus of this study.

\subsection{Link between Type II transonic buffet and LFO}
\label{subSecLFOVsHighReTB}

\begin{table}
\begin{center}
\def~{\hphantom{0}}
\begin{tabular}{cccccc}
$\alpha$    & $M$   & $Re$            & $L_z$ &  $St_b$ &  PSD$_b$ \\[6pt]
$4^\circ$   & 0.7\ & $5\times10^4$   & 0.1   &  0.033  &  0.214  \\
$4^\circ$   & 0.75  & $5\times10^5$   & 0.05  &  0.120  &  0.298   \\
$4^\circ$   & 0.75  & $1.5\times10^6$ & 0.05  &  --     &  0 \\
$6^\circ$   & 0.75  & $1.5\times10^6$ & 0.05  &  0.150  &  0.307 \\
\end{tabular}
\caption{Comparison of buffet and wake mode features for the NACA0012 aerofoil (all cases considered in \S\ref{subSecLFOVsHighReTB}; case, $\alpha = 4^\circ$, $M = 0.7$, $Re = 5\times 10^4$, $L_z = 0.05$, included for reference).}
\label{tableHighRe}
\end{center}
\end{table}


Type II transonic buffet, with shock waves occurring and oscillating exclusively on the suction side, was not observed for any of the cases simulated at $Re = 5\times10^4$. This is because offset conditions of Type II TBLO and fully -stalled conditions occur for $M$ below those at which shock waves appear. To link the Type II TBLO observed at $\alpha = 4^\circ$ and $M = 0.7$ at this $Re$ with Type II transonic buffet, a few more cases were simulated at higher $Re$. This is motivated by the observation in \citet{Zauner2018} that transonic buffet onset and offset can also occur with changes in $Re$. 
The parameters for the two additional cases simulated for which buffet occurs are $\alpha = 4^\circ$, $M = 0.75$ and $Re = 5\times10^5$, and $\alpha = 6^\circ$, $M = 0.75$ and $Re = 1.5\times 10^6$ (see table~\ref{tableHighRe}. We will refer to these two as high-$Re$ cases. The parametric values associated with these cases were found by trial and error since Type II transonic buffet occurs only in a narrow range of the parametric space. For these high-$Re$ cases, the grid used has the same distribution in the $x-y$ plane as Grid 1. However, the spanwise grid spacing is halved to $\delta z = 0.001$ to account for the increased $Re$. Considering the numerical expense, the narrower span of $L_z = 0.05$ is used. We emphasise that this configuration ($Re$, $L_z$ and $\delta z$) is similar to that used in a majority of simulations carried out in our previous studies \citep{Moise2022, Moise2022Trip, Zauner2022}, and for which various validations and verification have been performed with LES, DNS and experiments (see \S\ref{secMethod}). 

\begin{figure} 
\centerline{
\includegraphics[width=0.45\textwidth]{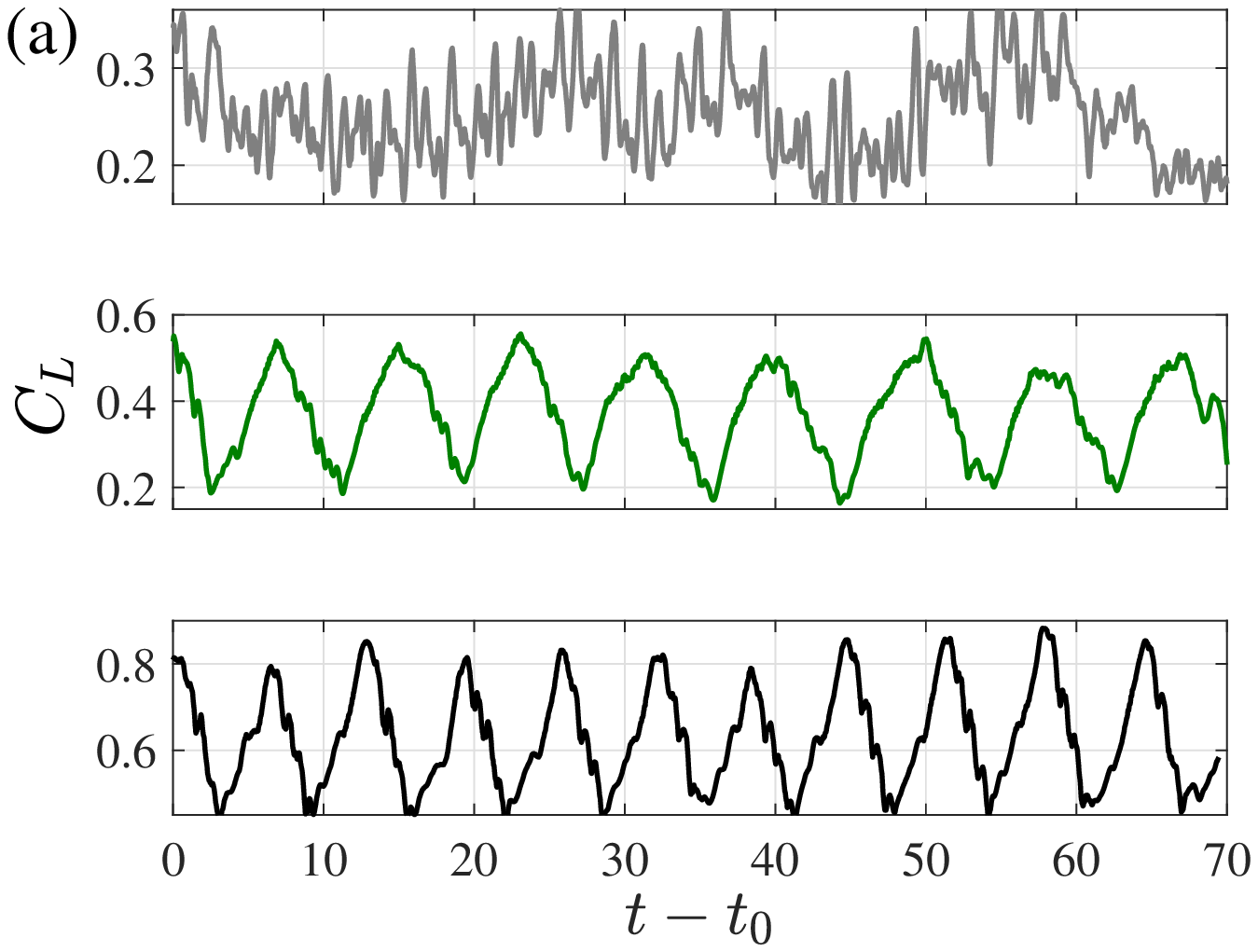}
\includegraphics[width=0.45\textwidth]{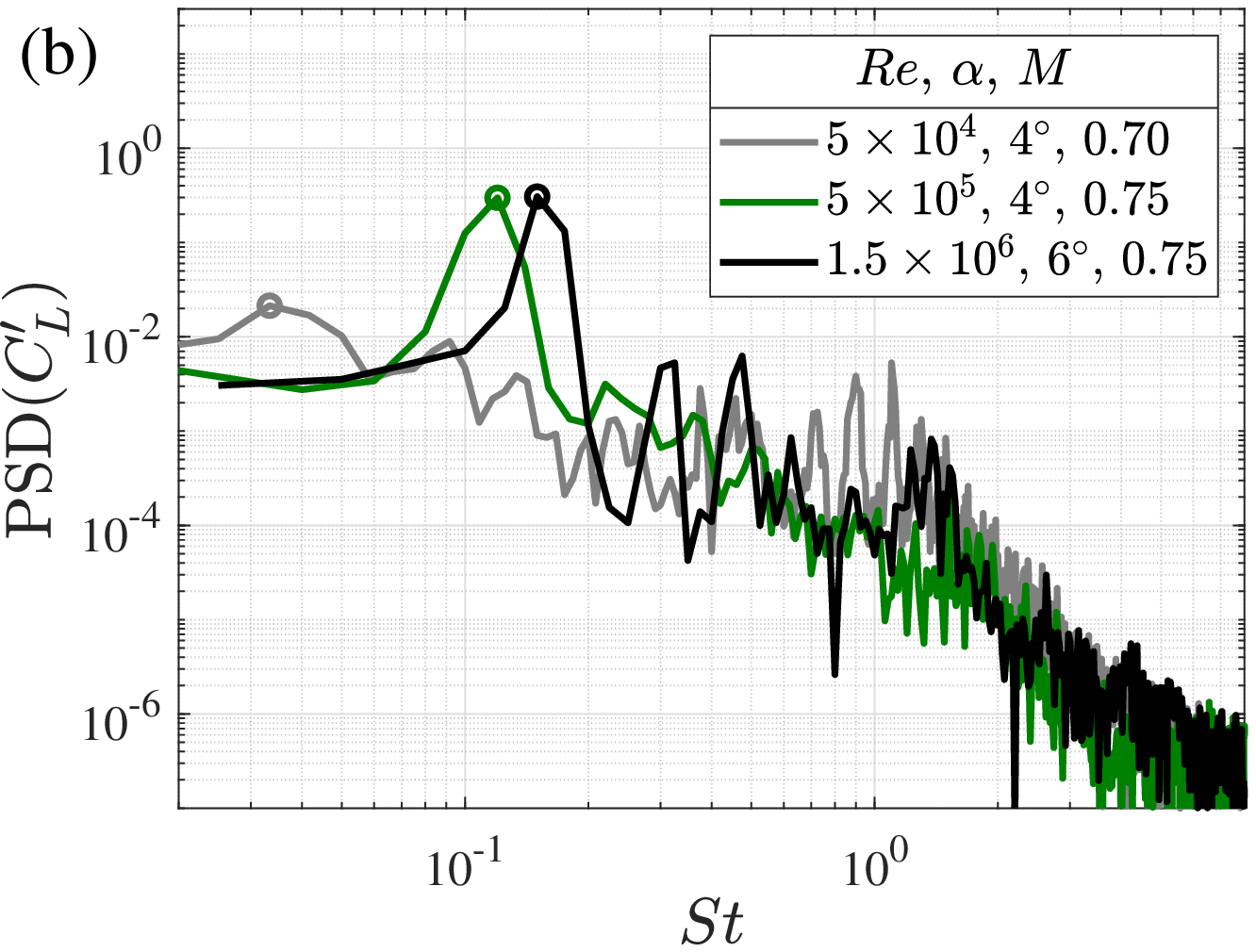}
}
\caption{(a) Temporal variation of lift coefficient past transients for $Re$, $\alpha$ and $M$ of $5\times10^4$, $4^\circ$ and $0.7$ (top), $5\times10^5$, $4^\circ$ and $0.75$ (middle), $1.5\times10^6$, $6^\circ$ and $0.75$ (bottom). (b) PSD of its fluctuating component as a function of the Strouhal number, $St$ when $Re$ is also varied.}
\label{fNACAClPSDLFOHighRe}
\end{figure}

The variation of the lift coefficient and the PSD of its fluctuating component are shown in figure~\ref{fNACAClPSDLFOHighRe} for the three different $Re$ examined. In contrast to the irregular temporal variation of the lift coefficient observed at subsonic conditions ($Re = 5\times10^4$), the high-$Re$ cases, for which the flow is transonic, exhibit a more regular temporal variation. The oscillation cycles for the latter are also dominated by the low-frequency content, as can also be inferred from the power spectra. A strong increase in the buffet frequency with $Re$ can be observed, from $St_b \approx 0.03$ at $Re = 5\times10^4$ to $St_b \approx 0.12$ at $Re = 5\times10^5$ and $St_b \approx 0.15$ at $Re = 1.5\times10^6$. Note that it is not only $Re$ that is increased between cases, but also $M$ (or $\alpha$), and the latter can have a strong influence in increasing $St_b$ \citep[\textit{e.g.}, figure~\ref{fNACAClPSD}\textit{a} here and Fig. 8,][]{Zauner2022}. 

\begin{figure} 
\centering
\includegraphics[trim={2cm 0cm 1.8cm 0cm},clip,width=.32\textwidth]{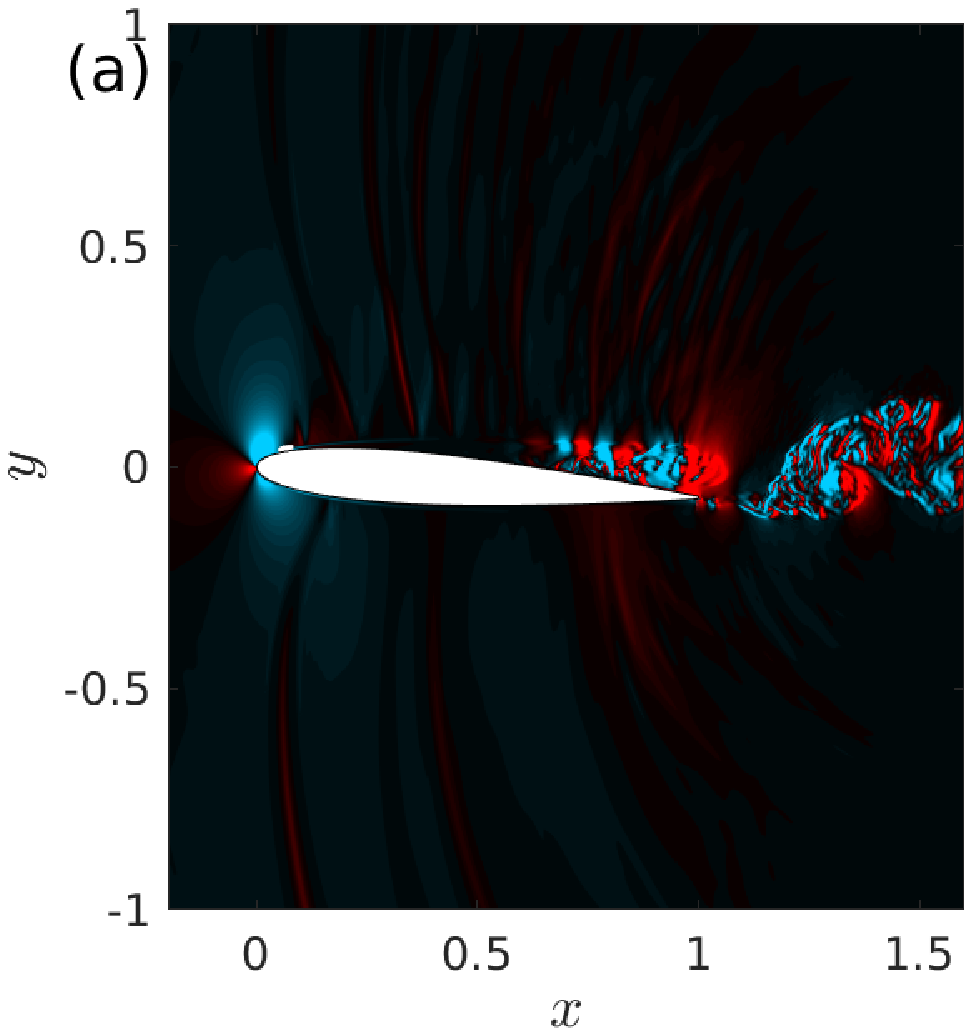}
\includegraphics[trim={2cm 0cm 1.8cm 0cm},clip,width=.32\textwidth]{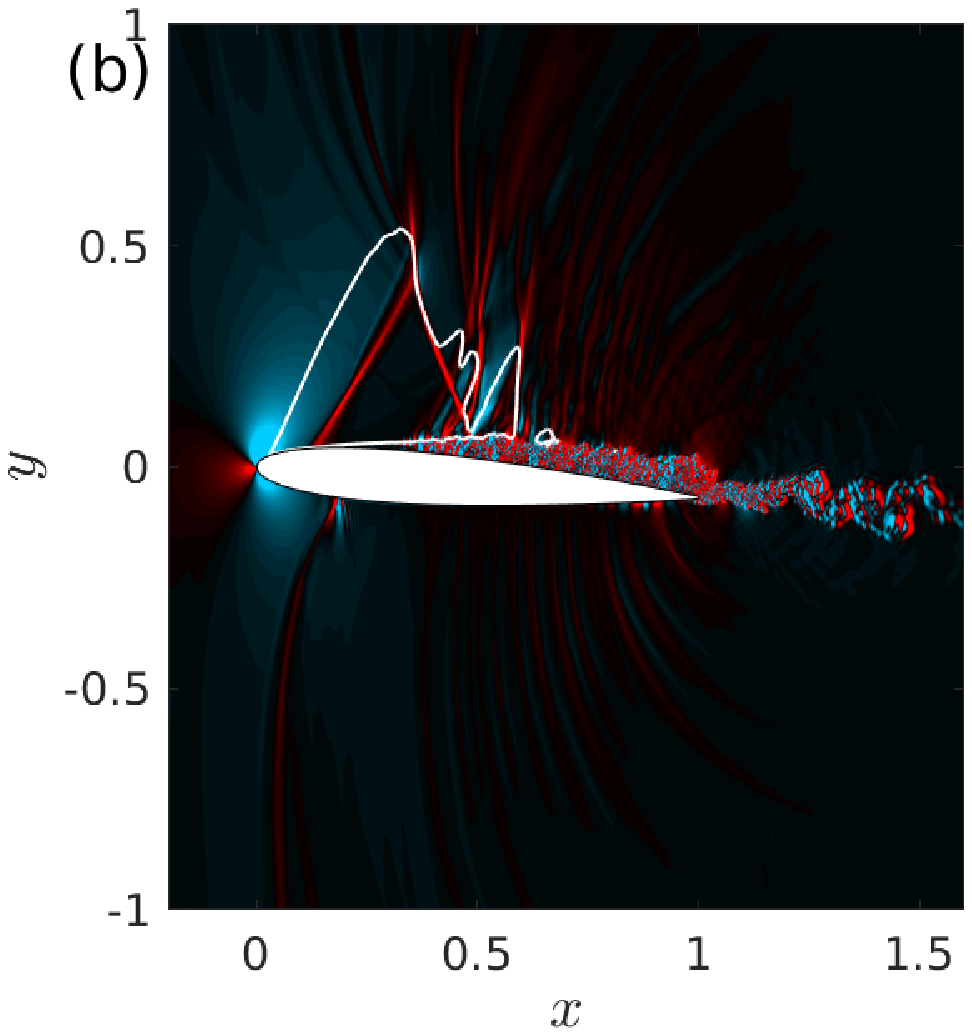}
\includegraphics[trim={2cm 0cm 1.8cm 0cm},clip,width=.32\textwidth]{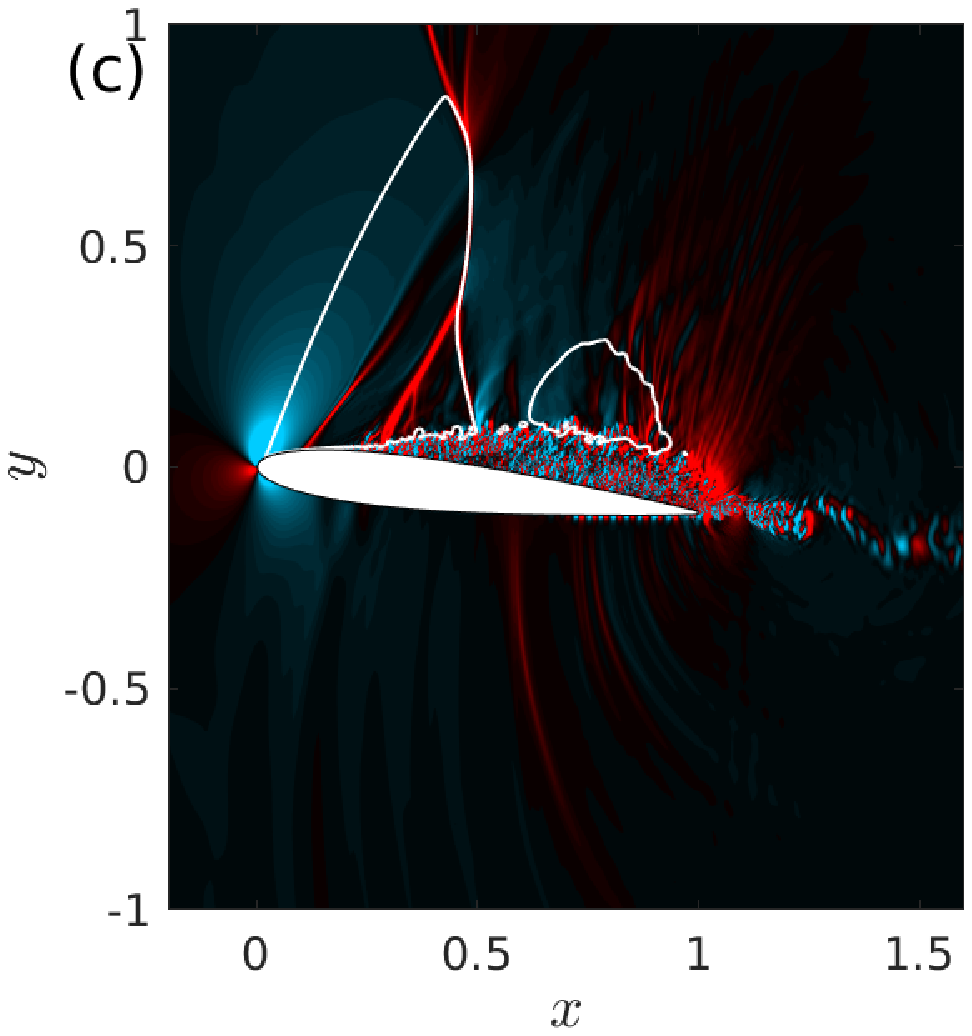}
\includegraphics[trim={2cm 0cm 1.8cm 0cm},clip,width=.32\textwidth]{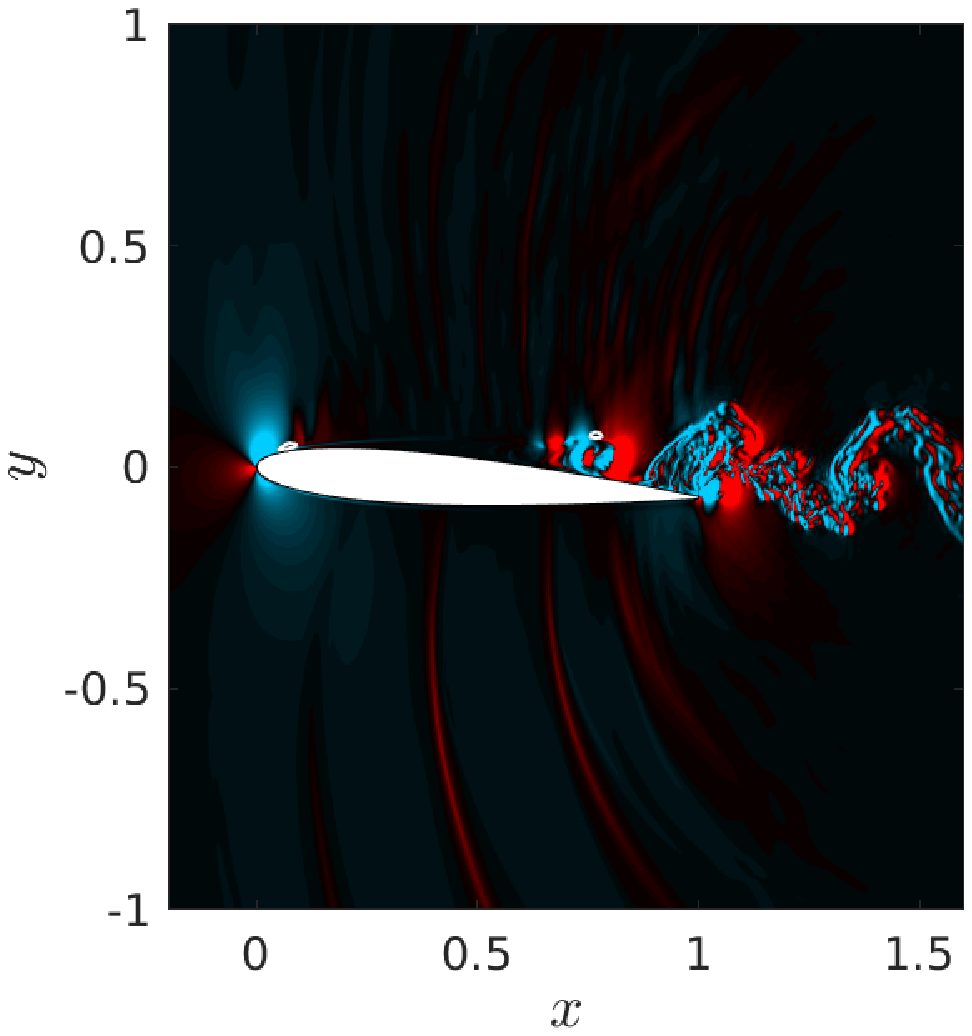}
\includegraphics[trim={2cm 0cm 1.8cm 0cm},clip,width=.32\textwidth]{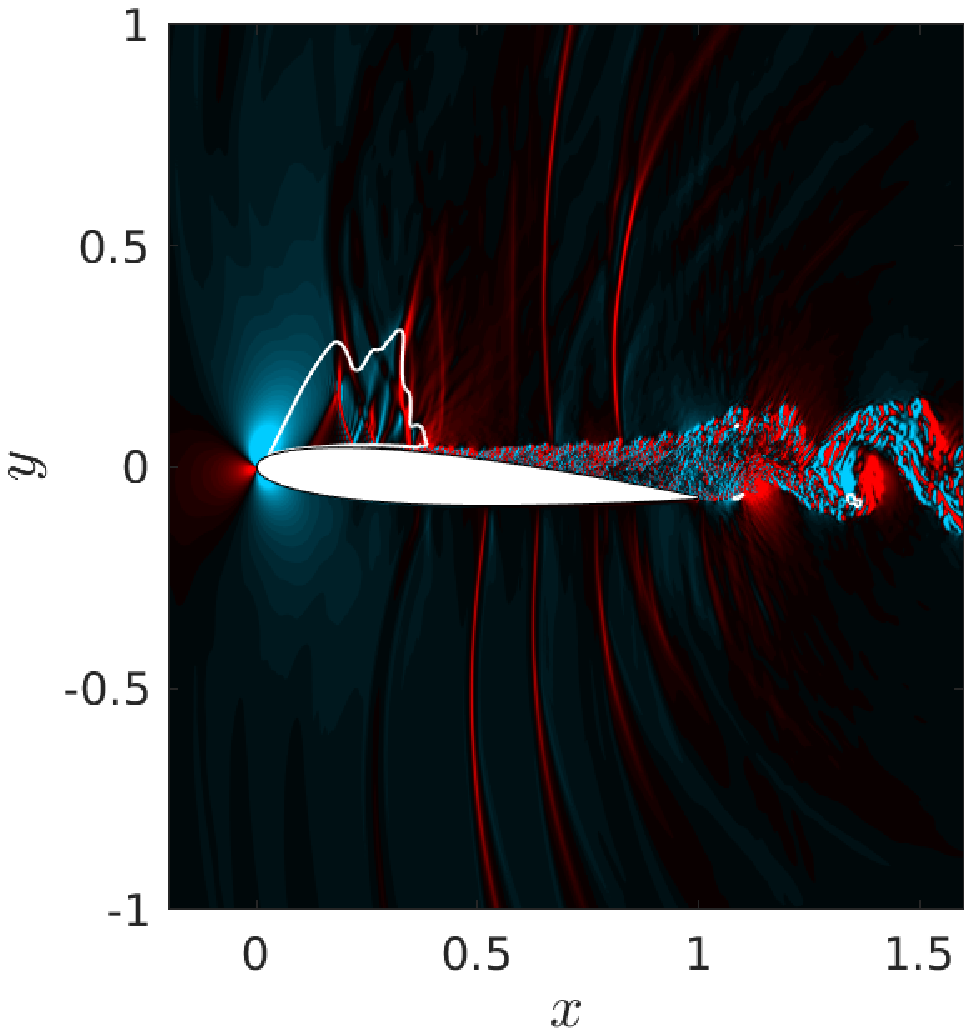}
\includegraphics[trim={2cm 0cm 1.8cm 0cm},clip,width=.32\textwidth]{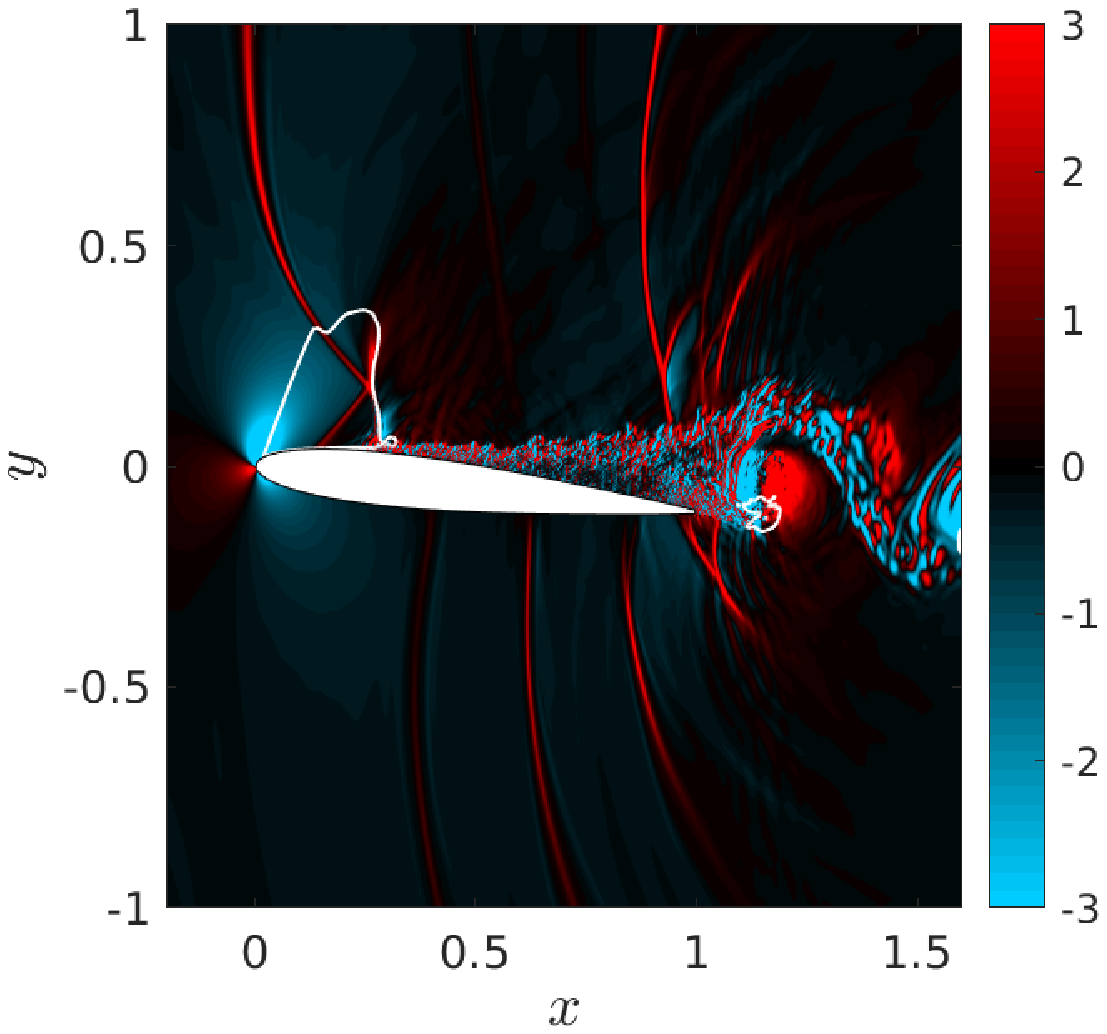}
\caption{Streamwise density gradient contours on the $x-y$ plane shown at the approximate high-(top) and low-(bottom) lift phases of the low-frequency cycle for (a) $\alpha = 4^\circ$, $M = 0.7$, $Re = 5\times10^4$, (b) $\alpha = 4^\circ$, $M = 0.75$, $Re = 5\times10^5$ and (c) $\alpha = 6^\circ$, $M = 0.75$, $Re = 1.5\times10^6$. The sonic line is highlighted using a white curve.}
    \label{fNACAHighReDensGrad}
\end{figure}

\begin{figure} 
\centering
\includegraphics[trim={2.5cm 0cm 3.5cm 0cm},clip,width=.32\textwidth]{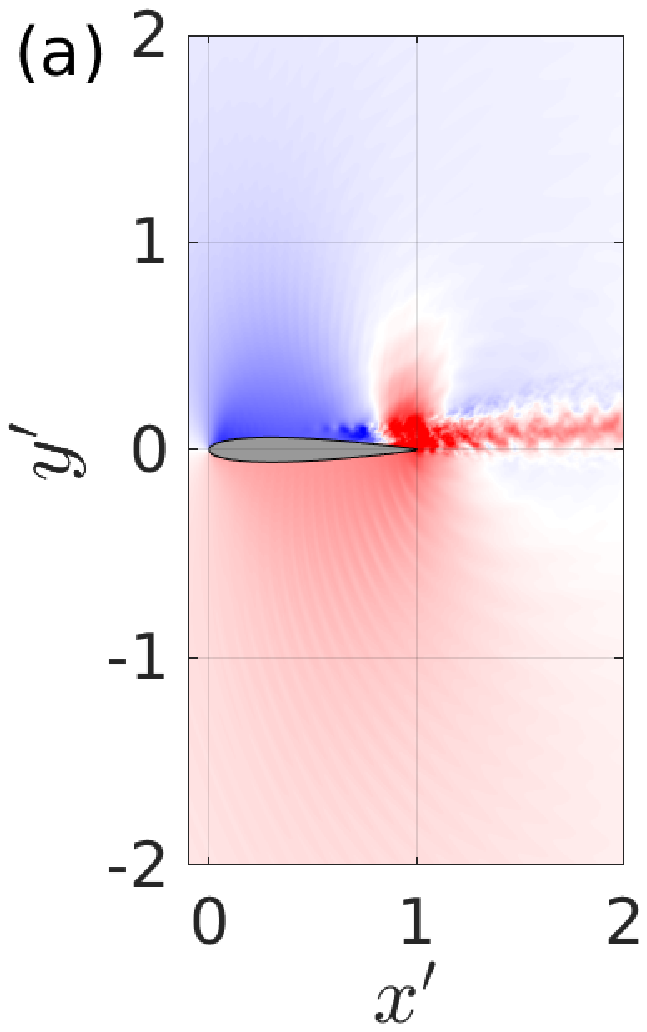}
\includegraphics[trim={2.5cm 0cm 3.5cm 0cm},clip,width=.32\textwidth]{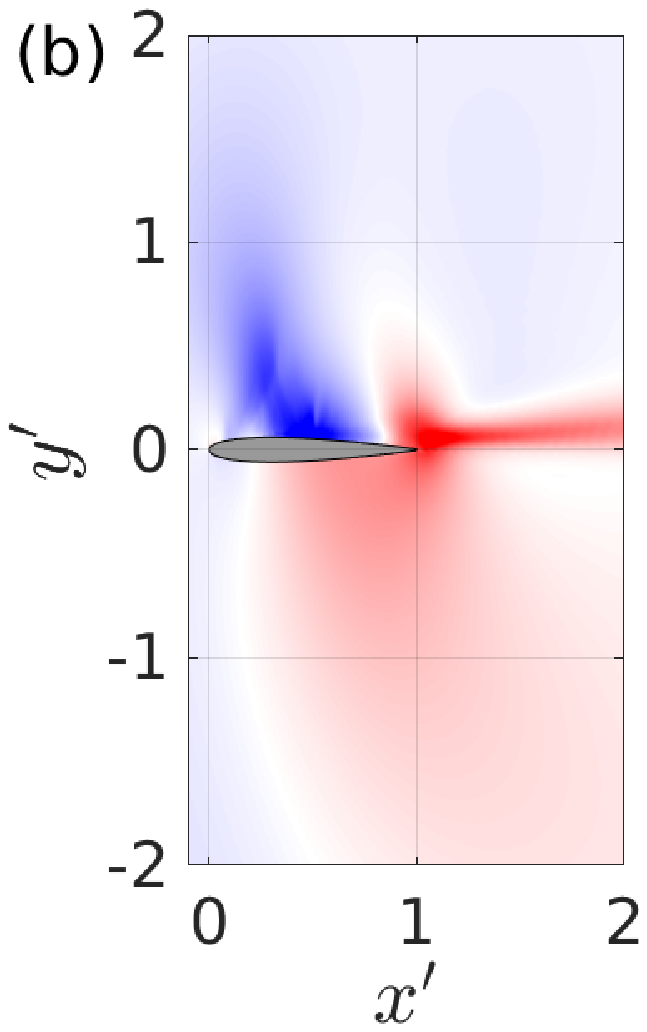}
\includegraphics[trim={2.5cm 0cm 3.5cm 0cm},clip,width=.32\textwidth]{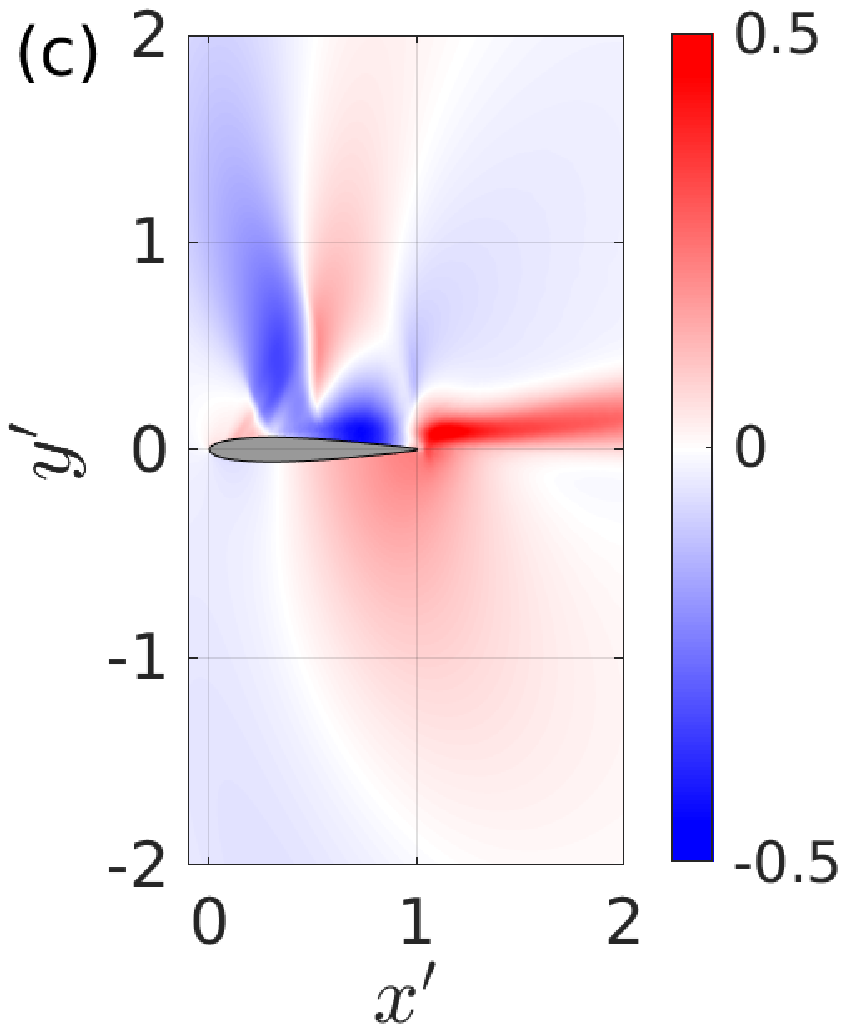}
\caption{Buffet modes from SPOD are shown using contour plots of the pressure field for cases (a) $\alpha = 4^\circ$, $M = 0.7$, $Re = 5\times10^4$, (b) $\alpha = 4^\circ$, $M = 0.75$, $Re = 5\times10^5$ and (c) $\alpha = 6^\circ$, $M = 0.75$, $Re = 1.5\times10^6$.} 
    \label{fNACAHighReSPODModes}
\end{figure}

The streamwise density-gradient fields in the high- and low-lift phases are compared in figure~\ref{fNACAHighReDensGrad}. The presence of shock waves which exhibit a fore-aft motion on the suction surface can be inferred for the high-$Re$ cases. Thus, these two cases can be considered as examples of Type II transonic buffet. The buffet modes obtained from SPOD are compared in figure~\ref{fNACAHighReSPODModes} for all cases. Comparing the cases of the lowest two $Re$ considered, it is evident that the spatial structure of the modes are similar, with a pattern of pressure reduction (blue) on the suction surface accompanying a pressure increase (red) near the trailing edge and the wake. This is essentially the same as that seen for LFO in figure~\ref{fNACALFOSPODModes}, except that the region associated with the pressure reduction extends further downstream here. For the highest $Re$, the buffet mode is similar to the other cases except for the presence of a small region associated with a pressure increase on the suction side. However, the animation provided for the spatio-temporal mode (movie 8, Supplementary material) shows that the spatial structures are closer at other time instants in the buffet cycle. This suggests that the choice of using the high-lift phase as the phase to compare modes from different cases is not always ideal, but sufficient to establish their similarities.   

\section{Discussion}
\label{secDisc}
\subsection{Link with other studies}
\label{subSecOtherTB}

\begin{figure} 
\centering
\includegraphics[trim={0cm 0.8cm 0cm 1.5cm},clip,width=.32\textwidth]{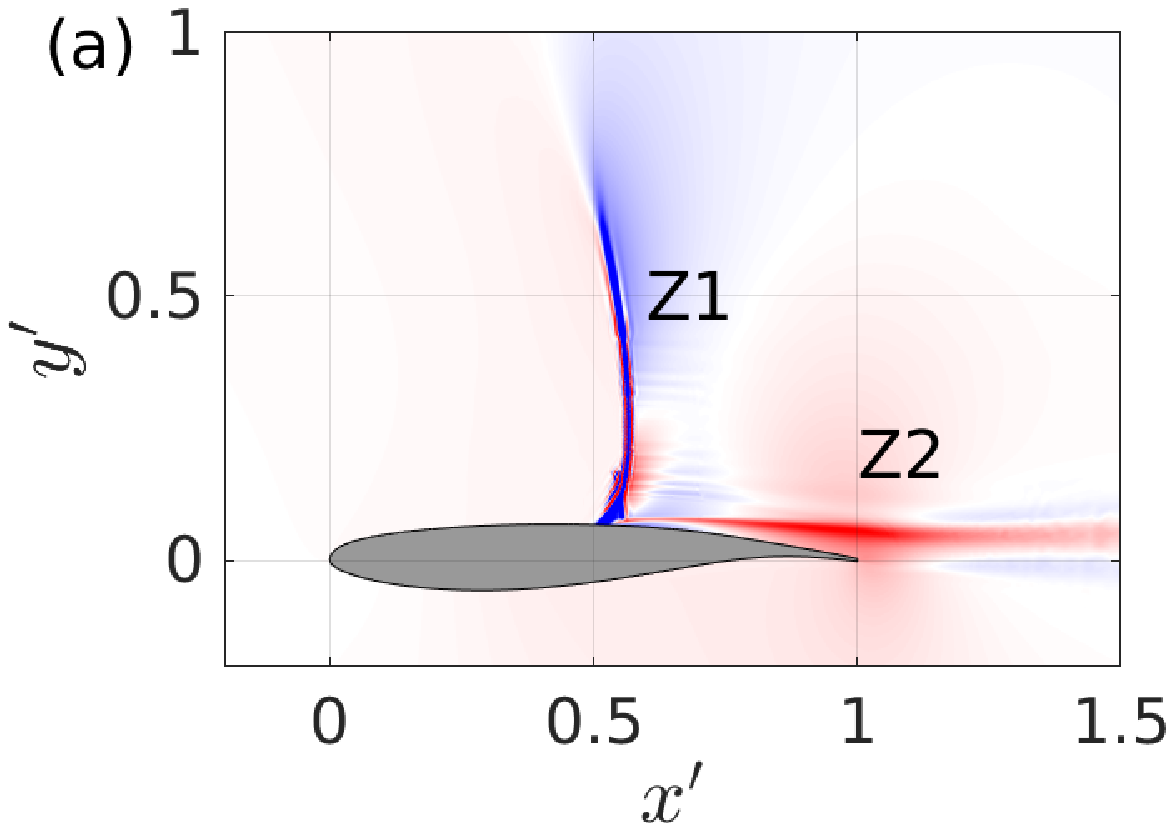}
\includegraphics[trim={0cm 0.8cm 0cm 1.5cm},clip,width=.32\textwidth]{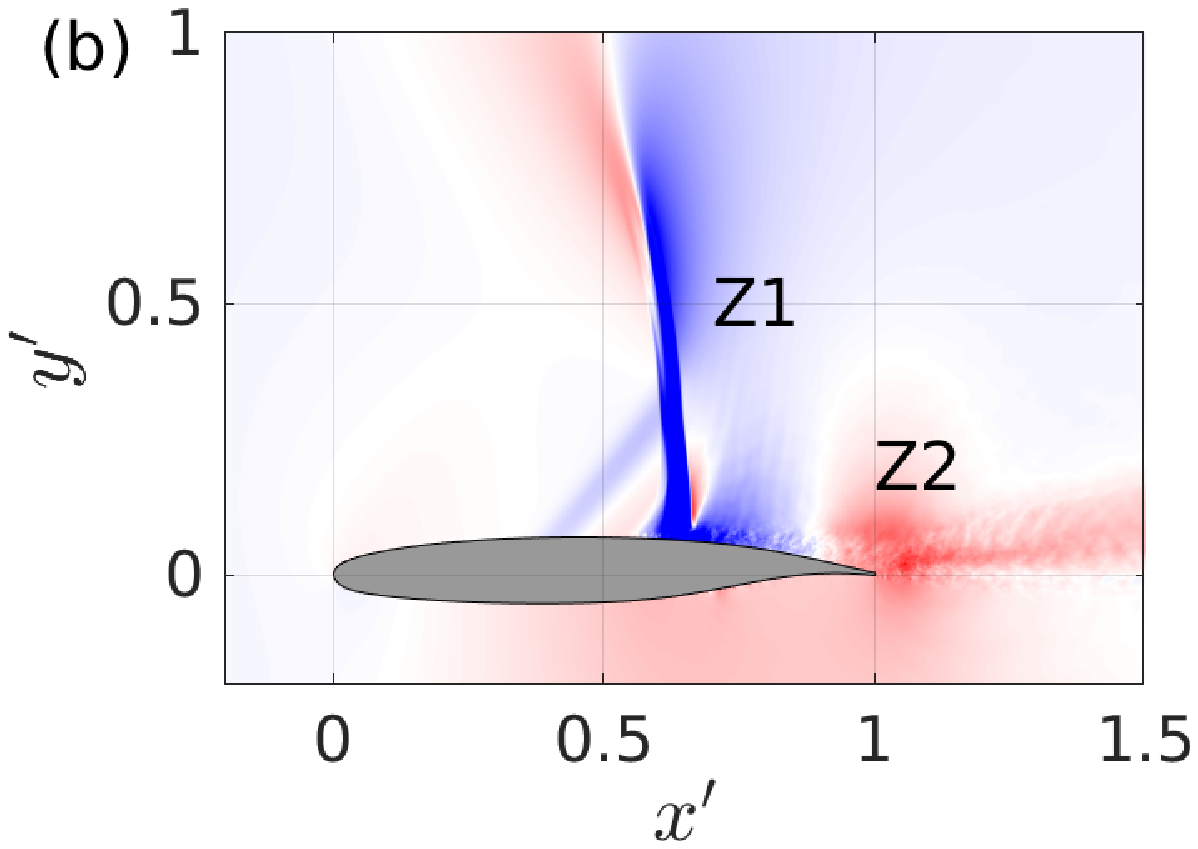}
\includegraphics[trim={0cm 0.8cm 0cm 1.5cm},clip,width=.32\textwidth]{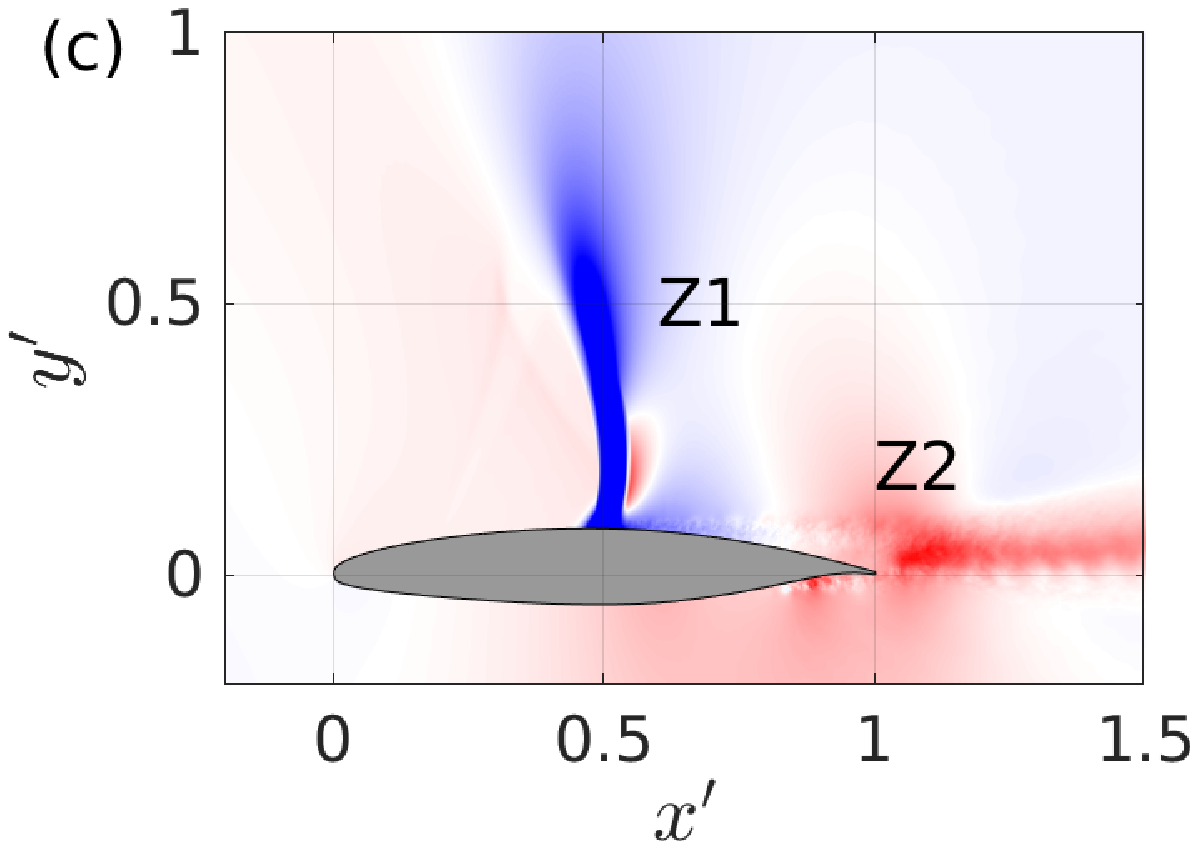}
\includegraphics[trim={0cm 0.8cm 0cm 1.5cm},clip,width=.32\textwidth]{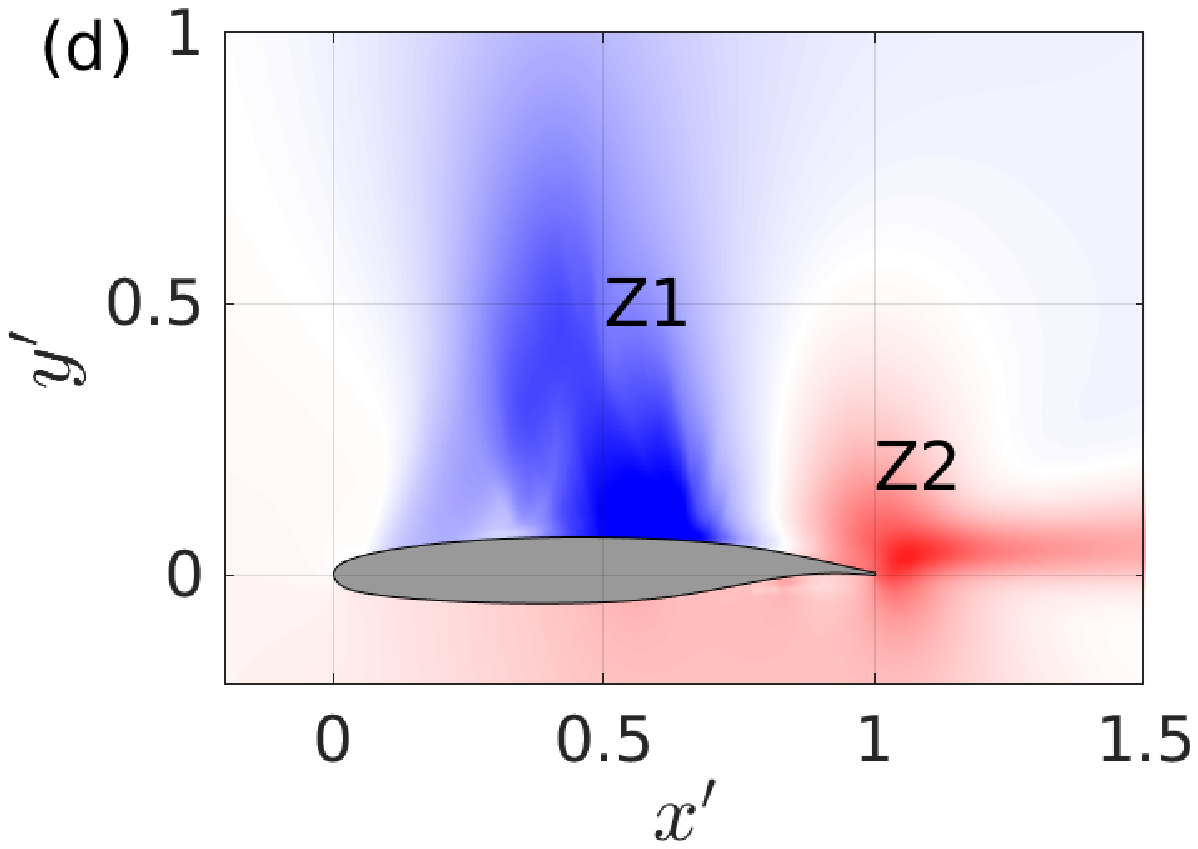}
\includegraphics[trim={0cm 0.8cm 0cm 1.5cm},clip,width=.32\textwidth]{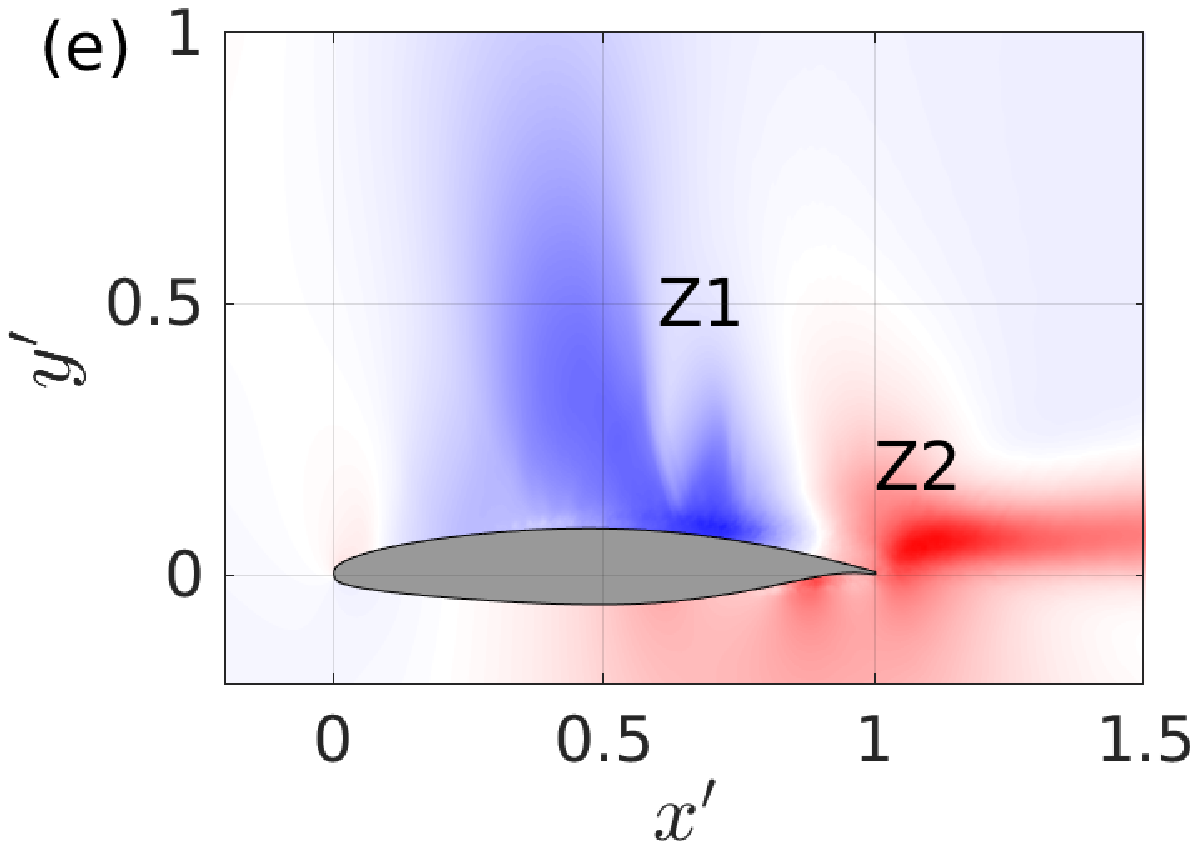}
\includegraphics[trim={0cm 0.8cm 0cm 1.5cm},clip,width=.32\textwidth]{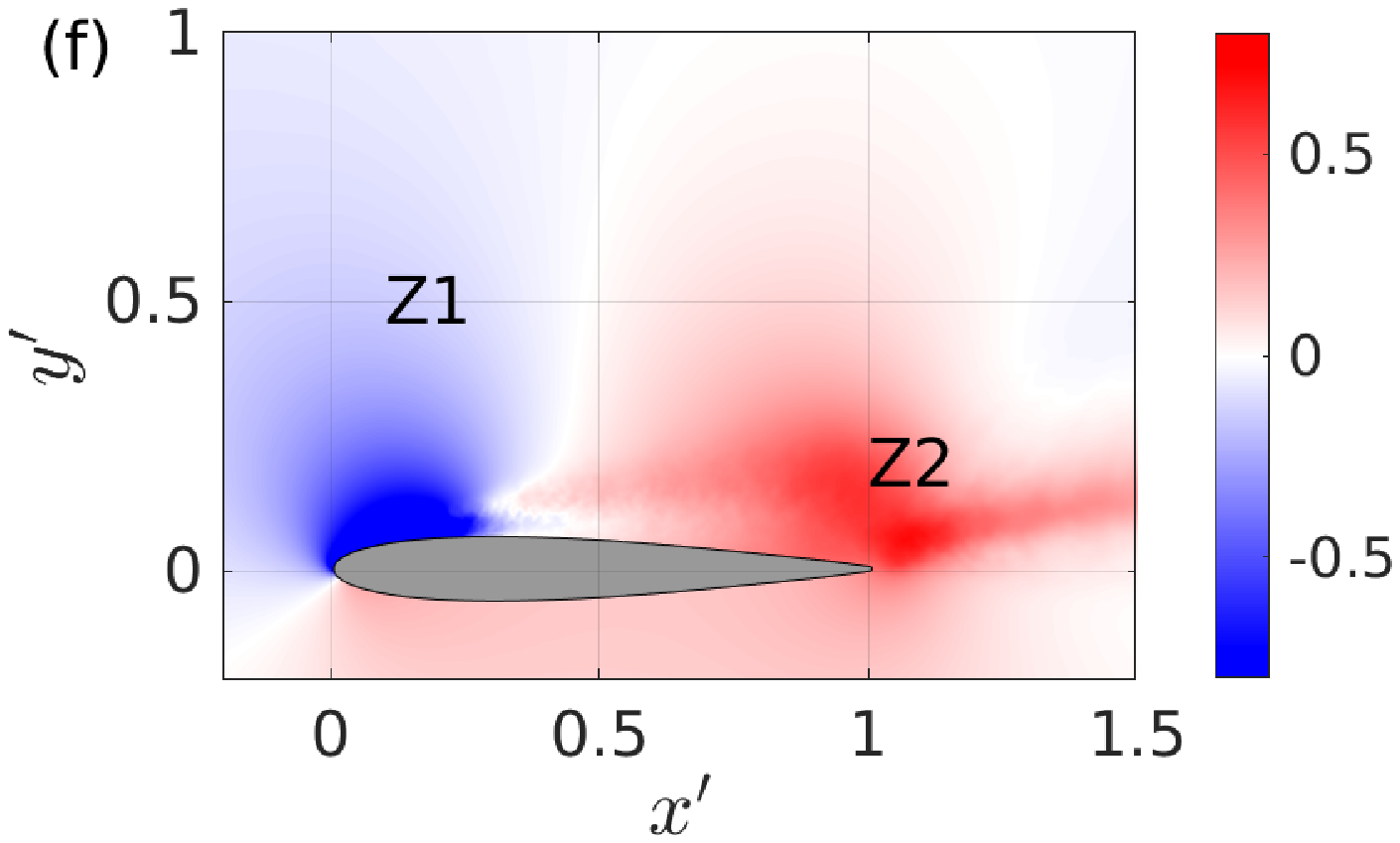}
\caption{Buffet modes for (a) OAT15A, fully turbulent conditions, $\alpha = 3^\circ$, $M = 0.74$, $Re = 3\times 10^6$, based on global linear stability analysis, and rest, based on SPOD, (b) OALT25, free transition, $\alpha = 4^\circ$, $M = 0.735$, $Re = 3\times10^6$, (c) V2C, forced transition (using a trip on the suction side at $x = 0.2$), $\alpha = 5^\circ$, $M = 0.735$, $Re = 5\times 10^5$, (d) same as (b), but with $Re = 5\times10^5$, (e) same as (c), but for free-transition conditions and (f) NACA0012, $\alpha = 9.4^\circ$, $M = 0.3$, $Re = 5\times 10^4$. The data for (a), (c) and (e) are from \citet{Moise2022Trip}, (b) and (d) from \citet{Zauner2022}, whereas (f) is from the present study.} 
    \label{fLinkDiffBuffetModes}
\end{figure}

The results from the preceding section suggest that LFO and Type II transonic buffet are similar for the symmetric NACA0012 aerofoil under conditions of free transition at low $Re$. However, most recent studies focus on Type II transonic buffet on supercritical aerofoils for forced-transition conditions and high $Re$ \citep{Giannelis2017}. To generalise the link between the two flow phenomena, we compare features of LFO to those of Type II transonic buffet from our previous studies where we have examined supercritical aerofoils at a variety of flow conditions. Note that these include cases with turbulent or laminar transonic buffet, with multiple shock waves or a single shock wave present. We consolidate these results on buffet mode features here so that a direct link between a typical transonic buffet mode and LFO can be established. However, to be consistent, we have presented the buffet mode features at the phase when the lift induced is maximum. This is different from the plots shown in previous studies, as they are for a different phase of the buffet cycle, based on features at the trailing edge. We start with figure~\ref{fLinkDiffBuffetModes}\textit{a}, where the dominant unstable mode obtained from a global linear stability analysis of results based on a RANS framework is shown. The results are for flow over ONERA's OAT15A aerofoil at $\alpha = 3^\circ$ for $M = 0.74$, $Re = 3\times 10^6$ and fully turbulent conditions. These results for Type II turbulent transonic buffet were originally presented in \citet{Moise2022Trip}. Note that this supercritical OAT15A aerofoil has been extensively studied for similar flow conditions in various studies \citep{Sartor2015, Fukushima2018, Garbaruk2021}. The SPOD mode at buffet frequency obtained from LES of flow around ONERA's laminar OALT25 profile at $\alpha = 4^\circ$ for $M = 0.735$, $Re = 3\times10^6$ and free-transition conditions is shown in figure~\ref{fLinkDiffBuffetModes}\textit{b}. These results on Type II laminar transonic buffet were originally presented in \citet{Zauner2022} and these flow conditions have also been studied using LES in \citet{Dandois2018} and experiments in \citet{Brion2020}. The SPOD mode at buffet frequency obtained from LES of flow around Dassault Aviation's laminar V2C profile at $\alpha = 5^\circ$ for $M = 0.735$, $Re = 5\times10^5$ and forced-transition conditions is shown in figure~\ref{fLinkDiffBuffetModes}\textit{c}. This is based on the results of Type II turbulent transonic buffet presented in \citet{Moise2022Trip}, where boundary layer transition to turbulence was triggered at $x = 0.2$ on the suction side using a synthetic unsteady jet. 

These three cases were all found to contain a single shock wave in the flow field that exhibited a fore-aft motion of small amplitude ($\Delta x < 0.1$).
Two zones of importance that are common to all three cases are marked as Z1 and Z2. The former (blue region) is present above the suction side close to the mid-chord and is concentrated around the mean shock location. The latter (red region, \textit{i.e.}, out of phase with Z1) occurs in the vicinity of the trailing edge and extends into the wake. In addition to these two prominent zones, another small zone of increased pressure near the shock at the current phase, \textit{i.e.} the unmarked red region at the shock foot which is bounded by Z1, can be identified. However, this zone might not be always visible as larger shock wave excursions could lead to Z1 masking this region. This is likely the case in Fig.~7a in \citet{Fukushima2018} and for the remaining cases considered. The other cases shown (figure~\ref{fLinkDiffBuffetModes}\textit{d} and \textit{e}) are the SPOD modes associated with Type II laminar transonic buffet for the OALT25 and V2C for the same respective flow conditions as that discussed before, but for a lower $Re = 5\times10^5$ for the former (\textit{i.e.}, $\alpha = 4^\circ$, $M = 0.735$ and free transition) and for free-transition conditions for the latter (\textit{i.e.}, $\alpha = 5^\circ$, $M = 0.735$ and $Re = 5\times10^5$). These two cases are characterised by multiple shock waves which exhibit large-amplitude streamwise oscillations ($\delta x \geq 0.1$). The zones, Z1 and Z2, can also be discerned here, although, in contrast to the previous cases, Z1 is not concentrated but extends over almost the entire suction surface due to the large-amplitude excursions associated with multiple shock waves. 

Finally, when comparing the SPOD mode of LFO obtained in this study (figure~\ref{fLinkDiffBuffetModes}\textit{f}) for which the flow remains essentially incompressible, the similarity is evident, with both Z1 and Z2 clearly identifiable, albeit with the extent of Z2 now extending upstream up to mid-chord. Thus, while the similarities between the global mode associated with transonic buffet on the commonly-studied OAT15A aerofoil and the SPOD mode associated with LFO in the present study might not be immediately apparent, examining the sequence of plots shown in figure~\ref{fLinkDiffBuffetModes} indicates that the spatial structure of these two modes is qualitatively the same. It consists of an out-of-phase coupling between a region upstream in the fore part of the aerofoil with a region located downstream in the aft part and extending into the wake. When considering this result in conjunction with the results that the frequencies of the two phenomena are close ($St \approx 0.1$ for the OAT15A case and $St \approx 0.04$) and the links provided in preceding sections (\S\S\ref{subSecLFOLowRe} and \ref{subSecLFOVsHighReTB}), we conclude that LFO and transonic buffet on aerofoils have similar characteristics and thus, their sustenance is likely to be driven by similar physical mechanisms. 

As noted in \S\ref{secIntro}, the results from previous studies show that both flow phenomena occur as global linear instabilities \citep{Crouch2007, Iorio2016, Busquet2021} and are associated with periodic boundary layer separation and reattachment (\textit{i.e.} switching between stalled and unstalled states), which further corroborate this conclusion. It is also interesting to note that in flight tests, buffet onset occurs during manoeuvres such as turns and pull-ups for flight Mach numbers in the range 0.1 to 0.8 (\textit{e.g.}, figure~4 in \citet{Skopinski1955} and figure~1 in \citet{Purser1951}). This buffet boundary varies continuously in plots of normal coefficient versus flight Mach number as the latter changes from subsonic to transonic regimes (referred to as stall and shock regimes), which is consistent with the present results. However, since it drops relatively steeply in the latter regime it has been suggested that ``Along the steep portion of the boundary the buffeting is thought to be primarily due to compressibility rather than to reaching the stall as in the low
Mach number portion of the boundary'' \citep{Purser1951}. However, the present results suggest that this assumption is incorrect and that variations in steepness might be a quantitative feature but that buffet's spatio-temporal features would still be qualitatively the same at all Mach numbers.


\subsection{Implications for buffet on aerofoils}
\label{subSecImplic2D}

There are several important consequences to the conclusion that transonic buffet and low-frequency oscillations are essentially the same phenomenon. Firstly, it implies that ``transonic" buffet is not exclusive to the transonic regime. Even in the transonic regime, it is well-known that buffet occurs only above an onset and below an offset value of $M$ \citep{Giannelis2017}. Thus, the transonic regime (or a shock wave) is neither necessary nor sufficient for buffet to occur. Hence, it is suggested that the etymology of referring to the phenomenon as ``transonic" buffet could be misleading or at the least, restrictive, as it can be interpreted to imply that it could occur only in the transonic regime. 

Secondly, many proposed physical models for transonic buffet require the presence of a shock wave. For example, the popular model proposed in \citet{Lee1990} for Type II transonic buffet assumes that waves that are generated at the shock foot travel downstream within the boundary layer and interact with the trailing edge, leading to the development of `Kutta' waves which then propagate upstream in the potential flow region and interact with the shock waves. The model then predicts buffet frequency based on the distance the waves travel, which is approximated as that between the shock wave and the trailing edge, which cannot be computed if no shock wave is present. 
Equivalent drawbacks also arise for other models that are similar to that of Lee such as those proposed in \citet{Jacquin2009} and \citet{Hartmann2013} for Type II transonic buffet and that proposed in \citet{Gibb1988} to explain Type I transonic buffet. Thus, we suggest that an acoustic wave-propagation-based feedback model for transonic buffet must be discarded or substantially modified to explain oscillations seen at subsonic and incompressible conditions.

Any improved model of buffet needs to recognise that the phenomenon arises with or without shock waves and for the upstream boundary layer being either laminar or turbulent. In this context the LFO has been previously observed \citep{Sandham2008} to be sustained based only on boundary-layer integral models and potential flow interaction. Hence both buffet and LFO could be associated with a global instability related to flow separation coupled to the external potential flow that occurs close to stall. Shock waves might then play a role in promoting stall-like conditions, but this is only secondary since stall occurs even in the absence of shock waves and conditions required for the same can be achieved at higher $\alpha$ by reducing $M$. Similarly, the state of the boundary layer, whether laminar or turbulent, would then affect the details but not the fundamental mechanism. Features like the sensitivity of the amplitude to spanwise domain size could then be linked to three-dimensional aspects of either separation or the potential flow response. Further examination is required to confirm this hypothesis, but if true, this would imply that flow control strategies for eliminating/mitigating buffet could shift focus away from shock waves to a consideration of the whole flow and aim at delaying stall.

\subsection{Implications for buffet on three-dimensional wings}
\label{subSecImplic3D}

The focus of this study has been on the aerofoil buffet on an unswept/straight infinite-wing section. However, as noted in \S\ref{secIntro}, when swept infinite-wing sections are considered, a new unsteady phenomenon, referred to as buffet cells, dominates and is characterised by the emergence of three-dimensional features \citep{Iovnovich2015}. Several studies have reported that the dominant unstable mode is no longer the two-dimensional mode associated with aerofoil buffet but a different three-dimensional mode \citep{Crouch2019, Paladini2019, he_timme_2021}. For convenience, we refer to the dominant modes on unswept and swept wing sections as 2D and 3D buffet modes, which are characterised by a zero and non-zero spanwise wavenumber, respectively. Note that the latter is the same as the ``intermediate-wavelength" modes noted in \citet{Crouch2019}. Studies on fully three-dimensional wings have shown that the dominant unstable mode is not the 2D but the 3D mode \citep{Timme2020}, and thus, the former has been suggested to be of less engineering significance than the latter. Given that this study's focus has been on the 2D mode, the general relevance of these results is considered next.  

Firstly, we note that although the dominance of the 3D modes over the 2D mode has been observed for both swept infinite-wing sections and fully three-dimensional wings, it has been done so only under assumptions of fully-turbulent conditions in a RANS framework for a narrow range of flow conditions. Indeed, examining the effect of sweep at low $Re$ using an LES framework and free-transition conditions, \citet{Moise2022} did not observe any 3D mode even for sweep angles as high as $40^\circ$, conditions at which the 2D mode was active. This implies that there are conditions at which the 2D mode can dominate over 3D modes, especially when boundary layer transition occurs naturally. Secondly, as noted in \citet{Crouch2019} (``the  intermediate-wavelength  modes  have  wavelengths  much  greater  than the  shear-layer  thickness,  and  are  concentrated  both  in  the  shock  and  the  shear  layer–  similar  to  the  long-wavelength  oscillatory  mode  linked  to  buffet."), the spatial structure of both 2D and 3D modes in the $x-y$ plane is qualitatively similar. Given this structural similarity, it is possible that the present results could also be applicable to the 3D modes. This would imply that oscillatory modes that resemble the 3D modes observed in the transonic regime could also be present in the subsonic regime and that a continuous variation of flow parameters can be used to link oscillations occurring in the two regimes. As noted in \S\ref{secIntro} the first aspect of this has already been established in \citet{Plante2020} where the authors have shown that the stall cells observed in the incompressible regime are similar to buffet cells observed in the transonic regime. This existence of equivalent phenomena in the incompressible regime, \textit{i.e.}, LFO for transonic buffet (aerofoils) and stall cells for buffet cells (wings) suggests that a continuous variation of parameters could also link wing buffet to its incompressible equivalent.     

\section{Conclusions}
\label{secConc}
In this study, the flow over an infinite-wing section based on the NACA0012 aerofoil profile has been studied for a wide range of flow conditions by performing simulations at various incidence angles, and freestream Mach and Reynolds numbers. At a low $Re = 5\times 10^4$ and zero incidence angle, Type I transonic buffet, characterised by shock waves oscillating out of phase on both sides of the aerofoil, was observed at high $M$. Decreasing $M$ with other parameters fixed, it was found that oscillations resembling transonic buffet persist, although the flow remains subsonic at all times. Extending the span of the domain strongly affected the amplitude of these oscillations, but the frequency was unaffected and the oscillations were sustained for all $L_z$ considered, with the highest simulated being $L_z = 1$. Using SPOD, the coherent spatio-temporal features in the flow field were scrutinised. For all cases, peaks were observed in the SPOD spectra at the buffet frequency. The spatial structure of the SPOD mode at this frequency and a given phase consisted of a zone associated with a pressure reduction on the surface of the aerofoil (Z1) which is accompanied by a zone associated with pressure increase in the wake of the aerofoil (Z2). This structure was preserved for all cases, irrespective of whether the flow is subsonic or transonic. Thus, it is concluded that shock waves and transonic flow are not necessary for buffet to sustain.

Increasing the incidence angle from $\alpha = 0^\circ$ while simultaneously reducing $M$, it was observed that buffet sustains, albeit with large variations in frequency and amplitude. However, examining the SPOD modes confirmed that the spatial structure is qualitatively the same on the suction side for all cases. At a high incidence and low freestream Mach number ($\alpha = 9.4^\circ$ and $M = 0.3$), compressibility effects were negligible, indicating that the oscillations can also be classified as LFO. Type II transonic buffet was also simulated by choosing higher $Re$, $M$ and $\alpha$.  Thus, it is demonstrated in this study that Type I and II transonic buffet, Type I and II TBLO and LFO, all have qualitatively the same spatial structure (SPOD) and frequency $St \ll 1$, indicating that a continuous variation in flow conditions can be used to link one with the others. Furthermore, using a consistent phase definition, the SPOD modes for LFO have been shown to be similar to the global mode obtained from a linear stability analysis for flow over ONERA's OAT15A aerofoil (exhibiting Type II transonic buffet). For all cases, zones Z1 and Z2 are the dominant spatial features and are clearly identifiable. Thus, it is concluded that transonic buffet and LFO are essentially similar. 

Since shock waves or transonic conditions are neither necessary nor sufficient for the oscillations to sustain, we suggest that the term `transonic buffet' is a misnomer from a physical perspective, although it could still be of practical relevance. Possible alternatives which drop the epithet, such as `low-frequency buffet', might be more apt in describing the phenomenon. It also implies that physical models for buffet such as those proposed in \citet{Lee1989}, for Type II, and \citet{Gibb1988}, for Type II, are likely to be incorrect or require to be adapted to explain LFO and transonic buffet. Alternatively, buffet might arise as a separation-related instability that occurs at conditions close to stall. If this is true, then flow control strategies that aim to delay stall could be beneficial in mitigating these oscillations. Further studies on physical mechanisms and control strategies are required to examine these aspects. Another important avenue of exploration is to see if transonic buffet on three-dimensional wings can also be linked to a similar phenomenon in the incompressible regime by a continuous variation in the parameter space.     

\backsection[Supplementary data]{\label{SupMat}Supplementary material will be made available}

\backsection[Acknowledgements]{We would like to acknowledge the computational time on ARCHER and ARCHER2 (UK supercomputing facility) provided by the UK Turbulence Consortium (UKTC) through the EPSRC grant EP/R029326/1. We also acknowledge the use of the IRIDIS High Performance Computing Facility, and associated support services at the University of Southampton, in the completion of this study. The V2C aerofoil geometry was kindly provided by Dassault Aviation. We thank ONERA for the OALT25 and OAT15A geometries. We also thank Dr. Timme and Dr. Wei for the data on the buffet mode from global linear stability analysis of RANS results (OAT15A).}

\backsection[Funding]{This study was funded by the Engineering and Physical Sciences Research Council (EPSRC) grant, ``Extending the buffet envelope: step change in data quantity and quality of analysis” (EP/R037167/1).}

\backsection[Declaration of interests]{The authors report no conflict of interest.}


\backsection[Author ORCID]{P. Moise, https://orcid.org/0000-0001-8007-4453; M. Zauner, https://orcid.org/0000-0002-6644-2990; N. Sandham, https://orcid.org/0000-0002-5107-0944}


\appendix


\section{Results for V2C aerofoil at zero incidence}
\label{secV2C}
\begin{figure} 
\centerline{
\includegraphics[width=0.495\textwidth]{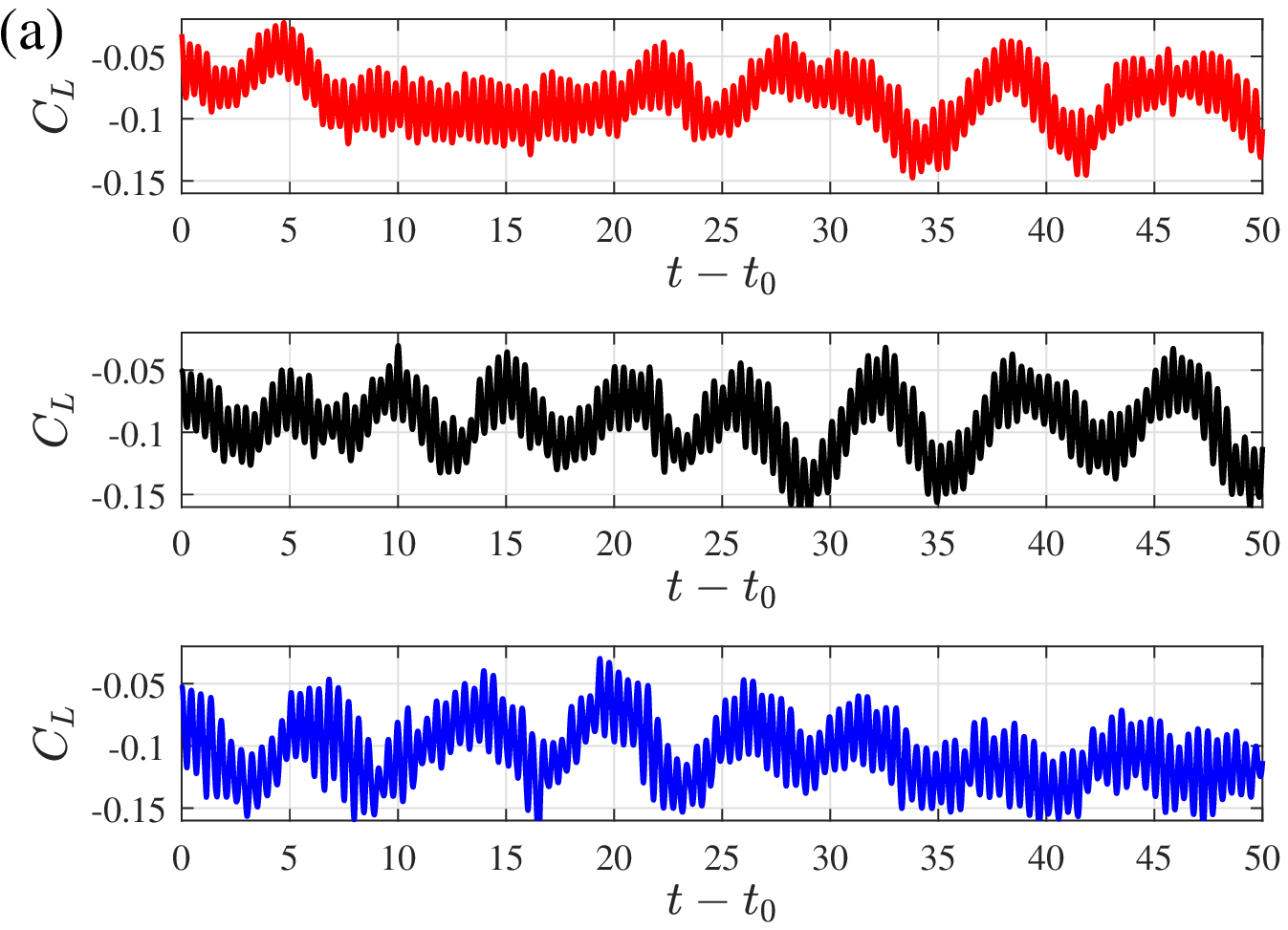}
\includegraphics[width=0.495\textwidth]{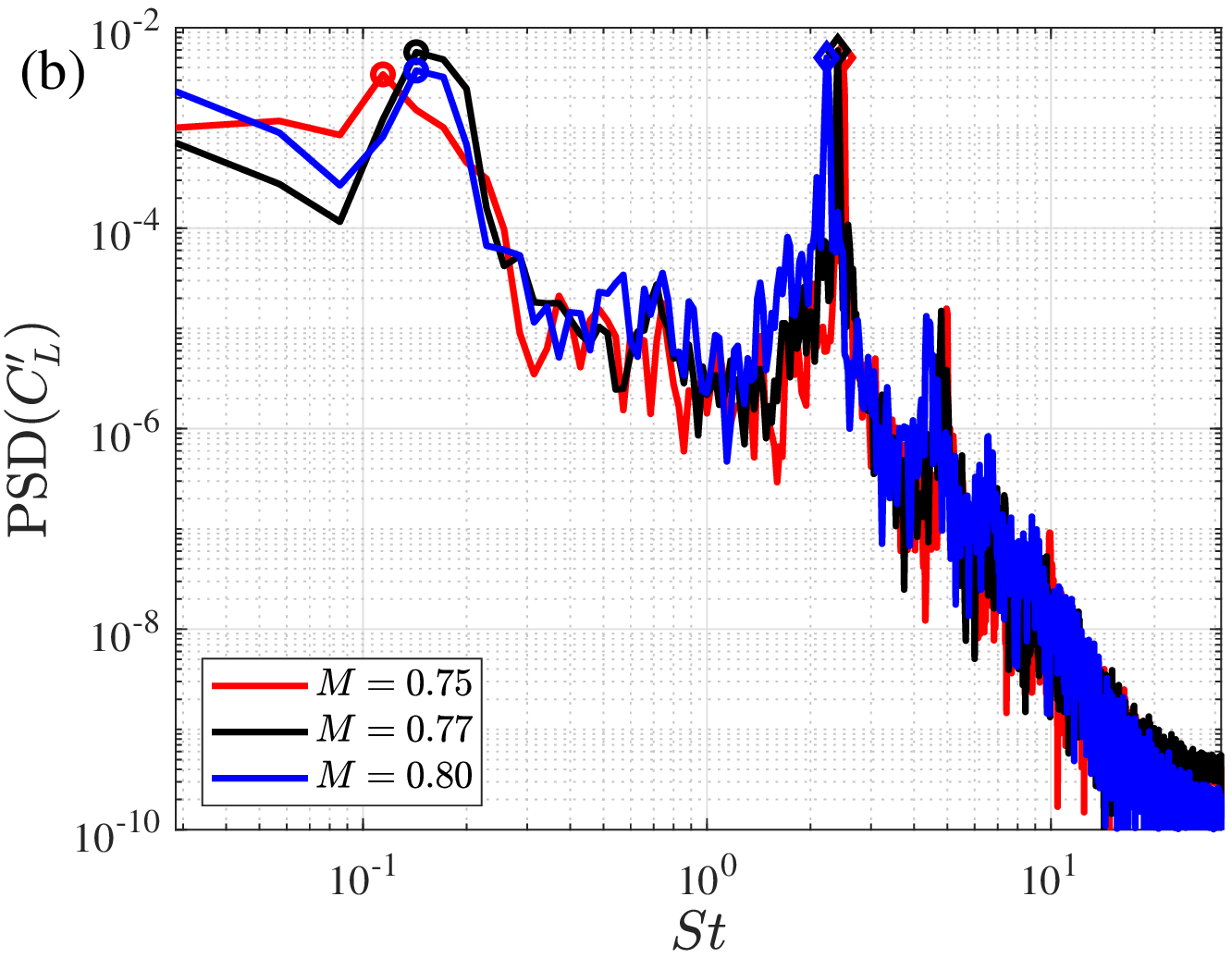}
}
\caption{(a) Temporal variation of lift coefficient past transients and 
(b) PSD of its fluctuating component for different freestream Mach numbers for the V2C aerofoil. Circles and diamonds highlight $St_b$ and $St_w$, respectively.}
\label{fV2C_ClMach}
\end{figure}

Type I TBLO were also observed for Dassault Aviation's laminar supercritical V2C aerofoil for $Re = 5\times 10^4$ and $L_z = 0.05$ (Grid 0). Lift characteristics are shown in figure~\ref{fV2C_ClMach} for three different freestream Mach numbers. While qualitatively similar, there are interesting quantitative differences between these results and those for the NACA0012 aerofoil at $\alpha = 0^\circ$ (cf. figures~\ref{fNACAClMach} and \ref{fNACAClPSD}). The lift oscillations are relatively irregular for the V2C aerofoil (similar to NACA0012 results for $\alpha \neq 0^\circ$), although the amplitude is similar for both aerofoils. The effect of spanwise width on this amplitude was not examined at this $Re$ due to the associated numerical expense (see \citet{Zauner2020PRF} for results at a higher $Re$). The buffet frequency (highlighted using circles in the PSD) is also relatively higher, with $St_b \approx 0.15$. This is similar to those reported at higher $Re$ for this aerofoil at zero incidence (\textit{e.g.}, $St_b = 0.13$ \citep{Moise2022} at $Re = 5\times10^5$), implying that the buffet frequency does not change significantly with $Re$ for this aerofoil when $\alpha = 0^\circ$. This should be contrasted with the results presented in \S\ref{subSecLFOVsHighReTB} where there is a strong variation in the buffet frequency with $Re$ for the NACA0012 aerofoil, albeit at $\alpha = 4^\circ$.

\begin{figure} 
\centering
\includegraphics[trim={1.2cm 0cm 0cm 0cm},clip,width=.32\textwidth]{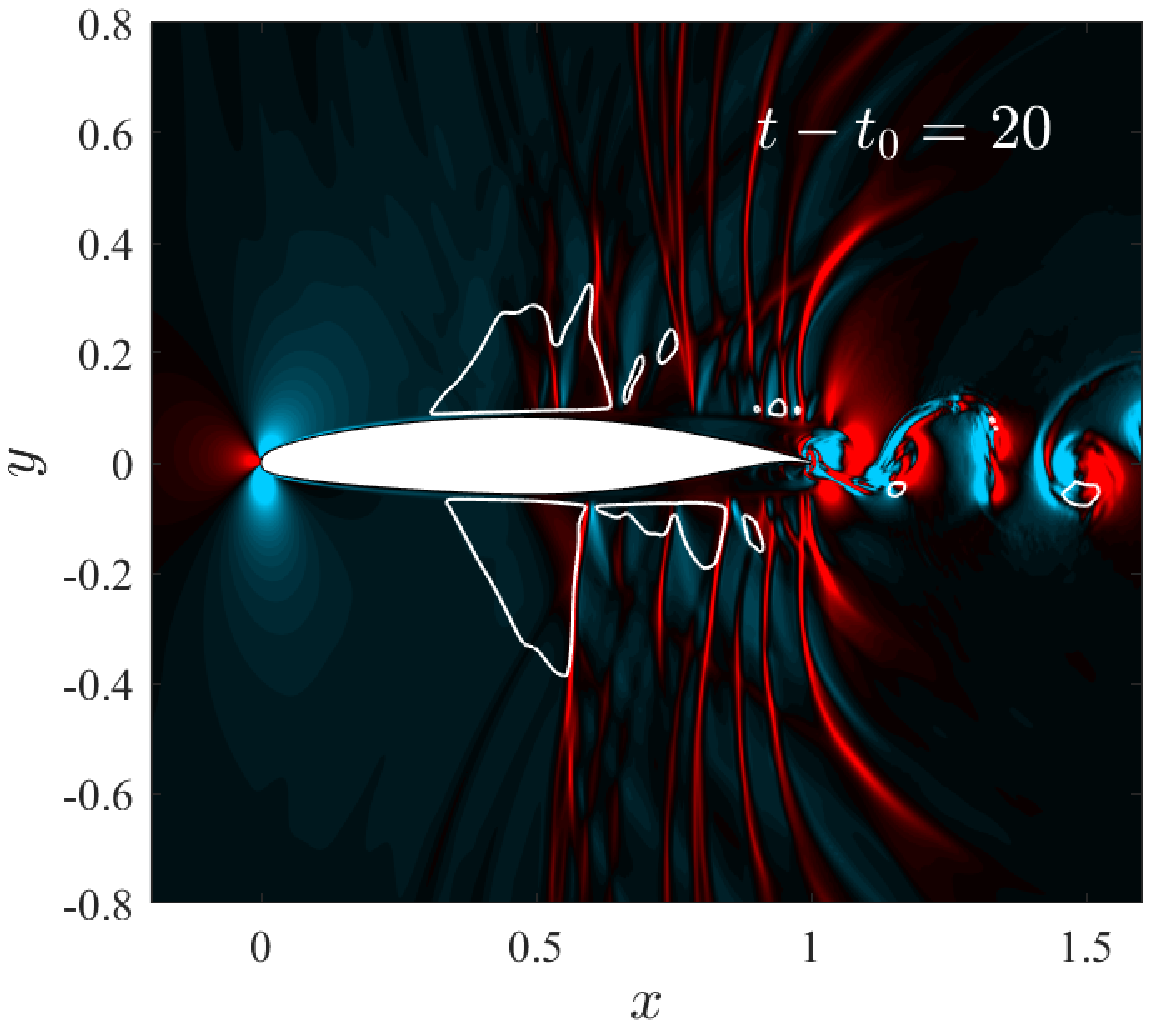}
\includegraphics[trim={1.2cm 0cm 0cm 0cm},clip,width=.32\textwidth]{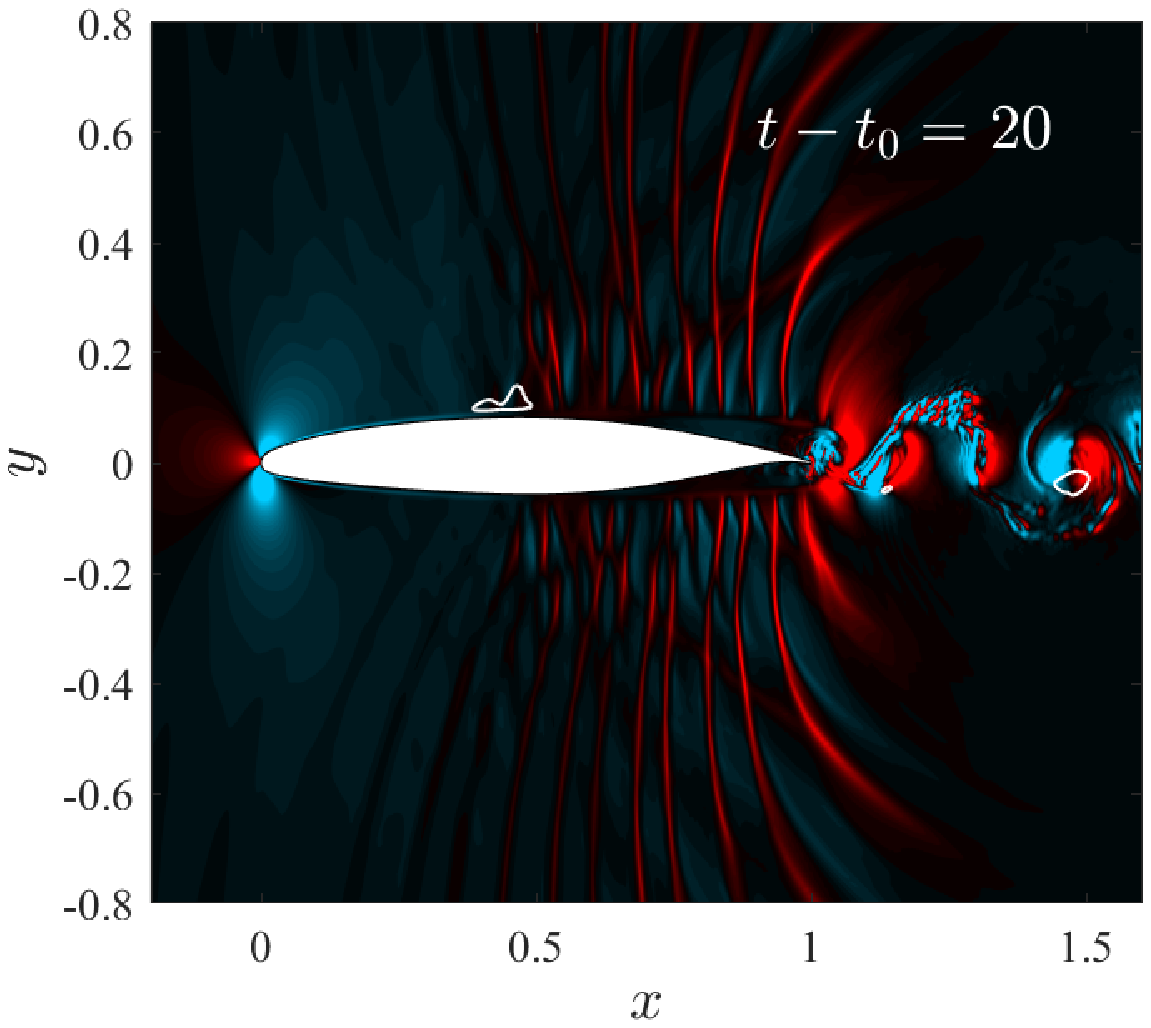}
\includegraphics[trim={1.2cm 0cm 0cm 0cm},clip,width=.32\textwidth]{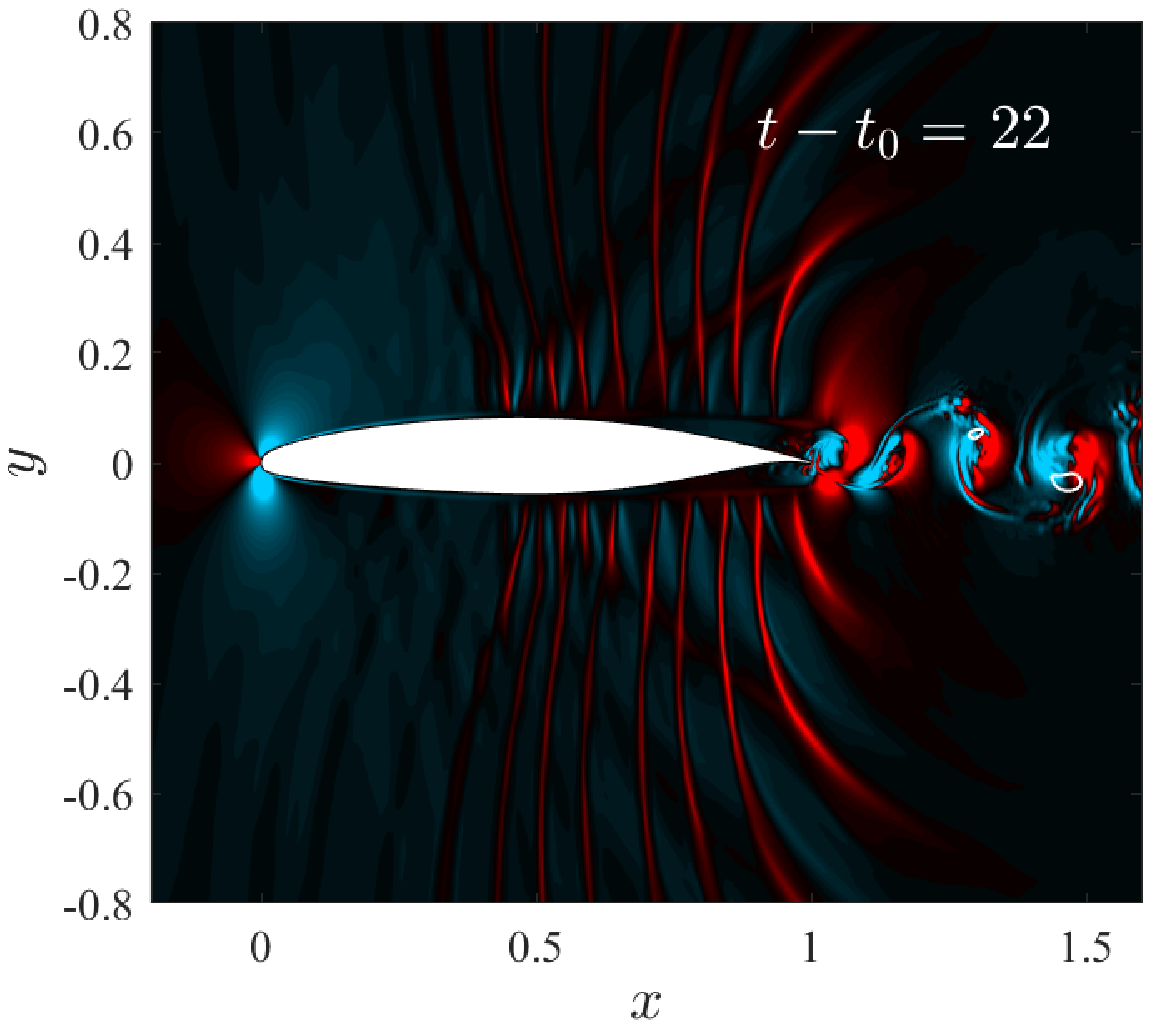}
\includegraphics[trim={1.2cm 0cm 0cm 0cm},clip,width=.32\textwidth]{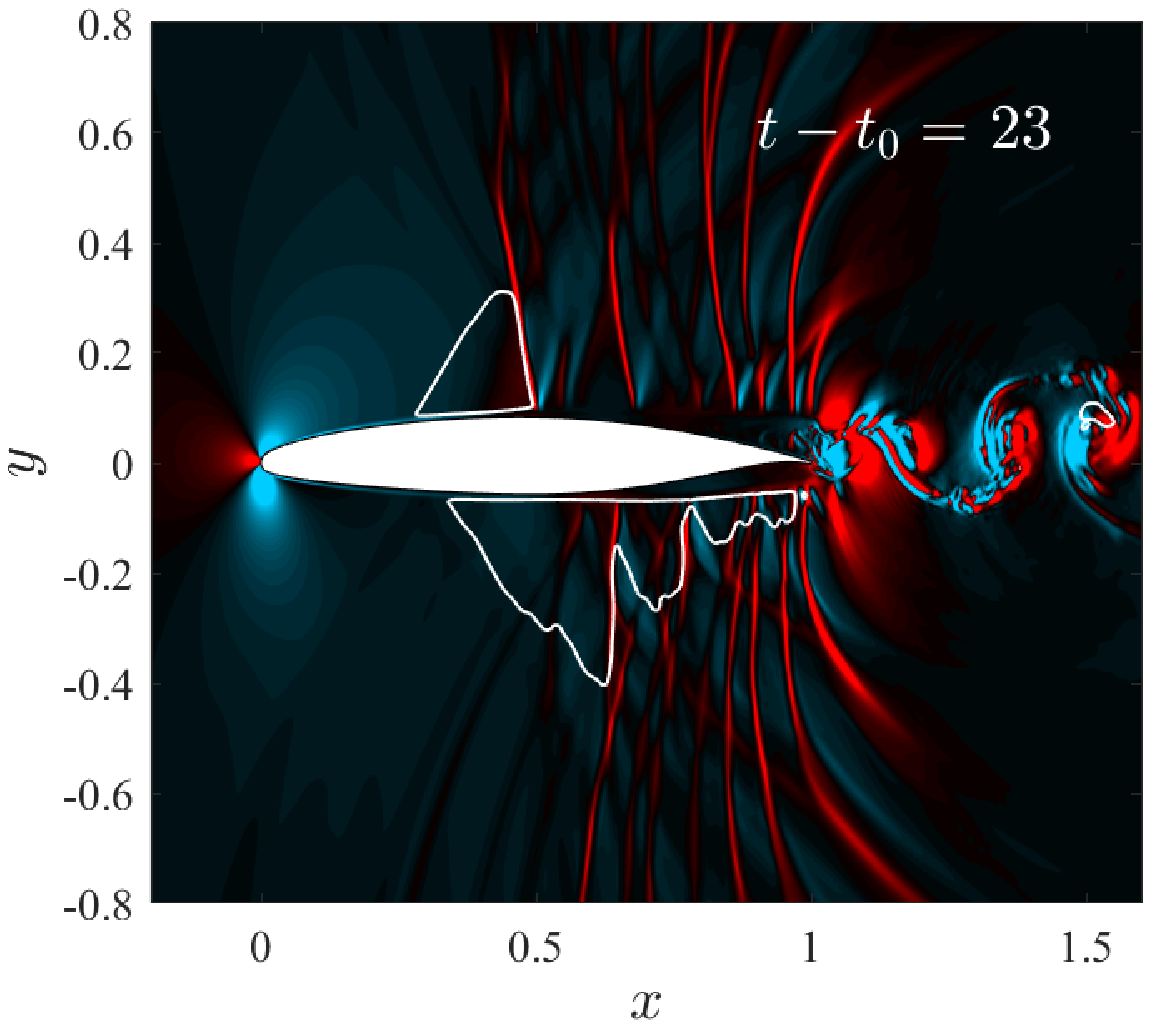}
\includegraphics[trim={1.2cm 0cm 0cm 0cm},clip,width=.32\textwidth]{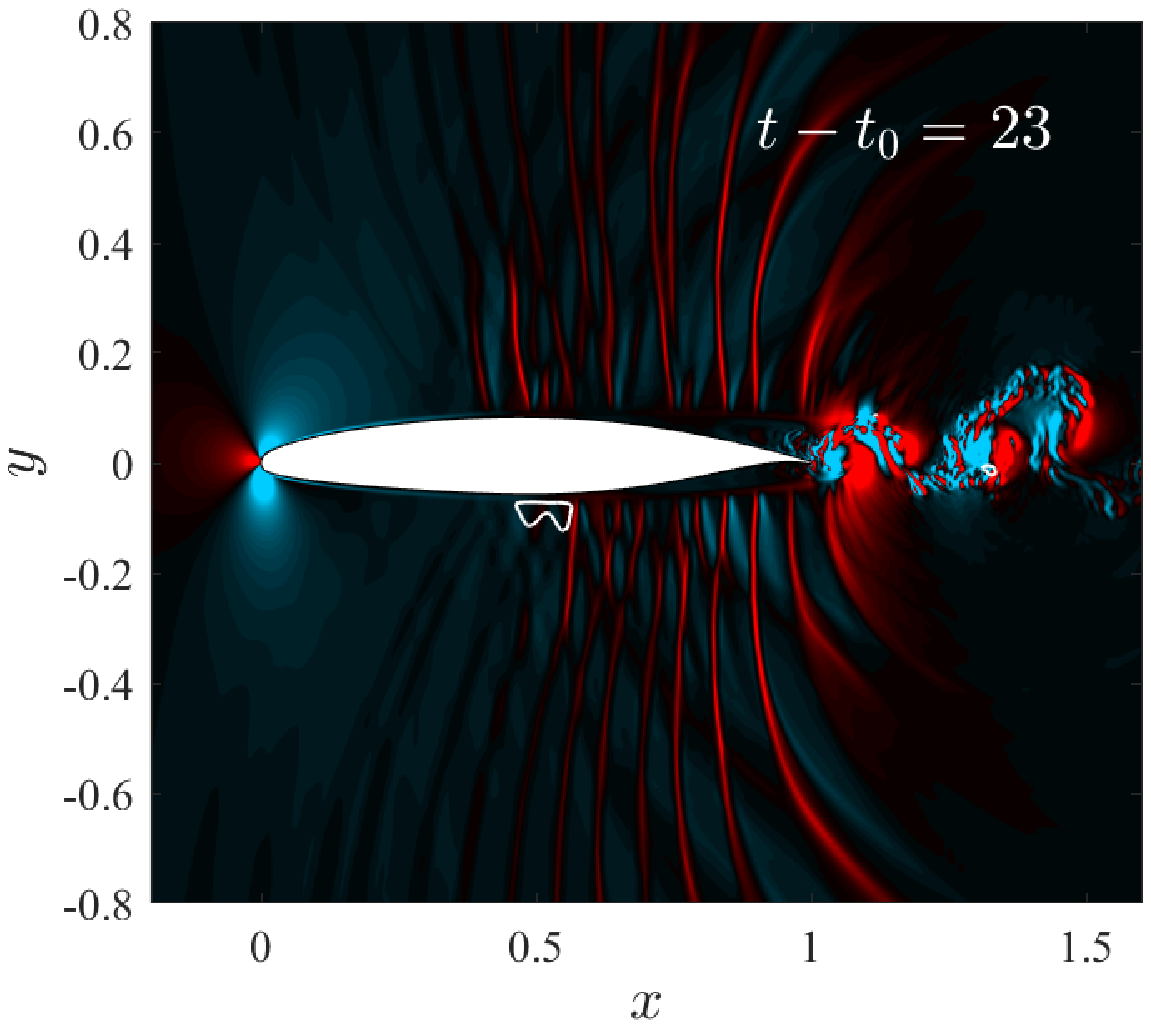}
\includegraphics[trim={1.2cm 0cm 0cm 0cm},clip,width=.32\textwidth]{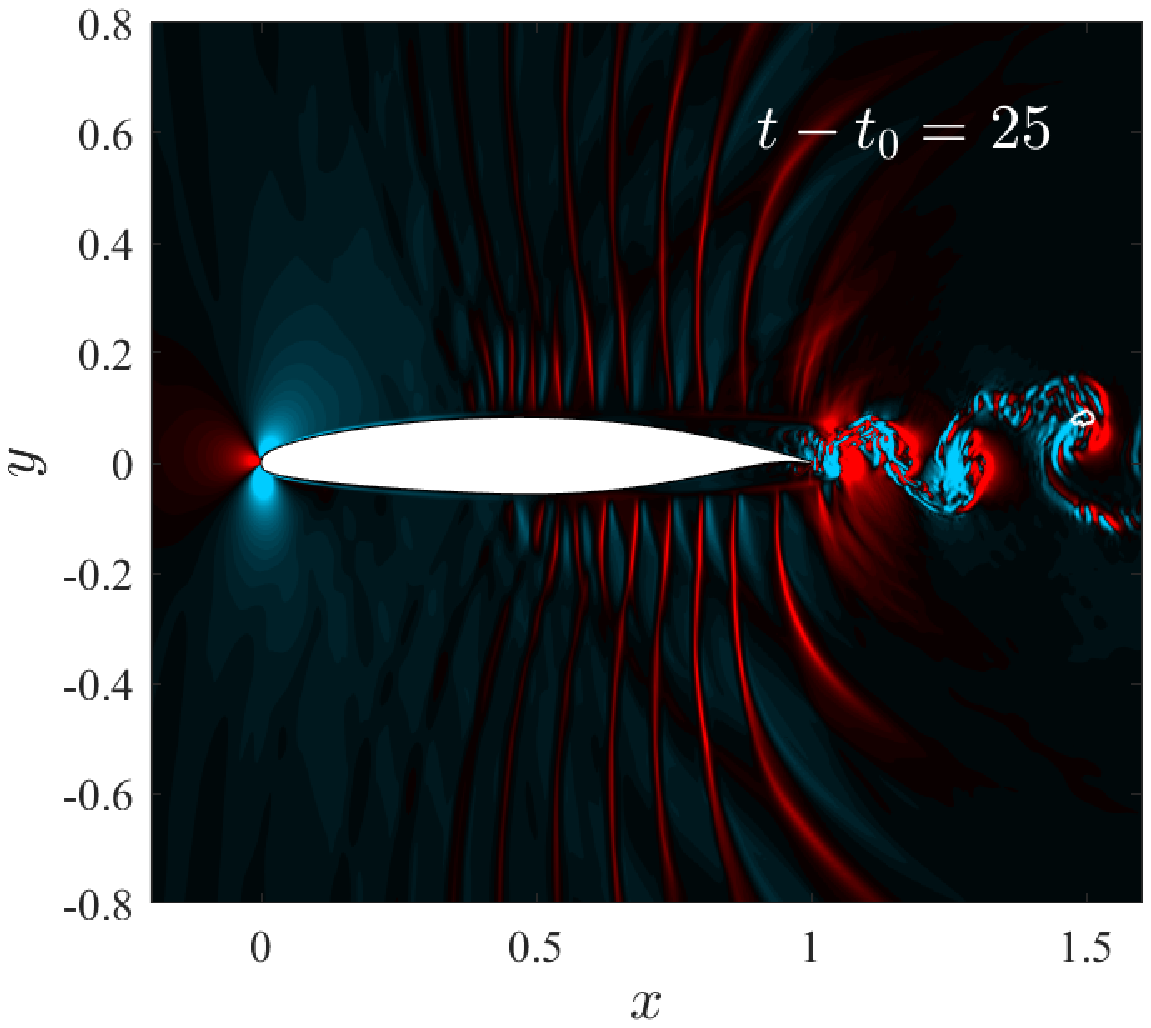}
\caption{Streamwise density gradient contours on the $x-y$ plane shown at approximate high-lift (top) and low-lift (bottom) phases for $M = 0.8$ (left), 0.77 (middle) and 0.75 (right), for the V2C aerofoil. The sonic line is highlighted using a white curve.}
    \label{fV2CDensGrad}
\end{figure}

\begin{figure} 
\centering
\includegraphics[trim={2.8cm 0cm 3.5cm 0cm},clip,width=.32\textwidth]{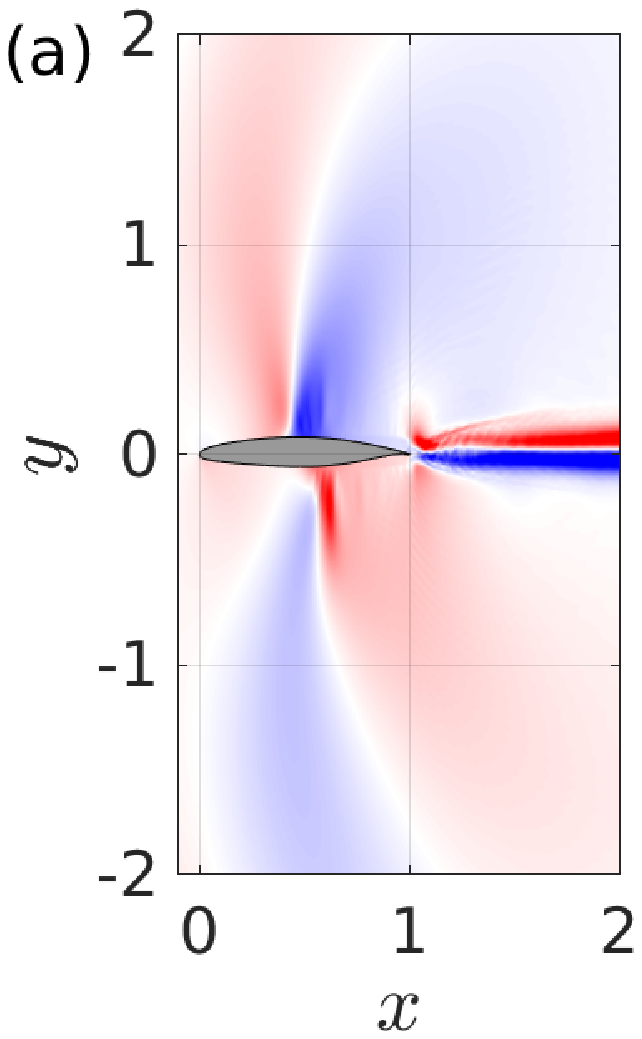}
\includegraphics[trim={2.8cm 0cm 3.5cm 0cm},clip,width=.32\textwidth]{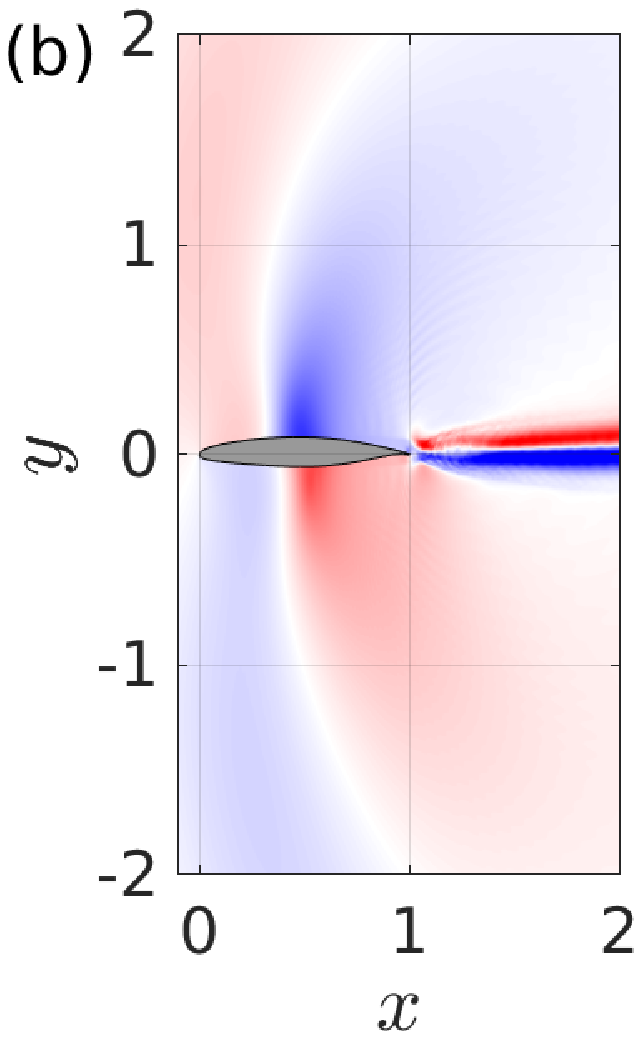}
\includegraphics[trim={2.8cm 0cm 3.5cm 0cm},clip,width=.32\textwidth]{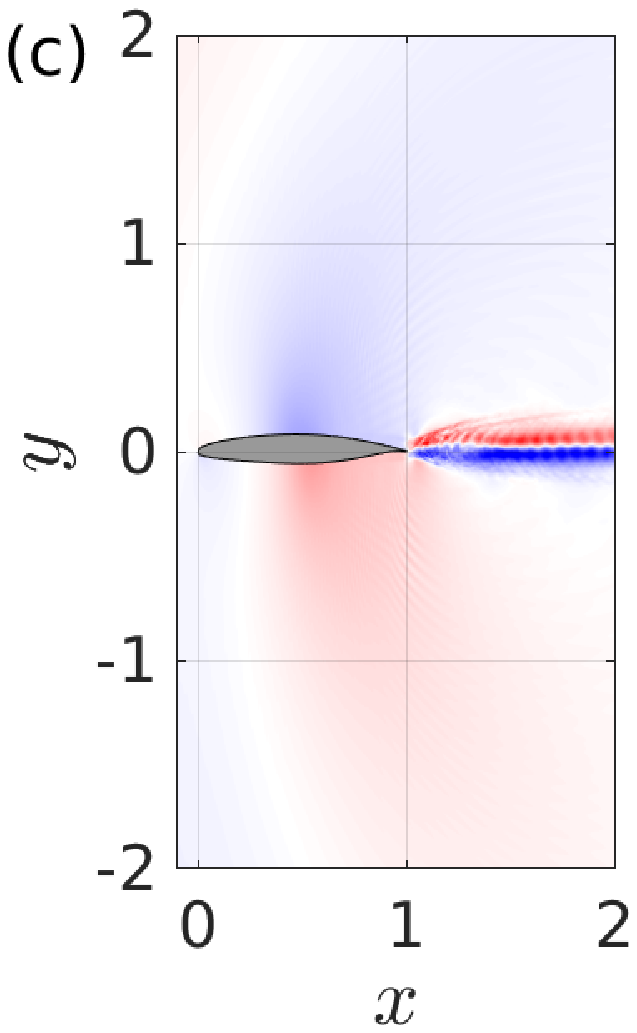}
\includegraphics[trim={2.8cm 0cm 3.5cm 0cm},clip,width=.32\textwidth]{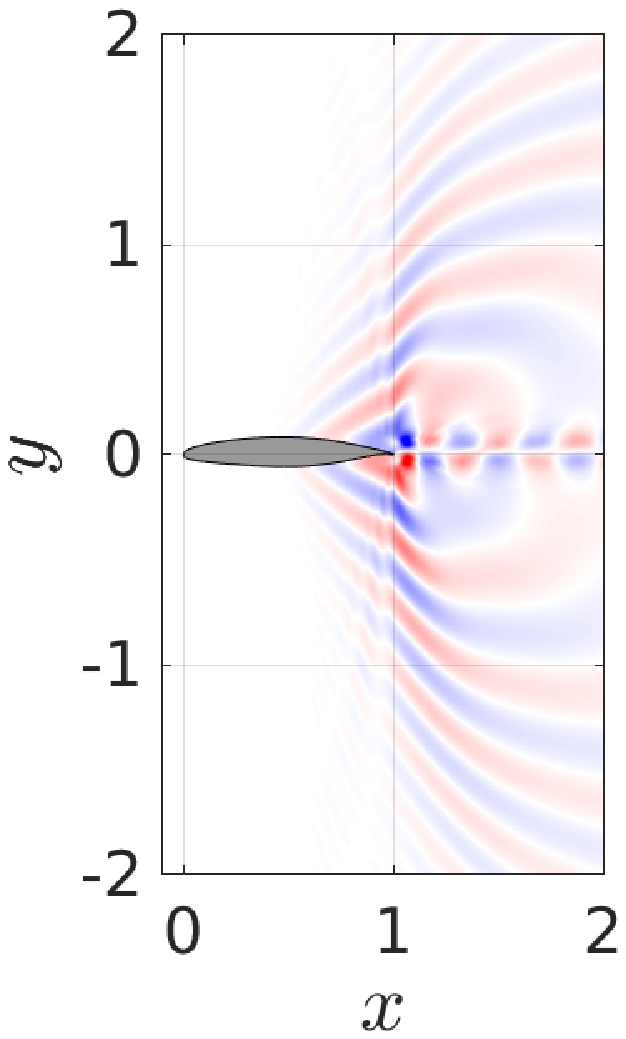}
\includegraphics[trim={2.8cm 0cm 3.5cm 0cm},clip,width=.32\textwidth]{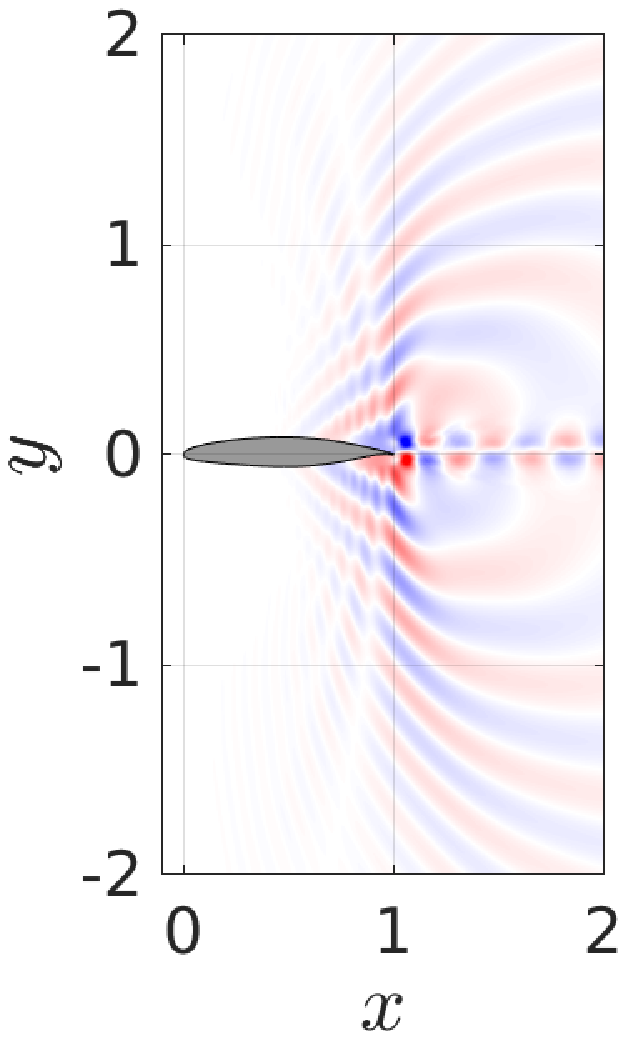}
\includegraphics[trim={2.8cm 0cm 3.5cm 0cm},clip,width=.32\textwidth]{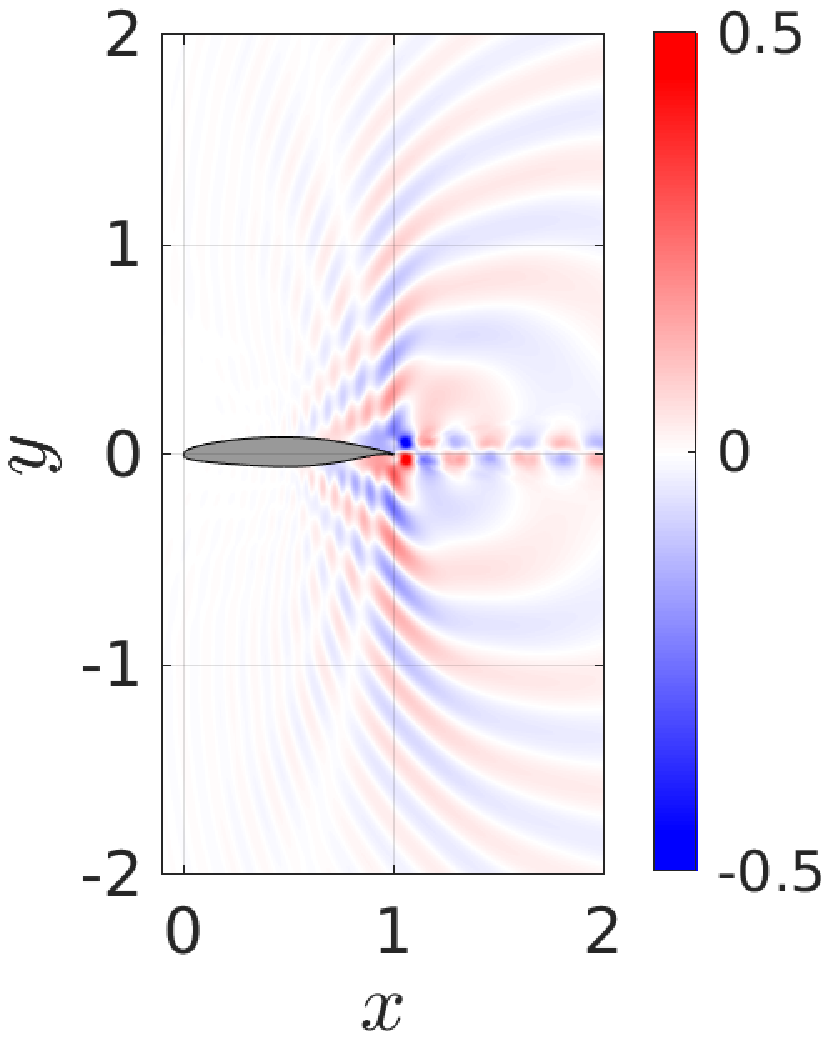}
\caption{Buffet (top) and wake (bottom) modes from SPOD for the V2C aerofoil are shown using contour plots of the pressure field at $Re = 5\times10^4$, $\alpha = 0^\circ$ for (a) $M = 0.8$, (b) $M = 0.77$ and (c) $M = 0.75$.}
    \label{fV2CSPODModes}
\end{figure}

Nevertheless, the trend seen for NACA0012 is also observed here, as seen from figure~\ref{fV2CDensGrad} which shows contours of streamwise density gradient on the $z = 0$ plane at high- and low-lift phases for the three $M$. At the highest $M = 0.8$, the flow exhibits Type I transonic buffet, with multiple shock waves and supersonic regions (delineated by the white isoline) which oscillate on either side of the aerofoil. At the intermediate $M = 0.77$, a small pocket of supersonic region occurs on the suction/pressure side in the high-/low-lift phase. At the lowest, $M = 0.75$, the flow field is entirely subsonic. Note that the V2C aerofoil is asymmetric, and thus, minor differences exist between the high- and low-lift phases, but there is a close match with trends seen for the NACA0012 aerofoil (\textit{e.g.}, the same trend is observed for the latter aerofoil at $M = 0.8$, 0.75 and 0.72 with other parameters the same). This is further corroborated by the buffet modes shown in figure~\ref{fV2CSPODModes} which also resemble those observed for the NACA0012 aerofoil, albeit with an additional region in the fore part of the aerofoil having an out-of-phase pressure variation with the aft part of the aerofoil. These results suggest that TBLO in the subsonic regime are a generic flow feature and could occur on other aerofoil profiles for similar flow conditions. 

\section{Effect of spanwise extent on Type I transonic buffet at $M=0.8$}
\label{appLzEffectOnM8}

\begin{figure} 
\centerline{
\includegraphics[width=0.45\textwidth]{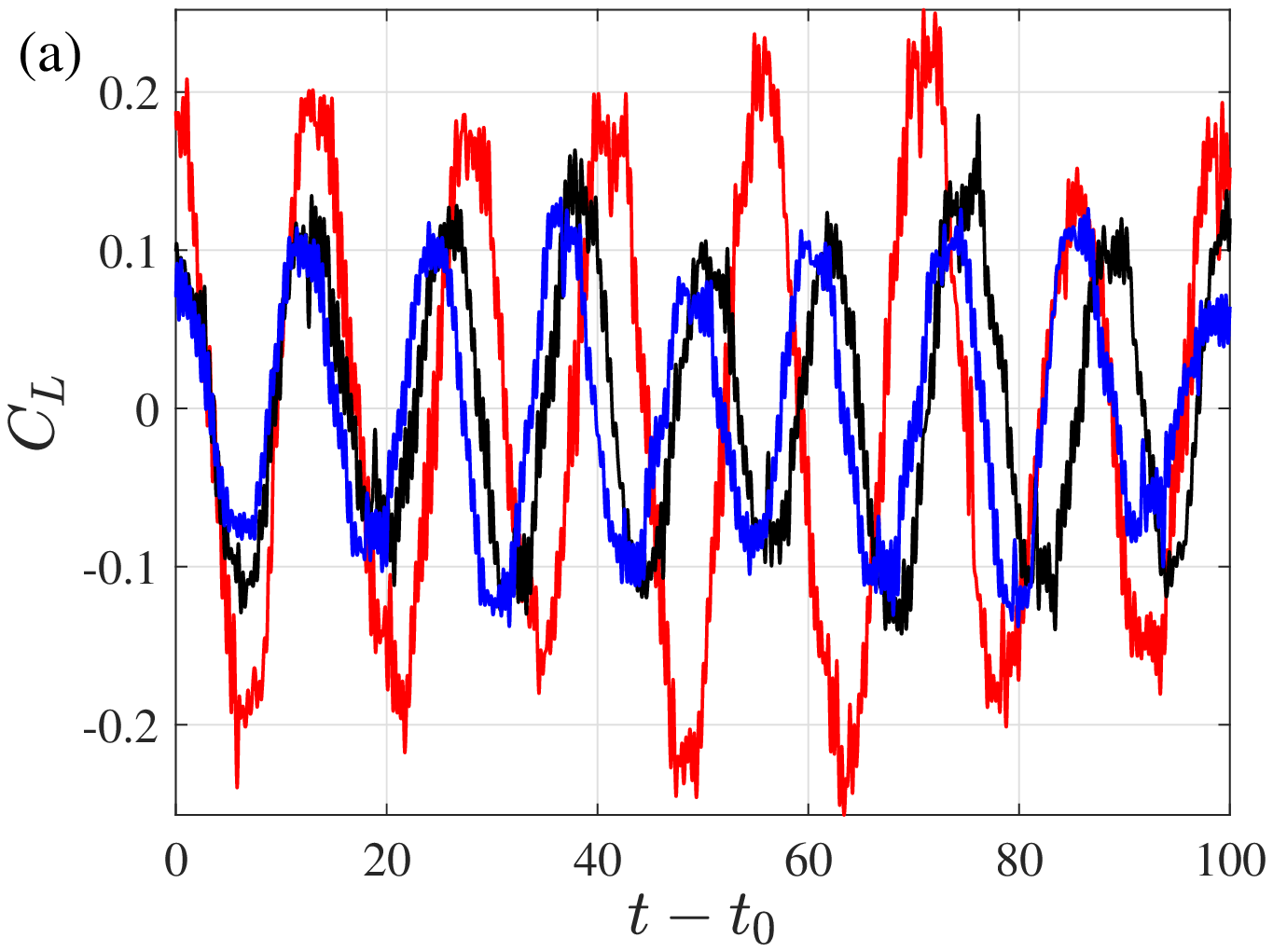}
\includegraphics[width=0.45\textwidth]{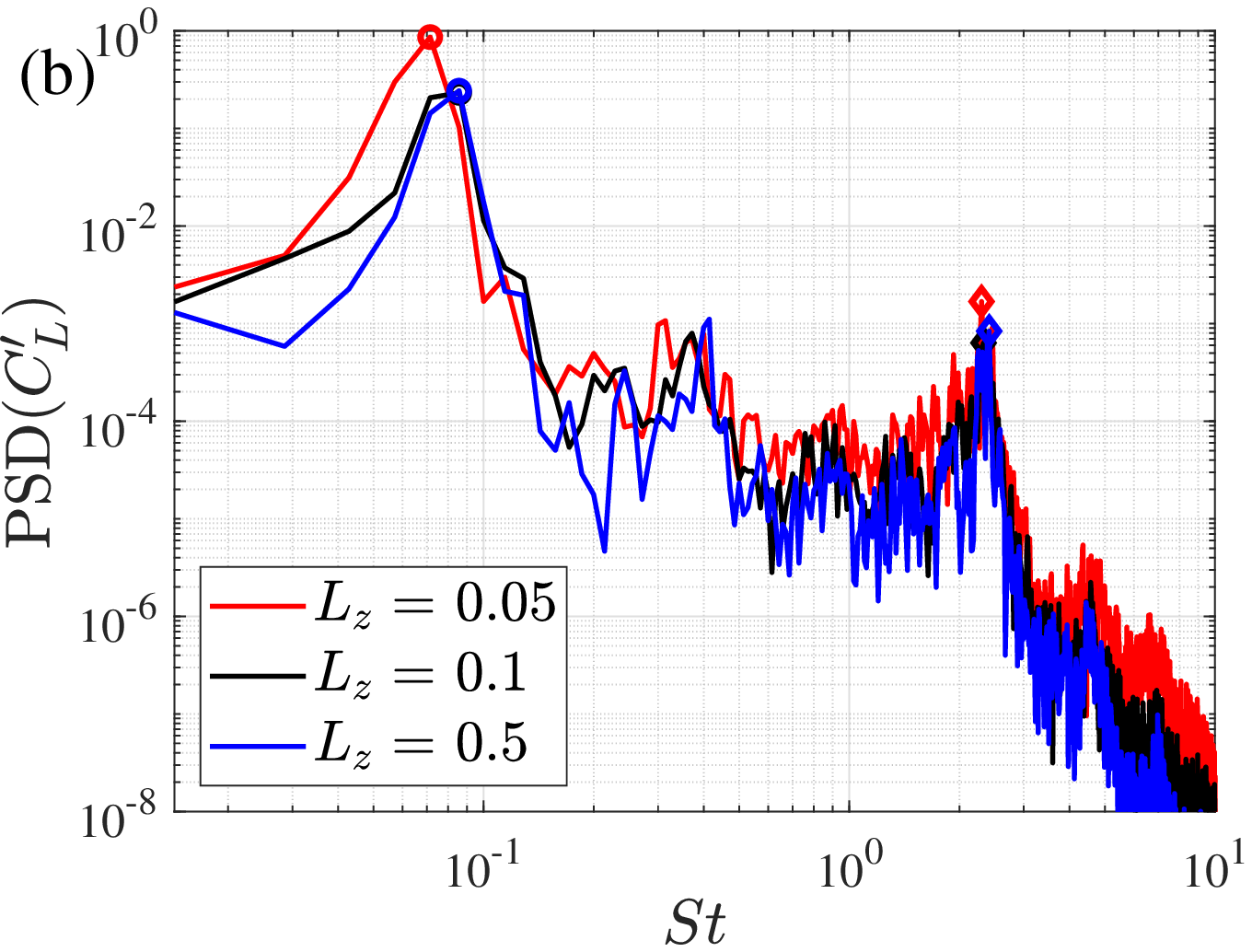}
}
\caption{(a) Temporal variation of lift coefficient past transients and (b) PSD of its fluctuating component as a function of the Strouhal number, $St$ for $M = 0.8$, $\alpha = 0^\circ$, $Re = 5\times10^4$ and NACA0012 aerofoil and different spanwise domain widths. Circles and diamonds highlight the buffet and wake mode Strouhal numbers ($St_b$ and $St_w$), respectively.}
\label{fNACAClPSDLzM8}
\end{figure}

The domain width, $L_z$, has been shown in \S\ref{subSecDomain} to strongly influence the buffet oscillations observed at $M = 0.75$, $\alpha = 0^\circ$ and $Re = 5\times10^4$ for the NACA0012 aerofoil. Note that at this $M$, the flow is only weakly transonic ($L_z = 0.05$, see figure~\ref{fNACADensGrad}\textit{b}) or entirely subsonic (higher $L_z$, see Supplementary Material). To see if this effect is also present when shock waves develop in the flow field, a few cases of $L_z > 0.05$ were simulated for $M = 0.8$ where Type I transonic buffet with shock waves was observed when $L_z = 0.05$, $\alpha = 0^\circ$ and $Re = 5\times10^4$ for the NACA0012 aerofoil. The temporal variation of the lift coefficient and its PSD are compared in figure~\ref{fNACAClPSDLzM8} for three different $L_z$. Due to associated numerical expense, other $L_z$ were not attempted. It is evident that the amplitude of the lift fluctuations is almost halved from approximately 0.2 to 0.1 when $L_z$ is doubled from 0.05 to 0.1. Thus, the trend of reducing buffet amplitude as $L_z$ is increased from 0.05 reported for $M = 0.75$ is also observed here. However, when $L_z$ is increased to 0.5, we see that the lift amplitude remains approximately the same implying that the effect of domain width is not significant for $L_z \geq 0.1$. This should be contrasted with the $M = 0.75$ case discussed in \S\ref{subSecDomain} where this approximately occurs for $L_z = 1$. Nevertheless, as with that case, the buffet frequency remains approximately the same for all $L_z$.

\bibliographystyle{jfm}
\bibliography{jfm}

\end{document}